\documentclass[final,3p,times]{elsarticle}

\usepackage{hyperref}
\usepackage{booktabs} 
\usepackage{amsmath,amsthm,mathtools}
\usepackage[inline]{enumitem}
\usepackage{tikz}
\usetikzlibrary{calc,positioning}
\usepackage{tikz-cd}
\usepackage{xcolor}
\usepackage{hyperref}
\usepackage{comment}
\usepackage{calc}
\usepackage{longtable}
\usepackage{todonotes}
\newtheorem{theorem}{Theorem} 
 \newtheorem{lemma}[theorem]{Lemma}
 \newtheorem{corollary}[theorem]{Corollary}
 \newtheorem{proposition}[theorem]{Proposition} 
 \newdefinition{remark}[theorem]{Remark}
 \newdefinition{definition}[theorem]{Definition}
 \newdefinition{example}[theorem]{Example}
 \newdefinition{fact}[theorem]{Fact}
 \newdefinition{notation}[theorem]{Notation}
 \newdefinition{framework}{Framework}
\newcommand{\FOL}{{\mathsf{FOL}}}

\newcommand{\HPL}{{\mathsf{HPL}}}
\newcommand{\FOHL}{{\mathsf{FOHL}}}
\newcommand{\FOHLR}{{\mathsf{FOHLR}}}
\newcommand{\FOHLS}{{\mathsf{FOHLS}}}
\newcommand{\RFOHL}{{\mathsf{RFOHL}}}
\newcommand{\Sig}{\mathtt{Sig}}
\newcommand{\Mod}{\mathtt{Mod}}
\newcommand{\Sen}{\mathtt{Sen}}

\newcommand{\nom}{\mathtt{n}}
\newcommand{\ac}{\mathtt{a}}
\newcommand{\bc}{\mathtt{b}}
\newcommand{\dc}{\mathtt{d}}
\newcommand{\rigid}{\mathtt{r}}
\newcommand{\flex}{\mathtt{f}}

\newcommand{\card}{\mathtt{card}}
\newcommand{\ext}{\mathtt{e}}
\newcommand{\ari}{\mathtt{ar}}
\newcommand{\Cat}{\mathbb{C}\mathsf{at}}
\newcommand{\Set}{\mathbb{S}\mathsf{et}}
\newcommand{\A}{\mathcal{A}}
\newcommand{\B}{\mathcal{B}}
\newcommand{\C}{\mathcal{C}}

\newcommand{\G}{\mathcal{G}}

\newcommand{\M}{\mathfrak{M}}
\newcommand{\N}{\mathfrak{N}}
\newcommand{\R}{\mathcal{R}}

\newcommand{\frag}{\mathcal{L}}

\newcommand{\at}[1]{@_{#1}\,}
\newcommand{\nec}[1]{[#1]}
\newcommand{\pos}[1]{\langle #1 \rangle}
\newcommand{\store}[1]{{\downarrow}#1\,{\cdot}\,}
\newcommand{\Forall}[1]{\forall #1\,{\cdot}\,}
\newcommand{\Exists}[1]{\exists #1\,{\cdot}\,}

\newcommand{\red}{\mathord{\upharpoonright}}
\usepackage{scalerel}

\setlist[enumerate,1]{
  font=\normalfont,
  label={\arabic*.},
  ref=\arabic*
}

\newlist{inlinenum}{enumerate*}{1}
\setlist[inlinenum,1]{
  font=\normalfont,
  label=(\emph{\alph*})
}

\setlist[itemize,1]{
}

\setlist[description]{
  font=\normalfont\em,
  leftmargin=\parindent
}

\newlist{plainlist}{itemize}{1}
\setlist[plainlist]{
  label={},
  leftmargin=0pt,
  itemsep=\parskip
}
\newcommand*{\pcformat}[1]{%
  [\;{\normalfont\itshape #1}\;]%
}

\usepackage{array,etoolbox}
\makeatletter

\newlength{\PS@lastparam}
\newlength{\PSlastparam}
\newcommand{\PSlp}{%
  \setlength{\PSlastparam}{\PS@lastparam}%
  \the\PSlastparam
}
\def\PS@sub@lastparam{}

\newcommand{\PS@numwidth}{99}
\newcommand{\PSnumwidth}[1]{%
  \renewcommand{\PS@numwidth}{#1}%
}

\newcommand{\PS@style}{\small}
\newcommand{\PS@numstyle}{\footnotesize}

\newlength{\PSindent}
\newlength{\PS@extraindent}
\newlength{\PSpre}
\newlength{\PSpost}
\newlength{\PS@Nwidth}
\newlength{\PS@Swidth}
\newlength{\PS@Ewidth}
\newlength{\PScolsep}
\newcommand{\PS@rownumber}{%
  \ifPS@subsubsteps
  \thePSsubstepc.%
  \the\numexpr\value{PSsubsubstepc}+1\relax
  \else
  \ifPS@substeps
  \thePSstepc.%
  \the\numexpr\value{PSsubstepc}+1\relax
  \else
  \the\numexpr\value{PSstepc}+1\relax
  \fi\fi
}
\newcommand{\PS@step}{%
  \ifPS@subsubsteps
  \refstepcounter{PSsubsubstepc}%
  \else
  \ifPS@substeps
  \refstepcounter{PSsubstepc}%
  \else
  \refstepcounter{PSstepc}
  \fi\fi%
}

\newif\ifPS@inprogress
\newif\ifPS@substeps
\newif\ifPS@subsubsteps
\newif\ifPS@continued
\newif\ifPS@subcontinued
\newcounter{PSc}
\newcounter{PSstepc}[PSc]
\newcounter{PSsubstepc}[PSstepc]
\renewcommand{\thePSsubstepc}{\thePSstepc.\arabic{PSsubstepc}}
\newcounter{PSsubsubstepc}[PSsubstepc]

\newenvironment{proofsteps}[1]{%
  \global\settowidth{\PS@lastparam}{\PS@style\hspace*{#1}}
  \ifPS@continued\else\refstepcounter{PSc}\fi
  \begingroup
  \setlength{\LTpre}{\PSpre}%
  \setlength{\LTpost}{\PSpost}%
  
  \setlength{\tabcolsep}{0pt}
  \noindent\PS@style
  \settowidth{\PS@Nwidth}{\PS@numstyle\PS@numwidth}%
  \setlength{\PS@Swidth}{#1}%
  \addtolength{\PS@Swidth}{-\PS@extraindent}%
  \setlength{\PS@Ewidth}{\linewidth}%
  \addtolength{\PS@Ewidth}{-\PSindent}%
  \addtolength{\PS@Ewidth}{-\PS@extraindent}%
  \addtolength{\PS@Ewidth}{-\PS@Nwidth}%
  \addtolength{\PS@Ewidth}{-\PScolsep}%
  \addtolength{\PS@Ewidth}{-\PS@Swidth}%
  \addtolength{\PS@Ewidth}{-\PScolsep}%
  \PS@inprogresstrue
  \longtable{%
    @{\hspace*{\PSindent}\hspace*{\PS@extraindent}\makebox[\PS@Nwidth][r]{\PS@rownumber}}%
    @{\hskip\PScolsep\PS@step}>{}p{\PS@Swidth}%
    @{\hskip\PScolsep}>{\footnotesize\raggedright\arraybackslash}p{\PS@Ewidth}%
  }%
}{%
  \ifPS@inprogress
  \addtocounter{table}{-1}%
  \endlongtable  
  \endgroup
  \PS@continuedfalse
  \PS@inprogressfalse
  \else\fi
}

\newcommand{\PSbreak}[1]{%
  \endproofsteps
  \par\medskip
  #1
  \medskip\par
  \PS@continuedtrue
  \proofsteps{\PS@lastparam}%
}

\newif\ifPS@sub@inprogress
\newenvironment{substeps}[1]{%
  \global\def\PS@sub@lastparam{#1}%
  \endproofsteps
  \settowidth{\PS@extraindent}{\PS@numstyle\PS@numwidth}%
  \addtolength{\PS@extraindent}{\PScolsep}%
  \addtolength{\PS@extraindent}{\widthof{\PS@numstyle.#1}}%
  \PS@substepstrue
  \PS@continuedtrue
  \PS@sub@inprogresstrue
  \proofsteps{\PS@lastparam}%
}{%
  \ifPS@sub@inprogress
    \endproofsteps
    \ifPS@laststep
      \PS@laststepfalse
      \PS@inprogressfalse
    \else
      \setlength{\PS@extraindent}{0pt}%
      \PS@substepsfalse
      \PS@continuedtrue
      \proofsteps{\PS@lastparam}%
    \fi
    \PS@sub@inprogressfalse
  \else
    \ifPS@laststep
      \PS@laststepfalse
      \PS@inprogressfalse
    \else
      \setlength{\PS@extraindent}{0pt}%
      \PS@substepsfalse
      \PS@continuedtrue
      \proofsteps{\PS@lastparam}%
    \fi
  \fi
}

\newif\ifPS@laststep
\newcommand{\laststep}{\global\PS@laststeptrue}
\newif\ifPS@lastsubstep
\newcommand{\lastsubstep}{\global\PS@lastsubsteptrue}

\newcommand{\adjustcol}[1]{%
  \global\advance\@colroom-#1%
}

\makeatother

\begin{document}

\begin{frontmatter}

\title{Forcing and interpolation in first-order hybrid logic with rigid symbols}

\author[a1]{Daniel G\u{a}in\u{a}}
\ead{daniel@imi.kyushu-u.ac.jp} 

\author[a2]{Go Hashimoto}%
\ead{go.427@s.kyushu-u.ac.jp}%

\address[a1]{Institute of Mathematics for Industry, Kyushu University} 
\address[a2]{Joint Graduate School of Mathematics for Innovation, Kyushu University} 

\begin{abstract}
In this paper, we establish an analogue of Craig Interpolation Property for a many-sorted variant of first-order hybrid  logic. We develop a forcing technique that dynamically adds new constants to the underlying signature in a way that preserves consistency, even in the presence of models with possibly empty domains. Using this forcing method, we derive general criteria that are sufficient for a signature square to satisfy Craig interpolation property.
\end{abstract}

\begin{keyword}
Institutional model theory \sep hybrid logic \sep forcing \sep Craig Interpolation

\MSC[2020] 03C25 \sep 03B45 \sep 03C99
\end{keyword}

\end{frontmatter}
\section{Introduction}
\paragraph{Motivation}
The many-sorted versions of logical systems (such as equational logic, first-order logic, etc.), as opposed to their unsorted counterparts, are recognized as particularly suitable for applications in computer science, especially in areas like formal methods.
In general, transitioning from the unsorted to the many-sorted case is far from trivial. Allowing for multiple sorts, and consequently multiple carriers in models -- some of which may be empty -- alters the properties of the logics and significantly increases the complexity of proofs.
These observations are well established in the literature of algebraic specification. 
They were already discussed by Goguen~\cite{journals/sigplan/GoguenM82} and have since been revisited in more recent works~\cite{gai-pop-rob,go-icalp24,hashimoto2025}.
An important example of a logical property that does not have a straightforward many-sorted generalization is Craig Interpolation Property (CIP)~\cite{Craig1957-CRALRA} and its equivalent formulation, Robinson's joint Consistency Property (RCP)~\cite{rob}. 
CIP is classically stated as follows: 
giving two first-order sentences $\phi_1$ in a language $\frag_1$ and $\phi_2$ in language $\frag_2$ such that $\phi_1\models\phi_2$ holds over the union language $\frag_1\cup\frag_2$, then there exists a first-order sentence $\phi$ in the intersection language $\frag_1\cap\frag_2$ such that $\phi_1\models \phi$ and $\phi\models \phi_2$.
Following the approach proposed in \cite{tar-bit}, 
we generalize the inclusion square  
\begin{center}
\begin{tikzcd}
\frag_2 \ar[r,hook]  & 
\frag_1 \cup \frag_2 \\
\frag_1\cap\frag_2 \ar[u,hook] \ar[r,hook] & \frag_1 \ar[u,hook]
\end{tikzcd} 
\end{center}
to a pushout of language translations/signature morphisms
\begin{center}
\begin{tikzcd}
\frag_2 \ar{r}[above]{\delta_2}  & 
\frag'\\
\frag \ar{u}[left]{\chi_2} \ar{r}[below]{\chi_1} & \frag_1 \ar{u}[right]{\delta_1}
\end{tikzcd} 
\end{center}
and replace the sentences $\phi_1$, $\phi$ and $\phi_2$ with sets of sentences $\Phi_1$, $\Phi$ and $\Phi_2$, obtaining the following formalization:
if $\delta_1(\Phi_1)\models\delta_2(\Phi_2)$ holds over $\frag'$, 
then there exists a set of sentences $\Phi$ in $\frag$ such that $\Phi_1\models \chi_1(\Phi)$ and $\chi_2(\Phi)\models \Phi_2$.
Consequently, this formulation accommodates logical systems that lack properties such as compactness, conjunction, and other classical connectives \cite{Tarlecki2024Fragility}.
The question of which pushout squares satisfy CIP has a straightforward answer in the unsorted setting: all pushout squares do.
On the other hand, characterizing the pushout squares that satisfy CIP in the general (sorted) case was an open problem, originally posed in \cite{tar-bit}. Nearly two decades later, a solution -- based on techniques from abstract model theory -- was finally provided in \cite{gai-pop-rob}.

In the foundations of system specification and formal development, applications of CIP primarily concern the combination and decomposition of theories. 
CIP formulated in terms of pushouts of signatures has played a central role in establishing fundamental modularity results. One of the earliest works highlighting its importance is \cite{DBLP:conf/fsttcs/MaibaumSV84}, where interpolation was used to guarantee the composability of successive implementation steps. This line of research was subsequently developed and refined through various versions of the modularization theorem~\cite{DBLP:journals/ipl/VelosoM95, DBLP:journals/ipl/Veloso96}.
A more widely recognized line of work is the development of module algebra \cite{DBLP:journals/jacm/BergstraHK90}, in which the interpolation theorem plays a central role in establishing key distributive laws for the export operator. 
Standard completeness proofs for proof calculi that handle consequences of structured specifications rely critically on interpolation \cite{Borzyszkowski2002}. In fact, no sound and complete proof calculus of this kind can exist without an appropriate interpolation property for the underlying logic~\cite{DBLP:journals/mscs/SannellaT14}.
\paragraph{Many-sorted first-order hybrid  logic with rigid symbols ($\FOHLR$)}
The logical system in which the results of this paper are established is $\FOHLR$, a logic whose models interpret a designated subset of signature symbols uniformly across all possible worlds. The semantics of $\FOHLR$, including its signatures and Kripke structures, coincides with that of First-Order Hybrid Logic with user-defined Sharing ($\FOHLS$)~\cite{dia-msc}.\footnote{Its Kripke structures can be viewed as multigraphs whose nodes are many-sorted first-order models.} 
The latter is obtained through the hybridization process~\cite{martins2011,dia-msc}, a generic method of hierarchical logic combination in which the characteristic features of hybrid logic are systematically added to an underlying base logic. In case of $\FOHLS$, the base logic is many-sorted first-order logic where models interpret sorts as sets, possibly empty.
A distinctive feature of $\FOHLS$ is that it allows users to specify which signature symbols are interpreted uniformly across all states and which may vary from one state to another.

The syntax of $\FOHLR$ is inspired by the logic introduced in~\cite{DBLP:conf/wollic/BlackburnMMH19}\label{ex:RFOHL}, where the retrieve operator $@$ is extended from formulas to terms.
While in standard hybrid logics this operator is used to change the point of evaluation of a formula, in $\FOHLR$ it can also be applied to terms, allowing them to refer explicitly to values at different worlds. 
It is worth noting that $\FOHLR$ introduced in this paper and $\FOHLS$ studied by Diaconescu and collaborators~\cite{martins2011,dia-msc,dia-qvh} have the same expressive power~\cite{gai-acm}. 
More precisely, the relationship between $\FOHLR$ and $\FOHLS$ is analogous to that between first-order logic and its unnested fragment, where terms are restricted to depth at most one~\cite{DBLP:books/daglib/0080659}.
Thus, $\FOHLR$ provides an alternative syntactic presentation of $\FOHLS$ while preserving its expressive capabilities.
This feature is essential for defining hybrid substitutions, which are a key ingredient in generalizing forcing from first-order logic to hybrid logical systems.
\paragraph{Forcing}
Forcing is a model-construction technique based on consistency that has proved to be a powerful tool for establishing important model-theoretic and proof-theoretic results. Originally introduced by Paul Cohen~\cite{cohen63,cohen64} to prove the independence of the Continuum Hypothesis from Zermelo-Fraenkel set theory, it was later generalized by Abraham Robinson~\cite{rob71} to a model-theoretic setting. Robinson's generalization was particularly successful for infinitary logics, where compactness, one of the central tools of first-order model theory, is unavailable. In institutional model theory, forcing was introduced in~\cite{gai-com} to prove a G\"odel Completeness Theorem, then extended to stratified institutions~\cite{gai-acm} to obtain completeness results for a large class of hybrid logics. It was subsequently adapted to logics combining hybrid and dynamic features, leading to an Omitting Types Theorem for the fragment of $\FOHLR$~\cite{DBLP:journals/apal/GainaBK23} obtained by restricting the semantics to models which interpret sorts as non-empty sets.

A common feature of these aforementioned developments is that forcing is formulated with respect to a single signature and is therefore most naturally suited to logical systems with non-empty carrier sets. To overcome this limitation, a new forcing technique was recently introduced in~\cite{go-icalp24,hashimoto2025}, where a forcing property is defined over a category of signatures rather than a fixed signature. This framework allows new constants to be introduced dynamically during model construction while preserving consistency, thereby enabling the construction of models with possibly empty carrier sets in many-sorted logical systems. In this paper, we extend this approach to hybrid logics by combining ideas from~\cite{gai-acm} and~\cite{go-icalp24}, and use the resulting forcing framework to obtain a direct proof of the CIP. To the best of our knowledge, this is the first forcing-based proof of the CIP in modal logic.
\paragraph{Contribution}
In \cite{ArecesBM01}, Areces et al.\ established the CIP for hybrid propositional logic, and in \cite{ArecesBM03} they extended this result to hybrid predicate logic with constant (rigid) domains. Although our results are closely related, they cannot be derived from theirs. The main reason is that the framework of \cite{ArecesBM03} is restricted to Kripke structures with constant domains, whereas our setting permits variable domains.
In addition, following standard practice in algebraic specification, we work in a many-sorted framework and consider arbitrary pushouts of signatures (see, e.g., \cite{tar-bit}), rather than only signature inclusions.
More recently, \cite{GBK22} established the RCP for a fragment of our hybrid logic using the Omitting Types Property, under the assumption that sorts are interpreted by non-empty sets. Subsequently, RCP was shown to be equivalent to CIP. 
In contrast, the present paper establishes CIP directly by means of forcing, as shown in Theorem~\ref{th:interpolation}, the central result of the present work.

An equally important contribution of the paper is the collection of counterexamples presented in Lemmas~\ref{lemma:preservation}--\ref{lemma:counter-empty}. 
By removing the assumptions of the main theorem one at a time, these examples demonstrate that each condition is essential for the validity of CIP. 
Taken together, they yield a precise criterion characterizing when a pushout square of signatures satisfies CIP. 
Furthermore, the example in Lemma~\ref{lemma:sort-inj} shows that allowing models with possibly empty domains invalidates the criterion of \cite{GBK22}, thereby highlighting a fundamental difference between the non-empty and possibly empty domain settings.
\paragraph{Institution model theory}
Our approach adopts the perspective of institution theory, an abstract framework for logical systems proposed in \cite{gog-ins}. This framework provides the level of generality required for abstract model theory while remaining well suited to practical applications, particularly in system specification and verification. Diaconescu's monograph~\cite{Diaconescu2008} offers a comprehensive overview of this line of research, while subsequent developments are dispersed across a wide range of articles.
For simplicity, we work within a concrete example of hybrid logic, applying the modularization principles of institution theory.
These principles revolve around signature morphisms and the satisfaction condition (ensuring that truth remains invariant under changes in notation). This approach introduces certain distinctive features, such as treating variables as special constants and the pervasive role of signature morphisms -- maps between signatures.
\paragraph{Structure of the paper}
Section~\ref{sec:FOL} recalls the definition of many-sorted first-order logic as formulated in institutional model theory~\cite{gog-ins}. Section~\ref{sec:FOHLR} introduces First-Order Hybrid Logic with Rigid Symbols ($\FOHLR$), originally presented in~\cite{gai-acm}. Section~\ref{sec:basic} develops the basic notions and fundamental results required for proving the main theorem, namely interpolation for pushouts of $\FOHLR$ signatures. Section~\ref{sec:forcing} generalizes forcing from the single-signature setting of~\cite{gai-acm} to categories of signatures, building on ideas from~\cite{go-icalp24} that are applicable to many-sorted logical systems with possibly empty domains. Section~\ref{sec:CIP} contains the main interpolation theorem together with a collection of counterexamples that lead to an optimal criterion characterizing when a pushout of $\FOHLR$ signatures satisfies CIP. Finally, Section~\ref{sec:conclusion} concludes the paper and outlines directions for future research.
\section{The base logic}\label{sec:FOL}
In this section, we recall the definition of many-sorted first-order logic in institutional model theory~\cite{gog-ins}, which provides the base logic from which first-order hybrid logic with rigid symbols is constructed.
\subsection{Many-sorted first-order logic ($\FOL$)}
This subsubsection establishes the notational conventions and reviews the fundamental definitions that will be needed in the subsequent sections.
\paragraph{Signatures}
Signatures are of the form $(S,F,P)$ where:
\begin{enumerate}
\item $S$ is a set of sorts;
\item $F$ is a set of function symbols, each written as $\sigma:\ari\to s$,
where $\sigma$ is the symbol name, $\ari\in S^*$ is the arity, and $s\in S$ is the sort;\footnote{$S^*$ denotes the set of all strings with elements from $S$.}
\item $P$ is a set of relation symbols, each written as $\pi:\ari$,
where $\pi$ is the symbol name and $\ari\in S^*$ is the arity.
\end{enumerate}
If $\ari=\varepsilon$, then a function symbol $\sigma:\ari\to s$ is called a
\emph{constant symbol}. Generally, $\ari$ ranges over arities, which are understood here as strings of sorts; in other words, an arity specifies the number of arguments together with their sorts.
Throughout this paper, we let $\Sigma$, $\Sigma'$ and $\Sigma_i$ range over first-order signatures of the form $(S,F,P)$, $(S',F',P')$ and $(S_i,F_i,P_i)$, respectively.
\paragraph{Signature morphisms}
A number of common techniques, such as adding constants and, importantly, introducing quantification, are regarded as signature expansions. Consequently, moving between signatures is a frequent operation. 
To make such transitions smooth, we use a notion of \emph{signature morphism}.
A signature morphism $\chi\colon \Sigma\to \Sigma'$ is a triple
$\chi=(\chi^{st},\chi^{op},\chi^{rl})$ of maps:
\begin{enumerate}
\item~$\chi^{st}\colon S\to S'$ is a function;
\item~$\chi^{op}:F\to F'$ is a function that maps each function symbol $\sigma:\ari\to s\in F$ to a function symbol $\sigma':\chi^{st}(\ari)\to \chi^{st}(s)\in F'$;
\item~$\chi^{rl}: P\to P'$ is a function that maps each relation symbol $\pi:\ari\in P$ to a relation symbol $\pi':\chi^{st}(\ari)\in P'$.
\end{enumerate}
When there is no risk of confusion, we may let $\chi$ denote any of
$\chi^{st}$, $\chi^{op}$, or $\chi^{rl}$.
\begin{fact}\label{fact:sig}
First-order signature morphisms form a category $\Sig^\FOL$ under componentwise composition of functions.
\end{fact}
\paragraph{Models} 
Given a signature $\Sigma$, a $\Sigma$-model is a triple
\begin{center}
$\mathfrak{A}=(\{\mathfrak{A}_s\}_{s\in S},\{\sigma^{\mathfrak{A}}\}_{\sigma:\ari\to s\in F},\{\pi^{\mathfrak{A}}\}_{\pi:\ari\in P})$
\end{center}
interpreting  each sort $s$ as a set $\mathfrak{A}_s$ (possibly empty), 
each function symbol $\sigma:\ari\to s$ as a function $\sigma^{\mathfrak{A}}: \mathfrak{A}_\ari\to \mathfrak{A}_s$ (where $\mathfrak{A}_\ari$ stands for $\mathfrak{A}_{s_1}\times\ldots\times \mathfrak{A}_{s_n}$ if $\ari=s_1\ldots s_n$), and
each relation symbol $\pi:\ari$ as a relation $\pi^{\mathfrak{A}}\subseteq \mathfrak{A}_\ari$.
Given a signature $\Sigma$, the $S$-sorted set of $\Sigma$-terms is denoted by $T_\Sigma$.
\footnote{An $S$-sorted set $\A$ is a family of sets $\{\A_s\}_{s\in S}$ indexed by the elements of the set $S$.
Similarly, we can define $S$-sorted functions $f:\A\to\B$ as family of functions $\{f_s:\A_s\to\B_s\}_{s\in S}$, one for each sort $s\in S$.
For all strings of sorts $s_1\dots s_n\in S^*$, we make the following notational conventions $\A_{s_1\dots s_n}\coloneqq\A_{s_1}\times\dots\times \A_{s_n}$ and $f_{s_1\dots s_n}\coloneqq f_{s_1}\times\dots\times f_{s_n}$.}
Moreover, $T_\Sigma$ can be organized naturally as a $\Sigma$-model.
Morphisms between models $h:\mathfrak{A}\to\mathfrak{B}$ are the usual $\Sigma$-homomorphisms, that is, 
$S$-sorted functions $\{h_s:\mathfrak{A}_s\to \mathfrak{B}_s\}_{s\in S}$ that preserve the structure:
\begin{enumerate}
\item $\sigma^\mathfrak{A};h_s=h_\ari;\sigma^\mathfrak{B}$ for all function symbols $\sigma:\ari\to s\in F$; and
\item $h_\ari(\pi^\mathfrak{A})\subseteq \pi^\mathfrak{B}$ for all relation symbols $\pi:\ari\in P$.
\end{enumerate} 
\begin{fact} \label{fact:mod}
For any signature $\Sigma$, the $\Sigma$-homomorphisms form a category $\Mod^\FOL(\Sigma)$ under the obvious composition as many-sorted functions. 
\end{fact}
For any signature morphism $\chi\colon \Sigma\to \Sigma'$, 
the reduct functor $\_\red_\chi\colon\Mod(\Sigma')\to\Mod(\Sigma)$ is defined as follows: 
\begin{enumerate}
\item The reduct $\mathfrak{A}'\red_\chi$ of a $\Sigma'$-model $\mathfrak{A}'$ is a defined by
$({\mathfrak{A}'\red_\chi})_s=\mathfrak{A}'_{\chi(s)}$ for each sort $s\in S$, and
$x^{\mathfrak{A}'\red_\chi}=\chi(x)^{\mathfrak{A}'}$ for each
function symbol $x:\ari\to s\in F$ or relation symbol $x:\ari\in P$.
Note that, unlike the single-sorted case, the reduct functor modifies the universes of models. 
For the universe of $\mathfrak{A}'\red_\chi$ is $\{\mathfrak{A}'_{\chi(s)}\}_{s\in S}$, which means that the sorts outside the image of $S$ are discarded. 
Otherwise, the notion of reduct is standard. 
\item The reduct $h'\red_\chi$ of a homomorphism $h'$ is defined by
  $(h'\red_\chi)_s=h'_{\chi(s)}$ for all sorts $s\in S$. 
\end{enumerate}
In the case of inclusions $\chi:\Sigma\hookrightarrow\Sigma'$, the corresponding reduct functor $\red_\chi$ is denoted by $\red_\Sigma$.
\begin{fact}
$\Mod^\FOL$ becomes a functor $\Sig^\FOL\to \Cat^{op}$, with $\Mod^\FOL(\chi)(h') = {h'\red_\chi}$ for each signature morphism $\chi\colon\Sigma\to\Sigma'$ and each $\Sigma'$-homomorphism $h'$.
\end{fact}
\paragraph{Sentences} 
We assume a countably infinite set of variable names $\{v_i\mid i<\omega\}$.
A variable for a signature $\Sigma$ is a triple $\pos{v_i,s,\Sigma}$, where $v_i$ is a variable name, and $s$ is a sort in $\Sigma$.
The set $\Sen^\FOL(\Sigma)$ of sentences over $\Sigma$ is given by the following grammar:
$$\phi \Coloneqq  t= t' \mid \pi(t_1,\ldots,t_n)\mid \neg\phi \mid \lor\Phi \mid \Exists{x}\phi_x$$ 
where
\begin{enumerate*}[label=(\alph*)]
\item~$t= t'$ is an equation with $t,t'\in T_{\Sigma,s}$ and $s\in S$,
\item~$\pi(t_1,\ldots,t_n)$ is a relational atom with $\pi\in P_{s_1\ldots s_n}$, $t_i\in T_{\Sigma,s_i}$ and $s_i\in S$, 
\item~$\Phi$ is a finite set of $\Sigma$-sentences,
\item~$x$ is a variable for $\Sigma$,
\item~$\phi_x$ is a $\Sigma[x]$-sentence, where $\Sigma[x]=(S,F[x],P)$, and $F[x]=F\cup\{x:\to s\}$, where $s$ is the sort of $x$.
\end{enumerate*}
Other Boolean connectives and the universal quantification can be defined as abbreviations of the above sentence building operators.
\paragraph{Sentence translations} 
Quantification comes with some subtle issues related to the translation of quantified sentences along signature morphisms that require a closer look.
The translation of a variable $\pos{v_i,s,\Sigma}$ along a signature morphism $\chi\colon\Sigma\to\Sigma'$ is $\pos{v_i,\chi(s),\Sigma'}$.
The sentence translations are defined by induction on the structure of sentences simultaneously for all signature morphisms $\chi\colon\Sigma\to\Sigma'$:
\begin{itemize}
\item $\chi(t=t') \coloneqq \chi(t)=\chi(t')$, 
where the function $\chi:T_\Sigma\to T_{\Sigma'}$ is formally defined by
$\chi(\sigma(t_1,\dots,t_n))=\chi(\sigma)(\chi(t_1),\dots,\chi(t_n))$ 
for all function symbols $\sigma:s_1\dots s_n\to s\in F$ and all terms $t_i\in T_{\Sigma,s_i}$, where $i\in\{1,\dots,n\}$.
\item $\chi(\pi(t_1,\dots,t_n))\coloneqq \chi(\pi)(\chi(t_1),\dots,\chi(t_n))$.
\item $\chi(\neg\phi)\coloneqq \neg\chi(\phi)$.
\item $\chi(\lor\Phi)\coloneqq\lor\chi(\Phi)$.
\begin{figure}[h]
\begin{center}
\begin{tikzcd}
\phi_x\ar[r,no head,dotted]& \Sigma[x] \ar[r,"\chi'"] & \Sigma'[x'] & 
\chi'(\phi_x) \ar[l,dotted,no head]\\
& \Sigma \ar[u,hook] \ar[r,"\chi",swap] & \Sigma' \ar[u,hook]
\end{tikzcd}
\end{center}
\caption{Quantification pushout}
\label{fig:quant-pushout}
\end{figure}
\item $\chi(\Exists{x}\phi_x)\coloneqq\Exists{x'}\chi'(\phi_x)$, where 
$x'$ is the translation of $x$ along $\chi$,
(that is, $x'=\pos{v_i,\chi(s),\Sigma'}$ when $x\pos{v_i,s,\Sigma}$)
and
$\chi':\Sigma[x]\to \Sigma'[x']$ is the extension of $\chi$ 
which maps  $x$ to $x'$.
Notice that $\chi'(\phi_x)$ is well-defined, 
as $\phi_x$ has a simpler structure than $\Exists{x}\phi_x$.
Notice that the purpose of Figure~\ref{fig:quant-pushout}  is to help the reader identify the signature over which the sentence $\phi_x$ is defined.
\end{itemize}
Consider the sentence $\Forall{x}\Exists{y} y=f(x)$ defined over a signature $\Sigma$ with two sorts $s_1$ and $s_2$ and a function symbol $f:s_1\to s_2$.
This sentence is a tautology.
Assume that $x$ is a variable for $\Sigma$ of the form $\pos{v_i,s_1,\Sigma}$, and notice that $\Exists{y}y=f(x)$ is a sentence over $\Sigma[x]$. 
Assume further that $y$ is a variable for $\Sigma[x]$ of the form $\pos{v_i,s_2,\Sigma[x]}$ which has the same name as $x$, and notice that $y=f(x)$ is a sentence over $\Sigma[x][y]$.
Now, define a signature morphism $\chi:\Sigma\to \Sigma'$, 
where $\Sigma'$ is a signature obtained from $\Sigma$ by replacing both $s_1$ and $s_2$ with a single sort $s$ and $\chi$ maps both $s_1$ and $s_2$ to $s$.
The translation of $\Forall{x}\Exists{y} y=f(x)$ along $\chi$ is $\Forall{x'}\Exists{y'} y'=f(x')$, where $x'=\pos{v_i,s,\Sigma'}$ and $y'=\pos{v_i,s,\Sigma[x']}$.
If we disregard the last components $\Sigma'$ and $\Sigma'[x']$ of  the variables $x'=\pos{v_i,s,\Sigma'}$ and $y'=\pos{v_i,s,\Sigma[x']}$, respectively, then we get $x'=y'$ and the resulting sentence $\Forall{x'}\Exists{x'} x'=f(x')$ is no longer a tautology.
In addition, this approach proves useful in preventing clashes between variables and constants within a given signature.
Furthermore, substitutions can be defined cleanly, avoiding the need for side conditions.
\begin{fact}
 $\Sen^\FOL$ is a functor $\Sig^\FOL \to \Set$, which commutes with the sentence building operators.
\end{fact}
\paragraph{Satisfaction relation} 
Satisfaction is the usual first-order satisfaction and it is defined using the natural interpretations of ground terms $t$ as elements $t^{\mathfrak{A}}$ in models $\mathfrak{A}$:
\begin{itemize}
\item $\mathfrak{A}\models_\Sigma t_1= t_2$ iff $t_1^{\mathfrak{A}}=t_2^{\mathfrak{A}}$.
\item $\mathfrak{A}\models_\Sigma \pi(t_1,\dots,t_n)$ iff $(t_1^{\mathfrak{A}},\dots,\mathfrak{A}_{t_n})\in \pi^{\mathfrak{A}}$.
\item $\mathfrak{A}\models_\Sigma \neg \phi$ iff $\mathfrak{A}\not\models_\Sigma\phi$.
\item $\mathfrak{A}\models_\Sigma\Exists{x}\phi_x$ iff $\mathfrak{A}^{x\leftarrow a}\models_{\Sigma'}\phi_x$ for some expansion $\mathfrak{A}^{x\leftarrow a}$ of $\mathfrak{A}$ along the inclusion $\iota:\Sigma\hookrightarrow\Sigma[x]$, that is, $\mathfrak{A}^{x\leftarrow a}\red_\Sigma=\mathfrak{A}$.
\end{itemize}
When there is no danger of confusion we may drop the subscript $\Sigma$ from the notation $\models_\Sigma$.
An expansion of $\mathfrak{A}$ to the signature $\Sigma[x]$ interpreting $x$ as the element $a$ is denoted by $\mathfrak{A}^{x\leftarrow a}$.
We have $\mathfrak{A}\models\Exists{x}\phi_x$ iff $\mathfrak{A}^{x\leftarrow a}\models\phi_x$ for some element $a$ from $\mathfrak{A}$.
The following proposition can be proved by induction on the structures of sentences.
\begin{proposition}[Satisfaction condition]
For all signature morphisms $\chi:\Sigma_1\to \Sigma_2$, all $\Sigma_2$-models $\mathfrak{A}$ and all $\Sigma_1$-sentences $\phi$ the following satisfaction holds:
\begin{center}
$\mathfrak{A}\models_{\Sigma_2}\chi(\phi)$  \quad iff \quad $\mathfrak{A}\red_\chi\models_{\Sigma_1}\phi$.
\end{center}
\end{proposition}

\paragraph{Working conventions}
When there is no risk of ambiguity, we omit the superscript $\FOL$ from the notations $\Sig^\FOL$, $\Sen^\FOL$ and $\Mod^\FOL$.
Furthermore, we identify a variable solely by its name and sort. 
However, situations arise where the third component of a variable cannot be omitted. 
One such case occurs when translating a sentence of the form $\Exists{x}\phi_x$ along a signature inclusion $\chi:\Sigma \hookrightarrow \Sigma[x]$. 
To avoid variable confusion, we rename $x$ to a fresh variable $y$, yielding $\Exists{y}\phi_{x \leftarrow y}$, where $\phi_{x \leftarrow y}$ is the sentence obtained from $\phi_x$ by substituting $y$ for $x$; 
we then translate this new sentence along $\chi$. 
This process illustrates how signature inclusions $\chi:\Sigma_1 \hookrightarrow \Sigma_2$ induce functions $\Sen(\chi): \Sen(\Sigma_1) \to \Sen(\Sigma_2)$, which can be viewed as inclusions on a subset of $\Sen(\Sigma_1)$ obtained by appropriately renaming variables.
Throughout this paper, given a signature inclusion $\chi:\Sigma_1 \hookrightarrow \Sigma_2$, we restrict attention to $\Sigma_1$-sentences $\phi$ for which the assumption $\Sen(\chi)(\phi) = \phi$ can be made without risk of confusion.
Also, we write 
\begin{itemize} 
\item $\Exists{x_1,\dots,x_n}\phi'$ instead of $\Exists{x_1}\dots\Exists{x_n}\phi'$, and 
\item $\Sigma[x_1,\dots,x_n]$ instead of $\Sigma[x_1]\dots[x_n]$.
\end{itemize}
These conventions simplify both the notation and the overall presentation.
\subsection{Many-sorted vs. single-sorted}
This subsection highlights subtle distinctions between many-sorted first-order logic and its single-sorted counterpart. Its primary purpose is motivational, providing background and context for the main interpolation result. Readers interested solely in the main results may restrict their attention to the definition of the Craig Interpolation Property given in Definition~\ref{def:CIP}.
\paragraph{Upward L\"owenheim-Skolem Property} 
The cardinality of a model $\mathfrak{A}$ over a signature $\Sigma$ is usually defined as the sum of the cardinalities of its carrier sets, that is, $\card(\mathfrak{A})=\sum_{s\in S}\card(\mathfrak{A}_s)$.
The following example, which can be found in \cite{hashimoto2025}, illustrates that the classical Upward L\"owenheim-Skolem (ULS) property does not hold when sorts are interpreted as distinct domains.
\begin{example}
Let $\Sigma$ be a signature with countably infinitely many sorts $\{s_n\}_{n<\omega}$, where each sort $s_n$ has exactly one constant symbol $c_n : \to s_n$, and there are no relation symbols.
Define the set of equations $\Gamma \coloneqq {\Forall{x_n} (x_n = c_n) \mid n<\omega}$, which asserts that each sort $s_n$ contains exactly one element.
\end{example}
Note that $\Sigma$ is a first-order signature, and $\Gamma$ is a set of $\Sigma$-sentences that admits an infinite model. However, all models of $\Gamma$ have cardinality at most $\omega$. Consequently, $\Gamma$ has no models of cardinality greater than $\omega$, and the traditional statement of the ULS property fails in this setting.

\paragraph{Soundness}
The following example is a simplified version of the one presented in \cite{journals/sigplan/GoguenM82}. Here, we focus on the essence to demonstrate that classical rules of first-order deduction are unsound.
\begin{example}\label{ex:sound}
Let $\Sigma=(S,F)$ be an algebraic signature consisting of:
\begin{itemize}
\item two sorts, that is, $S=\{Elt,Bool\}$, and 
\item two constants
$F=\{true:\to Bool, 
false:\to Bool\}$.
\end{itemize}
Let $\gamma=\Forall{x}true = false$, where $x$ is a variable of sort $Elt$.
\end{example}
Using the classical rules of first-order deduction, $\gamma\vdash_\Sigma true=false$.
However, $T_\Sigma\models\gamma$ but $T_\Sigma\not\models true =false$, which means that $\gamma\not\models true=false$.

Example~\ref{ex:sound} demonstrates that even proof rules that are sound in single-sorted first-order logic need not be sound in the many-sorted setting. 
This highlights a fundamental difference between the two frameworks and shows that results established for single-sorted logics cannot always be transferred directly to their many-sorted counterparts.
Rather, the introduction of multiple sorts can give rise to subtle semantic phenomena that invalidate arguments and proof techniques that are otherwise correct in the single-sorted case. 
\paragraph{Interpolation}
We recall the definition of interpolation as advanced in algebraic specifications~\cite{tar-bit, Diaconescu2008}.
\begin{definition}[Interpolation] \label{def:CIP}
Assume a pushout square of signature morphisms  $\mathcal{S}$ as depicted in the following diagram.
\begin{center}
\begin{tikzcd}
\Phi_2 \ar[r,dotted,no head]& \Sigma_2 \ar[r,"\delta_2"] & 
\Sigma' \\
& \Sigma \ar[r,"\chi_1",swap] \ar[u,"\chi_2"] & \Sigma_1 \ar[u,"\delta_1",swap] & \Phi_1 \ar[l,dotted,no head]
\end{tikzcd}
\end{center}
$\mathcal{S}$ has Craig Interpolation Property (CIP) whenever for all sets of sentences 
$\Phi_1\subseteq \Sen(\Sigma_1)$ and $\Phi_2\subseteq \Sen(\Sigma_2)$ 
such that 
$\delta_1(\Phi_1)\models_{\Sigma'} \delta_2(\Phi_2)$
there exists an interpolant $\Phi\subseteq \Sen(\Sigma)$ such that $\Phi_1\models_{\Sigma_1}\chi_1(\Phi)$ and $\chi_2(\Phi)\models_{\Sigma_2}\Phi_2$.
\end{definition} 
It is worth noting that the definition of interpolation is logic-independent, as it does not rely on the particular structure of sentences or signature morphisms. Consequently, it can be formulated for arbitrary logical systems. 

The following result shows that CIP fails in many-sorted first logic, in general.
\begin{lemma} \label{lemma:FOL-CIP}
The following pushout of first-order signature morphisms does not have CIP.
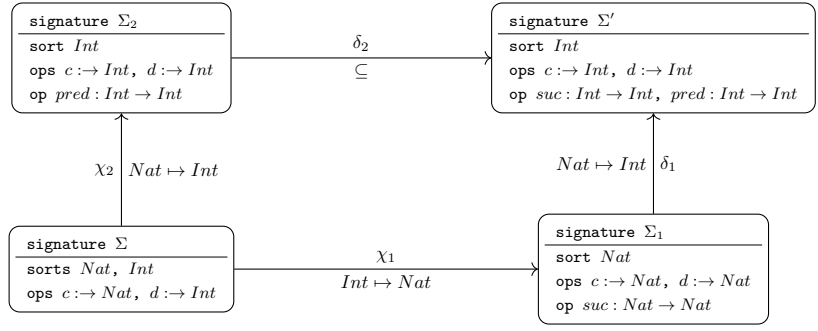
\begin{center}
 \begin{tikzpicture}[scale=.7, transform shape]
  \node(1)[draw,rectangle,rounded corners] {
   \tt\begin{tabular}{l}
    signature $\Sigma_2$  \\ \hline
    sort $Int$\\
    ops $c:\to Int$, $d:\to Int$
   \end{tabular}}; 
  \node(2)[draw,rectangle,rounded corners, right of = 1, node distance = 10 cm] {
   \tt\begin{tabular}{l}
    signature $\Sigma'$ \\ \hline
    sort $Int$\\ 
    ops $c:\to Int$, $d:\to Int$
   \end{tabular}};
  \node(3)[draw,rectangle,rounded corners, below of = 1, node distance = 4cm] 
   {\tt 
   \begin{tabular}{l}
   signature $\Sigma$ \\ \hline
   sorts $Nat$, $Int$\\
   ops $c:\to Nat$, $d:\to Int$
   \end{tabular}};
  \node(4)[draw,rectangle,rounded corners, below of = 2, node distance = 4cm] {
  \tt\begin{tabular}{l}
   signature $\Sigma_1$ \\ \hline
   sort $Nat$\\
   ops $c:\to Nat$, $d:\to Nat$
  \end{tabular}};
   \draw[->] 
    (1) to 
     node[above]{$\delta_2$} 
     node[below]{$\subseteq$}
    (2);
   \draw[->] (3) to node[above]{$\chi_1$} node[below]{$Int\mapsto Nat$} (4);
   \draw[->] (3) to node[left]{$\chi_2$} node[right]{$Nat\mapsto Int$} (1);
   \draw[->] (4) to node[right]{$\delta_1$} node[left]{$Nat \mapsto Int$} (2);
\end{tikzpicture}
\end{center}
\end{lemma}
\begin{proof}
Let $\phi_1\coloneqq(c=d)$ and $\phi_2\coloneqq (c=d)$.
Obviously $\phi_1\models_{\Sigma'}\phi_2$.
Suppose towards a contradiction that $\phi_1\models_{\Sigma_1}\chi_1(\phi)$ and $\chi_2(\phi)\models_{\Sigma_2}\phi_2$ for some $\Sigma$-sentence $\phi$.
\begin{itemize}
\item[] Let $\mathfrak{A}_1$ be the $\Sigma_1$-model interpreting
\begin{enumerate*}[label=(\alph*)]
\item $Nat$ as the set of integers $\mathbb{Z}$, and
\item both constants $c:\to Nat$ and $d:\to Nat$ as $1$.
\end{enumerate*}
Let $\mathfrak{A}$ be the $\Sigma$-model obtained from $\mathfrak{A}_1\red_{\chi_1}$ by changing the interpretation of $d:\to Int$ from $1$ to $-1$.
Let $h:\mathfrak{A}_1\red_{\chi_1}\to \mathfrak{A}$ be the isomorphism defined by $h_{Nat}=1_{\mathbb{Z}}$ and $h_{Int}(z)=-z$ for all integers $z\in \mathbb{Z}$.
Since $h_{Int}(d^{\mathfrak{A}_1\red_{\chi_1}})=h_{Int}(\chi_1(d)^{\mathfrak{A}_1})=h_{Int}(1)=-1=d^{\mathfrak{A}}$, the bijection $h$ is an isomorphism.
Since $\mathfrak{A}_1\models_{\Sigma_1}\phi_1$ and $\phi_1\models_{\Sigma_1}\chi_1(\phi)$, we have $\mathfrak{A}_1\models_{\Sigma_1}\chi_1(\phi)$.
It follows that $\mathfrak{A}_1\red_{\chi_1}\models_\Sigma \phi$.
Since $\mathfrak{A}$ and $\mathfrak{A}_1\red_{\chi_1}$ are isomorphic,
we have $\mathfrak{A}\models_\Sigma \phi$.
Let $\mathfrak{A}_2$ be the unique $\chi_2$-expansion of $\mathfrak{A}$.
It follows that $\mathfrak{A}_2\models_{\Sigma_2} \chi_2(\phi)$.
Since $\chi_2(\phi)\models_{\Sigma_2} \phi_2$, we get $\mathfrak{A}_2\models_{\Sigma_2} \phi_2$ which is a contradiction, 
since $c^{\mathfrak{A}_2}=c^{\mathfrak{A}}=1\neq -1=d^{\mathfrak{A}}=d^{\mathfrak{A}_2}$.
\end{itemize}
\end{proof}
The counterexample presented in Lemma~\ref{lemma:FOL-CIP} is due to \cite{journals/ipl/Borzyszkowski00}.
The following result provides conditions for the pushouts of signature morphisms under which interpolation holds. 
\begin{theorem}[First-order interpolation \cite{gai-pop-rob}] \label{th:FOL-CIP}
Consider a pushout square $\mathcal{S}$ of first-order signature morphisms $\{\Sigma_2\xleftarrow{\chi_2} \Sigma\xrightarrow{\chi_1}\Sigma_1,\Sigma_2\xrightarrow{\delta_2}\Sigma'\xleftarrow{\delta_1}\Sigma_1\}$ such as the one depicted in Definition~\ref{def:CIP}.
If  $\chi_1$ or $\chi_2$ is injective on sorts, then $\mathcal{S}$ has CIP.  
\end{theorem}
Theorem~\ref{th:FOL-CIP}, together with Lemma~\ref{lemma:FOL-CIP}, provides a criterion for determining when a pushout of first-order signature morphisms satisfies CIP. In $\FOL$, given a span of signature morphisms, the corresponding pushout has CIP whenever one of the morphisms is injective on sorts. The present work is devoted to establishing an analogous criterion for first-order hybrid logic with rigid symbols, which is introduced in the next section. As will become apparent in Section~\ref{sec:CIP}, the solution is not a straightforward generalization of the first-order result. In addition to injectivity on sorts, further conditions are required.

\section{The underlying hybrid logic}\label{sec:FOHLR}
In this section, we present a variant of First-Order Hybrid Logic with Rigid symbols ($\FOHLR$), introduced in~\cite{gai-acm}.
This logical system serves as the framework in which the main results of the paper are proved.
\paragraph{Signatures}
The signatures are of the form $\Delta=(\Sigma^\nom,\Sigma^\rigid,\Sigma)$:
\begin{itemize}
\item $\Sigma=(S,F,P)$ is a many-sorted first-order signature,
\item $\Sigma^\rigid=(S^\rigid,F^\rigid,P^\rigid)$ is a many-sorted first-order
signature of \emph{rigid} symbols such that $\Sigma^\rigid\subseteq \Sigma$.
\item $\Sigma^\nom=(S^\nom,F^\nom,P^\nom)$ is a single-sorted first-order signature such that
$S^\nom=\{\mathtt{nom}\}$ is a singleton, 
$F^\nom$ is a set of constants called \emph{nominals}, and 
$P^\nom$ is a set of binary relation symbols called \emph{modalities}.
\end{itemize}
We usually write  $\Delta=(\Sigma^\nom,\Sigma^\rigid\subseteq \Sigma)$ rather than $\Delta=(\Sigma^\nom,\Sigma^\rigid,\Sigma)$.
Throughout this paper, we let $\Delta$ and $\Delta_i$ 
range over signatures of the form $(\Sigma^\nom,\Sigma^\rigid\subseteq\Sigma)$ and $(\Sigma_i^\nom,\Sigma_i^\rigid\subseteq\Sigma_i)$, respectively, where $i$ is an index usually chosen from the set of natural numbers.
A \emph{signature morphism} $\chi:\Delta_1 \to \Delta_2$ consists of a pair of first-order signature morphisms
$\pos{\chi^{\nom} :\Sigma_1^{\nom} \to \Sigma_2^{\nom},
\chi:\Sigma_1 \to \Sigma_2}$
such that $\chi(\Sigma_1^{\rigid}) \subseteq \Sigma_2^{\rigid}$.
When there is no risk of confusion, we omit the superscript $\nom$ from the notation $\chi^\nom$.
\begin{fact}
$\FOHLR$ signature morphisms form a category $\Sig$ under the component-wise composition as first-order signature morphisms.
\end{fact}

\paragraph{Kripke structures}
  For every signature $\Delta$, the class of Kripke structures over $\Delta$
  consists of pairs $\M=(W,M)$, where
  
\begin{itemize}
\item $W$ is a first-order structure over $\Sigma^\nom$, called a \emph{frame}.
We denote by $|W|$ the universe of $W$ consisting of a set of possible worlds/states/nodes.
  
\item $M\colon|W|\to |\Mod^\FOL(\Sigma)|$ is a mapping from the universe of $W$ to the class of first-order $\Sigma$-structures such that the rigid symbols are interpreted in the same way across worlds:
\begin{center}
${M(w_1)\red_{\Sigma^\rigid}}={M(w_2)\red_{\Sigma^\rigid}}$ for all $w_1,w_2\in |W|$, 
\end{center}
where
 $M(w_i)\red_{\Sigma^\rigid}$ is the first-order reduct of $M(w_i)$ to the signature $\Sigma^\rigid$.
For any sort $s\in S$ and any possible world $w\in|W|$, 
the carrier set $M(w)_s$ of the first-order model $M(w)$ is a set, possibly empty.
\end{itemize}
We let $\M$ and $\N$ range over Kripke structures of the form $(W,M)$ and $(V,N)$, respectively.
A \emph{homomorphism} $h: \M \to \N$ over a signature $\Delta$ is also a pair 
\begin{center}
$\pos{W\stackrel{h}\to V, \{M(w)\stackrel{h_{w}}\to N(h(w))\}_{w \in |W|}}$
\end{center}
consisting of first-order homomorphisms $h:W\to V$ such that the mappings corresponding to rigid sorts are shared across the worlds, that is, $h_{w_{1}, s} = h_{w_{2}, s}$ for all possible worlds $w_{1}, w_{2} \in |W|$ and all rigid sorts $s \in S^\rigid$.

\begin{fact}
For any signature $\Delta$, the $\Delta$-homomorphisms form a category $\Mod(\Delta)$ under the component-wise composition.
Following standard terminology in $\FOL$, the objects of $\Mod(\Delta)$ are called $\Delta$-models.
\end{fact}

\paragraph{Reducts}
Every signature morphism $\chi: \Delta_1 \to \Delta_2$ induces a reduct functor $\_\red_\chi:\Mod(\Delta_2)\to \Mod(\Delta_1)$ defined by: 
\begin{enumerate}
\item Every $\Delta_2$-model $\M$ is reduced to a $\Delta$-model $\M\red_{\chi}=(W\red_{\chi^\nom},M\red_\chi)$, where
\begin{enumerate}
\item $W\red_{\chi^\nom}$ is the reduct of $W$ across $\chi^\nom:\Sigma_1^\nom\to \Sigma_2^\nom$ in single-sorted first-order logic,
\item $(M\red_\chi)(v)=M(w)\red_\chi$ for all $w\in|W|$, where $M(w)\red_\chi$ is the reduct of $M(w)$ across $\chi:\Sigma_1\to\Sigma_2$ in first-order logic.
\end{enumerate}
 \item The reduct $h\red_\chi:\M\red_\chi\to \N\red_\chi$ of a $\Delta_2$-homomorphism $h : \M \to \N$ is the following $\Delta_1$-homomorphism $\pos{h\red_{\chi^\nom} :W\red_{\chi^\nom} \to V\red_{\chi^\nom},\{h_w\red_\chi:M(w)\red_\chi\to \N(h(w))\red_\chi\}_{w\in|W|}}$.
\end{enumerate}
A $\chi$-expansion of a $\Delta_1$-model $\M$ is a $\Delta_2$-model $\M'$ such that $\M'\red_\chi=\M$. 
When $\chi$ is an inclusion, we usually denote $\M\red_\chi$ by $\M\red_\Delta$ -- in this case, the model reduct simply forgets the interpretation of those symbols in $\Delta_2$ that do not belong to $\Delta_1$.
\begin{fact}
For each signature morphism $\chi:\Delta_1\to\Delta_2$, the map  $\Mod:\Sig\to\Cat^{op}$, defined by
$\Mod(\chi)(h)  = h\red_\chi$ for all $\Delta_2$-homomorphisms $h$,
is a functor.
\end{fact}

\paragraph{Hybrid terms}
For any signature $\Delta$, we make the following notational conventions:
\begin{enumerate}
\item~$S^\ext\coloneqq S^\rigid\cup\{\mathtt{nom}\}$ the extended set of rigid sorts, where $\mathtt{nom}$ is the sort of nominals,
\item~$S^\flex \coloneqq S \setminus S^{\rigid}$ the subset of flexible sorts,  
\item~$F^\flex\coloneqq F\setminus F^\rigid$ the subset of flexible function symbols, 
\item~$P^\flex\coloneqq P\setminus  P^\rigid$ the subset of flexible relation symbols.
\end{enumerate}
Note that $(S^\flex,F^\flex,P^\flex)$ does not, in general, form a signature, as symbols in both $F^\flex$ and $P^\flex$ may involve rigid sorts.
\begin{definition}[Rigidification]
Let $\Delta=(\Sigma^\nom,\Sigma^\rigid\subseteq \Sigma)$ be a hybrid signature.
The \emph{rigidification} of $\Sigma$ is the first-order signature $@\Sigma=(@S,@F,@P)$ defined by:
\begin{enumerate}
\item $@S^\flex\coloneqq \{\at{k} s \mid k\in F^\nom \mbox{ and } s\in S^\flex\}$ and  $@S\coloneqq S^\rigid\cup  @S^\flex$,
\item $@F^\flex\coloneqq \{ \at{k}\sigma:at_k(\ari)\to at_k(s)\mid \sigma:\ari\to s\in F^\flex\}$ and $@F\coloneqq F^\rigid\cup @F^\flex$,
\item $@P^\flex\coloneqq \{\at{k} \pi: at_k(\ari) \mid k\in F^\nom \text{ and }(\pi\colon\ari)\in P^\flex\}$ and
$@P\coloneqq P^\rigid\cup  @P^\flex$, where
\end{enumerate}
\begin{itemize}
\item $at_k:S\to @S$ is defined by 
$at_k(s)=
\left\{\begin{array}{l l}
s & \text{ if } s\in S^\rigid\\
\at{k}{s} & \text{ if } s\in S^\flex,
\end{array}\right.$
and  
\item $at_k:S^*\to (@S)^*$ is defined by 
$at_k(s_1\dots s_n)=at_k(s_1)\dots ar_k(s_n)$ for all $(s_1\dots s_n)\in S^*$.
\end{itemize}
\end{definition}
For notational convenience, we introduce two additional functions:
\begin{itemize}
\item $at_k: F\to @F$ is defined by 
$at_k(\sigma:\ari\to s)=
\left\{\begin{array}{l l}
\sigma:\ari\to s & \text{ if } (\sigma:\ari\to s) \in F^\rigid\\
(\at{k}\sigma):at_k(\ari)\to at_k(s) & \text{ if } (\sigma:\ari\to s)\in F^\flex.
\end{array}\right.$
\item $at_k: P\to @P$ is defined by 
$at_k(\pi:\ari)=
\left\{\begin{array}{l l}
\pi:\ari & \text{ if } (\pi:\ari) \in P^\rigid\\
(\at{k}\pi):at_k(\ari) & \text{ if } (\pi:\ari)\in P^\flex.
\end{array}\right.$
\end{itemize}
The set of \emph{rigid $\Delta$-terms} is $T_{@\Sigma}$.
The set of \emph{open $\Delta$-terms} is $T_\Sigma$. 
The set of \emph{hybrid $\Delta$-terms} is $T_{\overline\Sigma}$, 
where the first-order signature $\overline\Sigma=(\overline{S},\overline{F},\overline{P})$ is defined by:
\begin{enumerate*}[label=(\alph*)]
\item $\overline{S}=S \cup @S^\flex$,
\item $\overline{F}=F \cup @F^\flex$, and
\item $\overline{P}=P \cup @P^\flex$.
\end{enumerate*}
\begin{definition}[Term evaluation]
A pair $(\M,w)$ consisting of a model $\M$ and a possible world $w\in|W|$ is called a \emph{pointed model}.
The interpretation of hybrid terms in pointed models is uniquely defined as follows:
\begin{enumerate}
\item $\sigma(t)^{(\M,w)}\coloneqq \sigma^{M(w)}(t^{(\M,w)})$, 
where $(\sigma:\ari\to s)\in F$, and $t\in T_{\overline\Sigma,\ari}$,
\footnote{If $t=(t_1,\dots,t_2)$ then $t^{(\M,w)}=(t_1,\dots,t_2)^{(\M,w)} = (t_1^{(\M,w)},\dots,t_n^{(\M,w)})$.}
\item $(\at{k} \sigma)(t)^{(\M,w)} \coloneqq \sigma^{M(k^W)} (t^{(\M,w)})$, 
where $(\sigma:\ari\to s)\in F^\flex$ and $t\in T_{\overline\Sigma,at_k(\ari)}$.
\end{enumerate}
\end{definition}
One of the advantages of working with sets instead of non-empty sets as the carrier sets of models is that the set of terms -- more precisely, the set of rigid terms -- can be always organized as a model.
\begin{lemma} \label{FOHLR-init}
Let $\Delta=(\Sigma^\nom,\Sigma^\rigid\subseteq\Sigma)$ be a signature. 
There exists an initial model of terms $T_\Delta=(W_\Delta,M_\Delta)$ defined as follows:
\begin{enumerate*}[label=(\alph*)]
\item $W_\Delta= T_{\Sigma^\nom}$, where modalities are empty relations;  and 
\item $M_\Delta\colon  F^\nom \to |\Mod^\FOL(\Sigma)|$
maps a nominal $k\in F^\nom$ to a first-order $\Sigma$-structure $M_\Delta(k)$  defined by:
\footnote{Given a category $\C$, the class of its objects is denoted by either $|\C|$ or $Ob(\C)$. 
In particular, $|\Mod^\FOL(\Sigma)|$ is the class of first-order $\Sigma$-models.}
\end{enumerate*}
\begin{enumerate}
\item $M_\Delta(k)_s = T_{(@\Sigma),at_k(s)}$ for all sorts $s\in S$.
\item $\sigma^{M_\Delta(k)} : T_{(@\Sigma),at_k(\ari)} \to T_{(@\Sigma),at_k(s)}$ is defined by 
$\sigma^{M_\Delta(k)}(t) = at_k(\sigma)(t)$ for 
all function symbols $\sigma:\ari\to s\in F$ and
all tuples of hybrid terms $t\in T_{(@\Sigma),at_k(\ari)}$.
\item $\pi^{M_\Delta(k)}$ is the empty set, for all relation symbols $(\pi\colon\ari)\in P$.
\end{enumerate}
\end{lemma}
The proof of Lemma~\ref{FOHLR-init} is based on the unique interpretation of
terms into models, and it is straightforward. 
\paragraph{Sentences}
The simplest sentences defined over a signature $\Delta$,
usually referred to as atomic, are given by
\begin{center}
$\rho \Coloneqq k  \mid t_{1} = t_{2} \mid \pi(t)$
\end{center}
where
\begin{enumerate*}[label=(\alph*)]
\item $k \in  F^\nom$ is a nominal,
\item $t_i \in T_{\overline\Sigma,s}$ are hybrid terms, $s\in \overline{S}$ is a hybrid sort,
\item $\pi:\ari\in\overline{P}$ and $t\in T_{\overline\Sigma,\ari}$.
\end{enumerate*}
We call 
\emph{hybrid equations} sentences of the form $t_1=t_2$, and  
\emph{hybrid relations} sentences of the form $\pi(t)$.

Assume a countable set $\{v_i\mid i<\omega\}$ of variable names. 
A variable for a signature $\Delta$ is a triple $\pos{v_i,s,\Delta}$,
where $v_i$ is a variable name and
$s$ is a sort from the extended set of rigid sorts $S^\ext$. 
The set $\Sen(\Delta)$ of \emph{full sentences} over $\Delta$ are given by the following grammar:
\begin{center}
$\phi \Coloneqq
\rho \mid
\lor \Phi \mid
\lnot \phi \mid
\at{k} \phi \mid
\pos{\lambda} \phi \mid
\store{z}\phi_z \mid
\Exists{x} \phi_x$
\end{center}
where 
\begin{enumerate*}[label=(\alph*)]
\item $\rho$ is an atomic sentence,
\item $k \in  F^\nom $ is a nominal,
\item $\Phi$ is a finite set of sentences over $\Delta$,
\item $z$ is a nominal variable for $\Delta$, and $\phi_z$ is a sentence over the signature $\Delta[z]$ obtained from $\Delta$ by adding $z$ as a new constant to $ F^\nom$,
\item $x$ is a nominal variable or a variable of rigid sort for $\Delta$, and $\phi_x$ is a sentence over the signature $\Delta[x]$, and
\item $(\lambda:\mathtt{nom}~\mathtt{nom})\in P^\nom$ is a binary modality.
\end{enumerate*}
Other than the first kind of sentences (\emph{atoms}), we refer to the sentence-building operators, as  
\emph{disjunction},
\emph{negation},
\emph{retrieve}, 
\emph{possibility},
\emph{store}, and
\emph{existential quantification},
respectively.
Other Boolean connectives and the universal quantification can be defined as abbreviations of the above sentence building operators.

Each signature morphism $\chi:\Delta_1\to\Delta_2$ induces a \emph{sentence translation}:
any $\Delta_1$-sentence $\phi$ is translated to a $\Delta_2$-sentence $\chi(\phi)$ by replacing, in an inductive manner, the symbols in $\Delta_1$ with symbols from $\Delta_2$ according to $\chi$.
It is worth noting that the translation of a quantified $\Delta_1$-sentence $\Exists{x}\phi_x$ along $\chi:\Delta_1\to \Delta_2$ is $\Exists{x'}\chi'(\phi_x)$, where 
\begin{enumerate}
\item $x'$ is the translation of $x$ along $\chi$, that is, $x'=\pos{v_i,\chi(s),\Delta_2}$ if $x=\pos{v_i,s,\Delta_1}$, and 
\item $\chi':\Delta_1[x]\to\Delta_2[x']$ is the extension of $\chi:\Delta_1\to\Delta_2$ which maps $x$ to its translation $x'$,
as depicted in Figure~\ref{fig:quant-pushout}.
\end{enumerate}
\begin{figure}[h]
\begin{center}
\begin{tikzcd}
\phi_x \ar[r,dotted,no head]& 
\Delta_1[x] \ar[r,"\chi'"]& 
\Delta_2[x'] & 
\chi'(\phi_x) \ar[l,dotted,no head] \\
\Exists{x}\phi_x\ar[r,dotted,no head]  & 
\Delta_1 \ar[r,"\chi",swap] \ar[u,hook]& 
\Delta_2 \ar[u,hook] & 
\Exists{x'}\chi'(\phi_x) \ar[l,dotted,no head]
\end{tikzcd}
\caption{Quantification pushout}
\end{center}
\label{fig:quantpushout}
\end{figure}
This extends the translation of quantified sentences along signature morphisms from $\FOL$ to the present setting. 
Similarly, the translation of any $\Delta_1$-sentence of the form $\store{x}\phi_x$ along $\chi:\Delta_1\to\Delta_2$ is $\store{x'}\chi'(\phi_x)$, where both $x'$ and $\chi'$ are as defined above. 
\begin{fact} \label{fact:sen-hyb}
$\Sen : \Sig\to\Set$ is a functor, where $\Set$ is the category of all sets.
\end{fact}
\paragraph{Local satisfaction relation}
Given a pointed model $(\M,w)$ over $\Delta$, we define the \emph{satisfaction of $\Delta$-sentences} by structural induction as follows:

\noindent\emph{For atomic sentences}:
\begin{itemize}
\item $(\M,w) \models_\Delta k $ iff $k^W = w$;
\item $(\M,w) \models_\Delta t_1 = t_2$ iff $t_1^{(\M,w)}= t_2^{(\M,w)}$;
\item $(\M,w) \models_\Delta \pi(t)$ iff $t^{(\M,w)} \in \pi^{(\M,w)}$.
\end{itemize}
\noindent\emph{For full sentences}:
\begin{itemize}
\item $(\M,w) \models_\Delta \lor \Phi$ iff $(\M,w) \models_\Delta \phi$ for some $\phi \in \Phi$; 
\item $(\M,w) \models_\Delta \neg \phi$ iff $(\M,w) \not\models_\Delta \phi$;
\item $(\M,w) \models_\Delta \at{k} \phi$ iff $(\M,k^W) \models_\Delta \phi$;
\item $(\M,w) \models_\Delta \pos{\lambda} \phi$ iff $(\M,v) \models_\Delta \phi$ for some $v \in \lambda^W(w)$.\footnote{$\lambda^W(w)=\{v\in|W| \mid (w,v)\in\lambda^W\}$.}
\item $(\M,w)\models_\Delta \store{z}{\phi_z}$ iff $(\M^{z\leftarrow w},w) \models_{\Delta[z]} \phi_z$,

where $\M^{z\leftarrow w}$ is the unique expansion $\M$ to the signature $\Delta[z]$ interpreting $z$ as $w$. 
\item $(\M,w)\models_\Delta \Exists{x}{\phi_x}$ iff $(\M^{x\leftarrow m},w) \models_{\Delta[x]} \phi_x$ for some $m\in \M(w)_s$, 

where $s$ is the sort of~$x$.
\end{itemize}
We adopt the following notational conventions:
\begin{itemize}
\item $(\M,w)\models_\Delta \Phi$ holds when $(\M,w)\models_\Delta \phi$ for all $\phi\in \Phi$.
\item $\Phi\models_\Delta \psi$ when $(\M,w)\models_\Delta\Phi$ implies $(\M,w)\models_\Delta \psi$ for all pointed models $(\M,w)$.
\item $\Phi\models^k_\Delta \psi$ when $(\M,w)\models_\Delta\Phi$ implies $(\M,k^W)\models_\Delta \psi$ for all pointed models $(\M,w)$.
\item $\Phi^\bullet$ denotes $\{\phi\in\Sen(\Delta)\mid \Phi\models_\Delta \phi\}$, 
the closure of $\Phi$ under the local satisfaction relation.
\end{itemize}
When there is no risk of confusion, we omit the subscript $\Delta$ from the notation $\models_\Delta$.
Global satisfaction relation is defined in the standard way: $M\models \phi$ if $(\M,w)\models\phi$ for all possible worlds $w\in |W|$.

The following \emph{satisfaction condition} can be proved by induction on the structure of $\Delta$-sentences.
The proof is essentially identical to those developed for several other variants of hybrid logic presented in the literature (see, e.g.~\cite{dia-qvh}).
\begin{proposition}[Local satisfaction condition] \label{prop:sat-cond}
Let $\chi:\Delta_1 \to \Delta_2$ be a signature morphism.
Then $(\M,w) \models \chi(\phi)$ iff $(\M\red_{\chi},w) \models \phi$,
for all pointed $\Delta_2$-models $(\M,w)$ and $\Delta_1$-sentences~$\phi$.\footnote{By the definition of reducts, $\M'$ and $\M'\red_{\chi}$ have the same possible worlds, which means that the statement of Proposition~\ref{prop:sat-cond} is well-defined.}
\end{proposition}
The local satisfaction condition for $\FOHLR$, stated in Proposition~\ref{prop:sat-cond}, is the modal counterpart of the satisfaction condition for $\FOL$. This property is fundamental to the institution-theoretic framework, as it enables a modular treatment of both theoretical results and their applications in formal methods.
As a direct consequence of the local satisfaction condition, logical consequence is preserved by signature morphisms. That is, for every signature morphism $\chi:\Delta_1 \to \Delta_2$, each set $\Phi$ of $\Delta_1$-sentences and any $\Delta_1$-sentence $\phi$, we have:
\[
\Phi \models_{\Delta_1} \phi \text{ implies }\chi(\Phi) \models_{\Delta_2} \chi(\phi).
\]
Thus, the satisfaction relation is preserved along signature morphisms, a key feature supporting modular specification and reasoning.
\begin{remark}
We adopt the standard convention of writing $\Diamond\phi$ for $\Delta$-sentence $\pos{\lambda}\phi$ when there is only one accessibility relation $\lambda$ among nominals.
\end{remark}
This notation emphasizes that the single-modal languages are special cases of the multimodal languages.
Other sentence building operators can be defined as abbreviations.
\begin{itemize}
\item \emph{False} is the disjunction of the empty set, that is, $\bot \coloneqq \lor \emptyset$.
\item \emph{True} is the negation of false $\top\coloneqq \neg\bot$.
\item \emph{Conjunction} is defined by $\land\Phi\coloneqq \neg\lor_{\phi\in \Phi}\neg\phi$.
\item \emph{Necessity} is defined by $\nec{\lambda}\phi\coloneqq \neg\pos{\lambda}\neg\phi$ and $\Box\phi\coloneqq \neg\Diamond\neg\phi$.
\item \emph{Universal quantification} is defined by $\Forall{x}\phi_x\coloneqq \neg\Exists{x}\neg\phi_x$. 
%
%
\item \emph{Until} is defined by 
$U(\phi, \psi) \coloneqq \store{x}\Diamond\store{y}(\phi \wedge \at{x}\Box(\Diamond y \Rightarrow \psi))$ over a signature whose set of binary relations on nominals is a singleton.
Note that $U(\phi, \psi)$ holds at a state $w$ if there exists a future state $v$ where $\phi$ holds, such that $\psi$ holds at every state between $w$ and $v$.
\end{itemize}
Note that store can be defined using existential quantification and Boolean connectives, since $\store{z}\phi_z$ is semantically equivalent to $\Exists{z}(z \land \phi_z)$. That is, $\store{z}\phi_z$ and $\Exists{z}(z \land \phi_z)$ are satisfied by exactly the same pointed models, and we therefore write $\store{z}\phi_z \equiv \Exists{z}(z \land \phi_z)$. Nevertheless, many of the results of this paper remain valid for fragments of $\FOHLR$ that do not allow quantification over nominal variables, provided that the store construct is treated as a primitive operator.
\paragraph{Fragments}
By a fragment of $\FOHLR$, we mean a logical system obtained by restricting the category of signatures or/and omitting certain sentence-building operators from the grammar used to define its sentences.
Fragments of $\FOHLR$ have been widely studied in the literature. We provide a few examples.
\begin{example}(First-Order Hybrid  Logic with user-defined Sharing ($\FOHLS$)) \label{ex:FOHLS}
This logic has the same signatures and Kripke structures as $\FOHLR$.
The sentences are obtained from atoms constructed with open terms only, that is,
if $\Delta=(\Sigma^\nom,\Sigma^\rigid\subseteq\Sigma)$, 
all (ground) equations over $\Delta$ are of the form $t_1=t_2$, where $t_1,t_2\in T_\Sigma$, and
all (ground) relation over $\Delta$ are of the form $\pi(t)$, where $(\pi:\ari)\in P$ and $t\in T_{\Sigma,\ari}$.
A version of $\FOHLS$ is the underlying logic of H system~\cite{cod-h}.
Other variants of $\FOHLS$ have been studied in \cite{martins2011,dia-msc,dia-qvh}. 
\end{example}
\begin{example}(Rigid First-Order Hybrid Logic ($\RFOHL$) \cite{DBLP:conf/wollic/BlackburnMMH19})\label{ex:RFOHL}
This logic is obtained from $\FOHLR$ by restricting the signatures $\Delta=(\Sigma^\nom,\Sigma^\rigid\subseteq\Sigma)$ such that 
\begin{enumerate*}[label=(\alph*)]
\item~$\Sigma^\nom$ has only one binary relation symbol,
\item~$\Sigma$ is single-sorted, 
\item~the unique sort is rigid,
\item~there are no rigid function symbols except variables, and 
\item~there are no rigid relation symbols.
\end{enumerate*}
\end{example}
\begin{example}(Hybrid Propositional Logic ($\HPL$)) \label{ex:HPL}
This is the quantified version of the most common form of hybrid logic (see, e.g., \cite{ArecesB01}).
$\HPL$ is obtained from $\FOHLR$ by restricting the signatures $\Delta=(\Sigma^\nom,\Sigma^\rigid\subseteq\Sigma)$ such that $\Sigma^\rigid$ is empty and the set of sorts in $\Sigma$ is empty.
Notice that if $\Sigma=(S,F,P)$ and $S=\emptyset$ then $F=\emptyset$ and $P$ contains only propositional symbols.
\end{example}
CIP holds for all fragments presented above.
\section{Basic definitions and results} \label{sec:basic}
In this section, we establish the terminology and we state some foundational results necessary for the present study.
\begin{definition}
Let  $\Phi$ be a set of sentences over a signature $\Delta$.
\begin{itemize}
\item $\Phi$  is \emph{(locally) consistent} if $\Phi\not\models_\Delta\psi$ for some $\Delta$-sentence $\psi$.
\item $\Phi$ is \emph{(locally) satisfiable} if $(\M,w)\models_\Delta\Phi$ for some pointed model $(\M,w)$.
\end{itemize}
Let $\phi_1$ and $\phi_2$ be two sentences over $\Delta$.
\begin{itemize}
\item $\phi_1$ is semantically equivalent to $\phi_2$, denoted by $\phi_1\equiv\phi_2$, if they are satisfied by the same pointed models:
$(\M,w)\models_\Delta \phi_1 \iff (\M,w)\models_\Delta \phi_2$, for all pointed models $(\M,w)$ over $\Delta$.
\end{itemize}
\end{definition}

\subsection{Presentations and theories}  \label{sec:presentation}
We adopt the terminology used in the algebraic specification literature.
A pair $(\Delta,\Phi)$ consisting of a signature $\Delta$ and a set of sentences $\Phi$ over $\Delta$ is called a \emph{presentation}.
A \emph{presentation morphism} $\chi:(\Delta_1,\Phi_1)\to (\Delta_2,\Phi_2)$ consists of a signature morphism $\chi:\Delta_1\to\Delta_2$ such that $\Phi_2\models\chi(\Phi_1)$.
Any presentation $(\Delta,T)$ such that $T=T^\bullet$ is called a \emph{theory}.
A \emph{theory morphism} is just a presentation morphism between theories.
\begin{definition}[Complete theories]
A complete theory is one that is maximally consistent, that is,  
$(\Delta,T)$ is a complete theory if
\begin{enumerate*}[label=(\alph*)]
\item $T\not\models_\Delta\bot$ and 
\item $T\cup\{\phi\}\models_\Delta\bot$ for any $\phi\in \Sen(\Delta)\setminus T$.
\end{enumerate*}
\end{definition}
By the same argument used in \cite[Lemma 2.20]{gai-acm}, $\FOHLR$ has the same expressive power as $\FOHLS$, which has been shown to be compact in \cite{DBLP:journals/logcom/Diaconescu17}.
It follows that $\FOHLR$ and its fragments are compact as well.
The following result is an analogue of Lindenbaum Lemma in $\FOHLR$ and it follows from compactness.
\begin{lemma}\label{lemma:lindenbaum}
Any consistent presentation can be extended to a complete theory.
\end{lemma}
\subsection{Amalgamation}
We recall the notion of amalgamation which is intensively used in institutional model theory to prove results in a structured and modular way.
\begin{definition}[Amalgamation]
Let us consider a commuting square of signature morphisms $\mathcal{S}$ such as the one depicted in the following diagram.
\begin{center}
\begin{tikzcd}
 \Delta_\bc \ar[r,"\chi_\bc"] & 
\Delta_\dc \\
\Delta \ar[r,"\chi",swap] \ar[u,"\delta"] & \Delta_\ac \ar[u,"\delta_\ac",swap] 
\end{tikzcd}
\end{center}
$\mathcal{S}$ is an amalgamation square if for all models $\M\in|\Mod(\Delta_\ac)|$ and $\N\in|\Mod(\Delta_\bc)|$ such that $\M\red_\chi=\N\red_\delta$ there exists a unique $\M'\in|\Mod(\Delta_\dc)|$ such that $\M'\red_{\delta_\ac}=\M$ and $\M'\red_{\chi_\bc}=\N$.
\end{definition}
Most of the institutions formalizing conventional or non-conventional logics have the amalgamation property~\cite{Diaconescu2008}.
These include many-sorted first-order logic $\FOL$ and first-order hybrid logic with rigid symbols $\FOHLR$.
\begin{lemma}[Amalgamation~\cite{dia-msc}]\label{lemma:amalgamation}
Both $\FOL$ and $\FOHLR$ have the amalgamation property, that is, all pushouts of signatures are amalgamation squares.
\end{lemma}
The concept of model amalgamation presented in this paper should not be confused with single
signature and much harder one from conventional model theory (e.g. \cite{DBLP:books/daglib/0030198}) which refers to the amalgamation of elementary extensions.
\subsection{Substitutions}  \label{sec:subst}
We recall the notion of hybrid substitution, hereafter referred to simply as substitution, introduced in \cite{gai-her} as part of the foundations of logic programming for an earlier version of $\FOHLR$.
In the present work, substitutions are essential for defining the forcing semantics of quantified sentences.

Let $\Delta$ be a signature, 
$C_1$ and $C_2$ two sets of new constants for $\Delta$ of sorts in $S^\ext$, the extended set of rigid sorts.
A substitution $\theta : C_1 \to C_2$ over $\Delta$ is a sort-preserving mapping from $C_1$ to the set of terms of the initial term model $T_{\Delta[C_2]}$ defined in Lemma~\ref{FOHLR-init}.
The following result is from \cite[Corollary 39]{gai-her}.

\begin{proposition}[Local satisfaction condition for substitutions]
A substitution $\theta : C_1\to C_2$ over $\Delta$ uniquely determines:
\begin{enumerate}
\item a sentence function $\theta:\Sen(\Delta[C_1])\to \Sen(\Delta[C_2])$, 
which preserves $\Delta$ and maps each constant $c\in C_1$ to a rigid term $\theta(c)$ over $\Delta[C_2]$, and

\item a reduct functor $\red_\theta\colon\Mod(\Delta[C_2])\to\Mod(\Delta[C_1])$,
which preserves the interpretation of $\Delta$ and such that 
for all constants $c\in C_1$ and all pointed $\Delta[C_2]$-models $(\M,w)$, 
we have $c^{(\M\red_\theta,w)}\coloneqq \theta(c)^{(\M,w)}$.
\end{enumerate}
such that the following local satisfaction condition holds:
$$(\M,w) \models \theta(\phi)\text{ iff }(\M\red_\theta,w)\models  \phi$$
for all $\Delta[C_1]$-sentences $\phi$ and all pointed $\Delta[C_2]$-models $(\M,w)$.
\end{proposition}
\subsection{Reachable and generated models}
The notion of reachable model was introduced in its current form within institutional model theory in \cite{Petria07}, and it also plays a central role in the present paper. 
An earlier appearance of reachable models in institutional model theory, although in a technically different form, can be found in \cite{DBLP:journals/tcs/Tarlecki85}.
\begin{definition}[Reachable models]
A Kripke structure $\M$ defined over a signature $\Delta$ is reachable 
iff 
the unique homomorphism from the initial Kripke structure $h:T_\Delta\to \M $ is surjective on $S^\ext$, 
that is,  
$h\colon W_\Delta\to W$ is surjective and 
$h_{w,s}: M_\Delta(k)_s\to M(h(k))_s$ is surjective for all nominals $k\in F^\nom$ and all rigid sorts $s\in S^\rigid$.
Similarly, a pointed model $(\M,w)$ is reachable if $\M$ is reachable.
\end{definition}
The following result gives an abstract characterization of the notion of reachable model based on substitutions.
\begin{proposition}[Reachable models] \label{prop:reachable}
Let $\Delta$ be a signature.
A $\Delta$-model $\M $ is reachable 
iff 
for each $S^\ext$-sorted set of new constants $C$
and any expansion $\M'$ of $\M $ to $\Delta[C]$, 
there exists a substitution  $\theta : C \to \emptyset$ over $\Delta$ such that $\M \red_\theta=\M'$.
\end{proposition}
The proof of Proposition~\ref{prop:reachable} is conceptually identical with the proof of  \cite[Proposition 49]{gai-her}.
\begin{definition}[Elementary equivalence \cite{TOCL2025}] \label{def:elementary}
Let $(\M,w)$ and $(\N,v)$ be two pointed models defined over a signature~$\Delta$.
\begin{enumerate}
\item $(\M,w)$ and $(\N,v)$ are \emph{elementarily equivalent} if they (locally) satisfy the same sentences: $(\M,w)\models\phi$ iff $(\N,v)\models\phi$, for all $\Delta$-sentences $\phi$. In this case, we write $(\M,w)\equiv(\N,v)$.
\item $\M$ and $\N$ are \emph{elementarily equivalent}  if they globally satisfy the same sentences: $\M\models\phi$ iff $\N\models\phi$, for all $\Delta$-sentences $\phi$. In this case, we write $\M\equiv\N$.
\end{enumerate}
\end{definition}
Definition~\ref{def:elementary} is a natural generalization of the classical notion of elementary equivalence from first-order logic to modal logics. As expected, it admits two variants: one based on the local satisfaction relation and another based on the global satisfaction relation.
This concept has already proved useful in other contexts. For example, it was used in~\cite{TOCL2025} as the foundation for hybrid-dynamic Ehrenfeucht--Fra\"{\i}ss\'{e} games. In the present paper, elementary equivalence plays a key role in establishing a criterion for CIP in $\FOHLR$.
\begin{lemma} \label{lemma:reach-flex}
Let $\M $ be a reachable model defined over a signature $\Delta$ and let $h:T_\Delta\to \M $ be the unique homomorphism from $T_\Delta$ to $\M$. 
For all nominals $k$ in $\Delta$ the pointed models $(\M,h(k))$ and $ (h(T_\Delta),h(k))$ are elementarily equivalent.
\end{lemma}
The proof of Lemma~\ref{lemma:reach-flex} is straightforward by induction on the structure of sentences.
The proof relies on the fact that quantification is only allowed over rigid domains, that is, variables used for quantification are of sorts from the extended set of rigid sorts.

\begin{definition}[Rigid first-order models]
Any Kripke structure $\M=(W,M)$ with a non-empty set of states defined over a signature $\Delta$ (uniquely) determines a first-order $\Sigma^\rigid$-model $\R_\M=M(w)\red_{\Sigma^\rigid}$, where $w$ is any state from $|W|$.
\end{definition}
Notice that $\R_\M$ is well-defined as rigid symbols are interpreted uniformly across states.
If $|W|=\emptyset$ then $\R_\M$ doesn't exist; hence, we need to impose the condition $|W|\neq\emptyset$.
\begin{definition}[Generated models]
Let $\M$ be a Kripke structure with a non-empty set of states defined over a signature $\Delta$.
The generated (sub-)model $\G_\M$ of $\M$ is defined by $\G_\M\coloneqq h(T_{\Delta_\M})\red_\Delta$, where
\begin{itemize}
\item $\Delta_\M$ is the signature obtained from $\Delta$ by adding 
\begin{enumerate*}[label=(\alph*)]
\item all elements of the $\Sigma^\nom$-model $W$ as new nominals, and
\item all elements of the $\Sigma^\rigid$-model $\R_\M$ as new rigid constants. 
\end{enumerate*}
\item $\M_\M$ is the expansion of $\M$ to $\Delta_\M$ interpreting each new constant as itself.
\item $h:T_{\Delta_\M}\to\M_\M$ is the unique homomorphism from $T_{\Delta_\M}$ to $\M_\M$.
\end{itemize}
Two Kripke structures $\M$ and $\N$ are \emph{quasi-isomorphic} if their generated substructures $\G_\M$  and $\G_\N$ are isomorphic.
Two pointed models $(\M,w)$ and $(\N,v)$ are quasi-isomorphic
if $(\G_\M,w)$ and $(\G_\N,v)$ are isomorphic, 
that is, there exists an isomorphism $h:\G_\M\to \G_\N$ which maps $w$ to $v$.
\end{definition}
The following result is a corollary of Lemma~\ref{lemma:reach-flex}.
\begin{corollary}\label{cor:gen-Kripke}
 $(\M,w)\equiv (\N,v)$  if there exists an isomorphism $h:\G_M\to\G_N$ such that $h(w)=v$.
\end{corollary}
\begin{proof}
It suffices to show that $(\G_\M,w)\equiv(\M,w)$.
Since $\M_\M$ is reachable,
by Lemma~\ref{lemma:reach-flex},
$(\M_\M,w)\equiv (h(T_{\Delta_\M}),w)$,
where $h:T_{\Delta_\M}\to\M_\M$ is the unique homomorphism from $T_{\Delta_\M}$ to $\M_\M$.
Hence, $\M\equiv \G_\M$.
\end{proof}
\begin{example}
Let $\Delta=(\Sigma^\nom,\Sigma^\rigid\subseteq \Sigma)$ be a signature, 
where
\begin{enumerate}
\item $\Sigma^\nom$ consists of one binary relation symbol $\lambda:\mathtt{nom}~\mathtt{nom}$,
\item $\Sigma^\rigid$ is empty, and $\Sigma$ is the signature of natural numbers with 
one sort $Nat$, 
one constant $0:\to Nat$, 
a successor operation $suc:Nat\to Nat$ and 
an addition operation  $\_+\_:Nat~Nat\to Nat$.
\end{enumerate}
Let $\M=(W,M)$ be the Kripke structure, where 
\begin{enumerate}
\item The set of possible worlds $|W|$ is $\{w_n \mid n<\omega\}$, and the binary relation $\lambda^W$ is $\{(w_n,w_{n+1})\mid n<\omega\}$.
\item $M(w_0)$ is $\mathbb{Z}$.
For all $n>0$, the model $M(w_n)$ is $\mathbb{Z}_n$ interpreting zero, successor and addition in the usual way.
\end{enumerate}
\end{example}
Notice that:
\begin{itemize}
\item $\Delta_\M=(\Sigma^\nom[W],\emptyset\subseteq \Sigma)$, and $\M_\M$ is the unique expansion of $\M$ to $\Delta_\M$ that interprets each nominal $w_n$ as itself.

\item Let $h:T_{\Delta_\M}\to\M_\M$ be the unique homomorphism from $T_{\Delta_\M}$ to $\M_\M$. Observe that the only elements of $\M_\M$ that do not belong to the image of $h$ are the negative integers in $M(w_0)=\mathbb{Z}$.

\item $\G_\M(w_0)$ is the standard model of the natural numbers, obtained from $M(w_0)=\mathbb{Z}$ by removing all negative integers. For every $n>0$, we have $\G_\M(w_n)=\M(w_n)$.
\end{itemize}
By Corollary~\ref{cor:gen-Kripke}, $(\M,w_n)\equiv(\G_M,w_n)$.

\subsection{Basic sentences}
We let $\Sen_0$ denote the subfunctor of $\Sen$ mapping each signature $\Delta$ to $\Sen_0(\Delta)$, the set of $\Delta$-sentences consisting~of
\begin{enumerate}
\item all atomic sentences,
\footnote{Recall that we have three types of atomic sentences: nominals $k$, hybrid equations $t_1=t_2$, and hybrid relations $\pi(t)$, where 
\begin{enumerate*}[label=(\alph*)]
\item $k \in  F^\nom$,
\item $t_i \in T_{\overline\Sigma,s}$ are hybrid terms, $s\in \overline{S}$ is a hybrid sort,
\item $\pi:\ari\in\overline{P}$ and $t\in T_{\overline\Sigma,\ari}$.
\end{enumerate*}}
and
\item all sentences of the form $\pos{\lambda}k$, where $(\lambda:\mathtt{nom}~\mathtt{nom})\in P^\nom$ is a relation symbol and $k\in F^\nom$ is a nominal.
\end{enumerate}
We denote by $\Sen_b$ the subfunctor of $\Sen$ which maps each signature $\Delta$ to $\Sen_b(\Delta)$, 
the set of sentences obtained from the sentences in $\Sen_0(\Delta)$ by applying exactly one time retrieve $@$.
\begin{definition}[Basic set of sentences]
\label{def:basic}
A set of sentences $\Psi$ defined over a signature $\Delta$ is \emph{basic} if there exists a $\Delta$-model $\M_\Psi$ such that 
\begin{center}
$\M \models \Psi$ iff there exists a homomorphism $h:\M_\Psi\to \M $
\end{center}
for all $\Delta$-models $\M$. 
We say that $\M_\Psi$ is a \emph{basic model} of~$\Psi$. 
If in addition the homomorphism $h$ is unique then the set $\Psi$ is called \emph{epi-basic}.
\end{definition}
The notion of basic set of sentences is from~\cite{dia-ult} where it was used to develop an institution-independent technique of the ultraproduct method for proving two important results for first-order logics cast as institutions: compactness and axiomatizability.
A trace of basic sentences can be found in \cite{DBLP:journals/tcs/Tarlecki85} under the name ``positive elementary sentence''.
According to \cite{dia-ult}, in first-order logic, any set of atomic sentences is basic. 
In hybrid logics, the atomic sentences are not basic because homomorphisms do not preserve the global satisfaction of atomic sentences.
However, homomorphisms preserve the local satisfaction of atomic sentences.

\begin{proposition} [Basic set of sentences] \label{prop:loc-basic}
Given a signature $\Delta$,
any set of sentences $\Psi\subseteq \Sen_b(\Delta)$ is epi-basic and its basic model $\M_\Psi$ is reachable.
\end{proposition}
The proof of Proposition~\ref{prop:loc-basic} is conceptually identical with the proof of \cite[Proposition 3.33]{gai-acm}.
The basic model of $\Psi$ is obtained from $T_\Delta$ by a factorization to some congruence relation.
\section{Forcing} \label{sec:forcing}
This section introduces a forcing technique that non-trivially generalizes forcing for Kripke structures with non-empty domains studied in \cite{gai-acm,GainaBK23}.
To illustrate the motivation behind our approach, we give the following example.
\begin{example}\label{ex:classic}
Let $\Delta=(\Sigma^\nom,\Sigma^\rigid\subseteq\Sigma)$ be a signature defined as follows:
\begin{itemize}
\item $\Sigma^\nom$ consists of two nominals $k_1$ and $k_2$ and one modality $\lambda$;
\item $\Sigma$ consists of a finite number of rigid sorts $\{s_1,\dots,s_n\}$, where $n\geqslant 2$. 
\end{itemize}
For each $i\in\{1,\dots, n\}$, 
let $x_i$ be a variable of sort $s_i$ and define
\begin{center}
$\gamma_i\coloneqq (\Exists{x_i}\top) \implies \neg(\at{k_1}\pos{\lambda^i}k_2)$,
\footnote{The sentence $\pos{\lambda^i} \phi$ is an abbreviation for $\underbrace{\pos{\lambda}\dots\pos{\lambda}}_{i\text{-times}}\phi$.}
\end{center}
which states that if the sort $s_i$ has at least one element then there is no path of edges labeled $\lambda$ of length $i$ from $k_1$ to $k_2$.
We define 
\begin{center}
$\gamma\coloneqq \bigvee_{i=1}^n \at{k_1}\pos{\lambda^i}k_2$, 
\end{center}
which asserts that there is a path of edges from $k_1$ to $k_2$ labeled $\lambda$ of length i, for some $i\in\{1,\dots,n\}$.
Finally, we define $\Gamma\coloneqq \{\gamma\}\cup\{ \gamma_i\mid 1\leqslant i\leqslant n\}$. 
\end{example}
Let $\Delta$ and $\Gamma$ be as defined in Example~\ref{ex:classic}, and let $C = \{C_s\}_{s\in S^\ext}$ be a set of constants such that $\card(C_s) = \omega$ for all $s\in S^\ext$. These constants, traditionally referred to as Henkin constants, are introduced to facilitate the construction of models of $\Gamma$ over the extension $\Delta[C]$ of $\Delta$. 
Although $\Gamma$ is consistent over $\Delta$, it becomes inconsistent over $\Delta[C]$:
\begin{itemize}
\item[] Assume that $(\M,w)\models_{\Delta[C]}\Gamma$.
For all $i\in \{1,\dots,n\}$, since $(\M,w)\models_{\Delta[C]} \Exists{x_i}\top$, we have $(\M,w)\models_{\Delta[C]}\neg(\at{k_1}\pos{\lambda^i}k_2)$.
On the other hand, since $(\M,w)\models_{\Delta[C]} \bigvee_{i=1}^n \at{k_1}\pos{\lambda^i}k_2$, we have $(\M,w)\models_{\Delta[C]}\at{k_1}\pos{\lambda^i}k_2$ for some $i\in \{1,\dots,n\}$.
Hence, we have obtained a contradiction.
\end{itemize}
This observation motivates the development of a forcing technique that permits the dynamic introduction of Henkin constants during the construction, rather than requiring them to be fixed in advance.

There is another important respect in which the present work differs from the forcing developments of \cite{gai-acm,DBLP:journals/apal/GainaBK23}. The forcing methods proposed in those papers are based on the global satisfaction relation. Consequently, given a sentence $\phi$, they can be used to construct a model $\M$ that globally satisfies $\phi$. In contrast, the forcing method developed here is based on the local satisfaction relation. Its objective is to construct a pointed model $(\M,w)$ such that $\phi$ holds at the designated world $w$. To achieve this, we extend the signature $\Delta$ of $\phi$ with a fresh nominal $c$ and construct a model $\M$ over the expanded signature $\Delta[c]$ satisfying $\M \models \at{c}\phi$. The world named by $c$ then serves as the designated point at which $\phi$ holds, yielding the desired pointed model. 

The distinction between the global and local approaches is essential. Consider, for instance, the sentence \[ \phi \coloneqq k_1 \land \neg \at{k_2} k_1. \] No model can globally satisfy $\phi$, since at the world denoted by $k_2$ the formula $k_1$ must be false. Nevertheless, $\phi$ is locally satisfiable. Indeed, any pointed model $(\M,w)$ with at least two worlds $w$ and $v$, such that $w = k_1^\M$ and $v = k_2^\M$, satisfies $\phi$ at $w$. Thus, while global forcing cannot be used to construct a model for $\phi$, a forcing method based on local satisfaction can. 
An illustration of this idea is provided by the construction of the generic set in Theorem~\ref{th:interpolation}.
\subsection{Concepts and results}
We develop a model-theoretic notion of forcing for $\FOHLR$, extending the forcing construction of \cite{gai-acm} from individual signatures to categories of signatures. 
It is worth mentioning that the results of this section apply to any fragment $\frag$ of $\FOHLR$ that is closed under the retrieve $@$ and negation $\neg$. 
In particular, $\frag$ may be taken to be $\HPL$ or $\RFOHL$.
The main ingredients of the theory are: 
\begin{enumerate} \item \emph{forcing properties}, namely collections of abstract conditions compatible with the local satisfaction of basic sentences; 
\item \emph{forcing relations}, canonically induced by forcing properties and intended to mimic satisfaction on reachable models; 
\item \emph{generic sets}, that is, increasing chains of conditions from which reachable models can be constructed; 
\item \emph{generic models}, which are concrete model-theoretic realizations of generic sets. 
\end{enumerate}
\begin{definition}[Forcing property] \label{def:forcing}
\emph{A forcing property} is a tuple $\mathbb P=(P,\leqslant,\Omega,f)$, where:

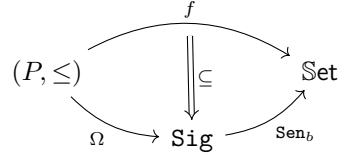
\begin{center}
\begin{tikzcd}[row sep=small]
(P,\leqslant) \ar[dr,bend right=20,"\Omega",swap] \ar[rr,bend left,"f", ""{below,name=U}] &  & \Set\\
& \Sig \ar[ur,bend right=20,"\Sen_b",swap]  \ar[Rightarrow,from = U,"\subseteq"] & 
\end{tikzcd}
\end{center}

\begin{enumerate}
\item $(P,\leqslant)$ is a partially ordered set with a least element $0$.
		
The elements of $P$ are traditionally called conditions.
		
\item $\Omega: (P,\leqslant) \to \Sig$ is a functor, which maps each arrow $(p\leqslant q)\in (P,\leqslant)$ to an inclusion $\Omega(p)\subseteq \Omega(q)$.
\item $f:(P,\leqslant)\to\Set$ is a functor from the small category $(P,\leqslant)$ to the category of sets $\Set$  such that $f \subseteq \Omega;\Sen_b$ is a natural transformation, that is:
\begin{enumerate*}[label=(\alph*)]
\item~$f(p)\subseteq \Sen_b(\Omega(p))$ for all conditions $p\in P$, and
\item~$f(p)\subseteq f(q)$ for all relations $(p\leqslant q)\in (P,\leqslant)$.
\end{enumerate*}
\item If $f(p)\models \varphi$ then $\varphi\in f(q)$ for some $q\geqslant p$, for all sentences $\varphi\in \Sen_b(\Omega(p))$.
\end{enumerate}
\end{definition}
A classical forcing property is a particular case of forcing property such that $\Omega(p)=\Omega(q)$ for all conditions $p,q\in P$.
We extend our notational conventions by assuming that any signature $\Omega(p)$ is of the form $(\Sigma_p^\nom,\Sigma_p^\rigid\subseteq \Sigma_p)$, 
where 
$\Sigma_p^\nom=(\{\mathtt{nom}\},F_p^\nom,P_p^\nom)$, 
$\Sigma_p^\rigid=(S_p^\rigid,F_p^\rigid,P_p^\rigid)$ and 
$\Sigma_p=(S_p,F_p,P_p)$.
Also, we write $\Omega_p$ instead of $\Omega(p)$.
As usual, forcing properties determine suitable relations between conditions and sentences.
\begin{definition} [Forcing relation] \label{def:forcing-relation}
Let $\mathbb P=\langle P,\leqslant,\Omega,f \rangle$ be a forcing property. \emph{The forcing relation} $\Vdash$ between pairs $(p,k)$, where $p\in P$ is a condition and $k\in F_p^\nom$ is a nominal, and sentences $\phi\in\Sen(\Omega_p)$, that is, 
\[ \Vdash \;\subseteq\; \bigcup_{p\in P}\bigl(\{p\}\times F_p^\nom \bigr) \times \Sen(\Omega_p), \] 
is defined inductively on the structure of the sentence $\phi$ as follows:
\begin{description}
\item[$\phi\in\Sen_0(\Omega_p)$:] $(p,k)\Vdash \phi$ if $\at{k}\phi \in f(p)$.
\item[$\lor \Phi$:] $(p,k)\Vdash \lor \Phi$ if $(p,k)\Vdash \phi$ for some $\phi\in \Phi$.
\item[$\neg \phi$:] $(p,k)\Vdash \neg \phi$ if $(q,k)\not\Vdash \phi$ for all $q\geqslant p$.
\item[$\at{\ell}\phi$:] $(p,k)\Vdash \at{\ell}\phi$ if $(p,\ell)\Vdash \phi$.
\item[$\pos{\lambda}\phi$:] $(p,k)\Vdash\pos{\lambda}\phi$ if $\at{k}\pos{\lambda}\ell\in f(p)$ and $(p,\ell)\Vdash \phi$ for some nominal $\ell$ in $\Omega_p$.
\item[$\store{x}\phi_x$:] $(p,k)\Vdash \store{x}\phi_x$ if $(p,k)\Vdash \phi_{x\leftarrow k}$, 
where $\phi_{x\leftarrow k}$ is obtained from $\phi_x$ by substituting $k$ for $x$.
\item[$\Exists{x}\phi_x$:] $(p,k)\Vdash \Exists{x}\phi_x$ if $(p,k)\Vdash \theta(\phi_x)$ for some $\Omega_p$-substitution $\theta:\{x\} \to \emptyset$.
\end{description}
The relation $(p,k)\Vdash \phi$ in $\mathbb{P}$, is read as $p$ forces $\phi$ at $k$.
A condition $p$ weakly forces $\phi$ at $k$, in symbols, $(p,k)\Vdash^w\phi$ if for all $q\geqslant p$ there exists $r\geqslant q$ such that $(r,k)\Vdash \phi$.
\end{definition}
It is not difficult to show that $(p,k)\Vdash \neg\phi$ iff $(p,k)\Vdash^w \neg\phi$.
Other basic properties of forcing are presented below.
\begin{lemma}[Forcing properties] \label{lemma:forcing-property}
Let $\mathbb P=(P,\leqslant,\Omega,f)$ be a forcing property. 
For all conditions $p\in P$ and all sentences $\phi\in\Sen(\Omega_p)$ we have: 
\begin{enumerate}
\item \label{fp-1} $(p,k)\Vdash \neg\neg \phi$ iff for each $q\geqslant p$ there is a condition $r\geqslant q$ such that $r\Vdash \phi$.
		
\item \label{fp-2} If $p\leqslant q$ and $(p,k)\Vdash \phi$ then $(q,k)\Vdash \phi$.
		
\item  \label{fp-3} If $(p,k)\Vdash \phi$ then $(p,k)\Vdash \neg\neg \phi$.
		
\item \label{fp-4} We can not have both $(p,k)\Vdash \phi$ and $(p,k)\Vdash \neg \phi$.
\end{enumerate}
\end{lemma}
The second property stated in the above lemma shows that the forcing relation is preserved along inclusions of conditions.
The fourth property shows that the forcing relation is consistent, that is, a condition cannot force all sentences.
The remaining conditions are about negation.
The proof of Lemma~\ref{lemma:forcing-property} is conceptually identical with the proof of \cite[Lemma 4.4]{gai-acm}.
\begin{definition}[Generic set] \label{def:gs}
Let $\mathbb P=(P,\leqslant,\Omega,f)$ be a forcing property.
A non-empty subset of conditions $G\subseteq P$ is generic if
\begin{enumerate}
\item\label{gs-1} $G$ is an ideal, that is:
\begin{enumerate*}[label=(\alph*)]
\item~for all $p\in G$ and all $q\leqslant p$ we have $q\in G$, and
\item~for all $p,q\in G$ there exists $r\in G$ such that $p\leqslant r$ and $q\leqslant r$; 
\end{enumerate*} and
\item\label{gs-2} for all conditions $p\in G$, all sentences $\phi \in \Sen(\Omega_p)$ and all nominals $k$ in $\Omega_p$,
there exists a condition $q\in G$ such that $q\geqslant p$ and either $(q,k)\Vdash \phi$ or $(q,k)\Vdash \neg \phi$ holds.
\end{enumerate}
We write $(G,k)\Vdash\phi$ if $(p,k)\Vdash\phi$ for some $p\in G$.
\end{definition}
A generic set $G$ describes a reachable model which satisfies all sentences forced by the conditions in $G$.
\begin{remark}\label{rem:gsr}
Let $\Omega_G\coloneqq \bigcup_{p\in G}\Omega_p$ and observe that
$\mu: \Omega \Rightarrow \Omega_G$, where $\mu_p:\Omega_p\hookrightarrow\Omega_G$ is an inclusion for all $p\in G$,
is  the co-limit of the functor $\Omega:(G,\leqslant)\to \Sig$.
\end{remark}
The results are developed over the signature $\Omega_G=(\Sigma^\nom_G,\Sigma^\rigid_G\subseteq \Sigma_G)$ which is the vertex of the co-limit $\mu: \Omega \Rightarrow \Omega_G$. 
We refer to $\Omega_G$ as the signature of $G$.
\begin{definition} [Generic model]
Let $\mathbb P=(P,\leqslant,\Omega,f)$ be a forcing property and $G\subseteq P$ a generic set.
A model $\M $ defined over $\Omega_G$ is a \emph{generic model} for $G$ iff 
for every $\Omega_G$-sentence $\phi$ and any nominal $k$ in $\Omega_G$, 
we have $(\M ,k^W)\models \phi \mbox{ iff } (G,k)\Vdash \phi$.
\end{definition}
The notion of generic model is the semantic counterpart of the definition of generic set.
The following result shows that every generic set has a generic model.
\begin{theorem}[Generic Model Theorem] \label{th:gm}
Let $\mathbb P=(P,\leqslant,\Omega,f)$ be a forcing property and $G\subseteq P$ a generic set.
Then there is a generic model for $G$ which is reachable.
\end{theorem}
\begin{proof}
Let $\Psi\coloneqq\{\at{k}\phi\in\Sen_b(\Omega_G) \mid (G,k)\Vdash\phi\}$ be the basic set of sentences forced by the generic set $G$.
Let $\M_\Psi=(W_\Psi,M_\Psi)$ be a basic model of $\Psi$ that is reachable.
We show that $\M_\Psi$ is a generic model for $G$.
\begin{description}
\item[$\phi\in \Sen_0(\Omega_G)$:]  
For the forward implication, assume that $(\M_\Psi,k^{W_\psi})\models \phi$.
\begin{proofsteps}{20em}
there exists a homomorphism $\M_{\{\at{k} \phi\}}\to \M_\Psi$ &
as $(\M_\Psi,k^{W_\psi})\models\phi$ and $\at{k}\phi$ is basic \\
$\Psi\models\at{k}\phi$ & 
since both $\Psi$ and $\at{k}\phi$ are basic\\ 
$\Psi'\models\at{k}\phi$ for some finite $\Psi' \subseteq \Psi$ &
by compactness\\
$\Psi'\subseteq f(p)$ for some $p\in G$ & 
since $(G,\leqslant)$ is generic and $\Psi'$ is finite\\
$f(p)\models\at{k}\phi$ & 
since $\Psi'\models\at{k}\phi$ and $\Psi'\subseteq f(p)$\\ 
$(G,k)\Vdash\phi$ or $(G,k)\Vdash \neg\phi$ & 
since $G$ is generic\\
\rlap{suppose towards a contradiction that $(G,k)\Vdash\neg\phi$} &
\substeps{9}
$(q,k)\Vdash \neg\phi$ for some $q\in G$ &
since $(G,k)\Vdash \neg\phi$\\
$r\geqslant q$ and $r\geqslant p$ for some $r\in G$ &
since $(G,\leqslant)$ is an ideal\\
$(r,k)\Vdash\neg\phi$ & 
since $(q,k)\Vdash\neg\phi$ and $q\leqslant r$\\
$f(r)\models\at{k}\phi$ & 
since $f(p)\models\at{k}\phi$ and $f(p)\subseteq f(r)$ \\
$\at{k}\phi\in f(u)$ for some $u\geqslant r$ & by the definition of forcing property\\
$(u,k)\Vdash\phi$ & 
by the definition of forcing property\\
$(u,k)\Vdash\neg\phi$ & 
since $u\geqslant r$ and $(r,k)\Vdash\neg\phi$ \\
contradiction & 
since $(u,k)\Vdash\phi$ and $(u,k)\Vdash\neg\phi$
\endsubsteps
$(G,k)\Vdash\phi$ & since $(G,k)\not\Vdash\phi$
\end{proofsteps}

For the backward implication, assume that $(G,k)\Vdash\phi$.
By the definition of $\Psi$, we have $\at{k}\phi\in \Psi$.
Since $\M_\Psi\models \Psi$, we get $(\M_\Psi,k^{W_\Psi})\models\phi$.
\item[$\lor\Phi$:]
The following are equivalent: 
\begin{proofsteps}{23em}
  $(\M_\Psi,k^{W_\Psi})\models \lor\Phi$ & 
  \\
  $(\M_\Psi,k^{W_\Psi})\models \phi$ for some $\phi\in \Phi$ &
  by the semantics of disjunction\\
  $(G,k)\Vdash \phi$ for some $\phi\in \Phi$  & 
  by induction hypothesis\\
  $(G,k)\Vdash \lor \Phi$ & 
  by the definition of $\Vdash$
\end{proofsteps}
\item[$\neg\phi$:]
The following are equivalent:
\begin{proofsteps}{22em}
$(\M_\Psi,k^{W_\Psi})\models \neg \phi$ & 
\\
 $(\M_\Psi,k^{W_\Psi})\not \models \phi$ & 
by the semantics of negation\\
 $(G,k)\not \Vdash \phi$ & 
by induction hypothesis \\
 $(p,k)\not \Vdash \phi$ for all $p\in G$ & 
by the definition of $\Vdash$\\
 $(p,k)\Vdash \neg \phi$ for some $p\in G$ & 
since $G$ is generic\\
 $(G,k)\Vdash \neg \phi$
\end{proofsteps}
\item[$\at{\ell} \phi$:]
This case is straightforward since $\at{k} \at{\ell} \phi$ is semantically equivalent to $\at{\ell} \phi$.
 \item[$\pos{\lambda}\phi$:]
The following are equivalent:
\begin{proofsteps}{25em} 
$(\M_\Psi,k^{W_\Psi})\models\pos{\lambda}\phi $ & \\
$(\M_\Psi,k^{W_\Psi})\models \pos{\lambda}\ell$ and $(\M_\Psi,\ell^{W_\Psi})\models\phi$ 
for some $\ell\in F_G^\nom$ &
since $\M_\Psi$ is reachable \\ 
  
$(G,k)\Vdash\pos{\lambda}\ell$ and $(G,\ell)\Vdash \phi$ for some $\ell\in F_G^\nom$ &
by induction hypothesis \\
  
$(G,k)\Vdash\pos{\lambda}\phi$ & since $G$ is generic
\end{proofsteps}
\item [$\store{x}\phi_x$:] Straightforward as $\at{k}\store{x}\phi_x$ is semantically equivalent to $\at{k}\phi_{x\leftarrow k}$.
 \item[$\Exists{x}\phi_x$:]
The following are equivalent:

\begin{proofsteps}{25em}
$(\M_\Psi,k^{W_\Psi})\models \Exists{x}\phi_x$ & 
\\
$(\M_\Psi^{x\leftarrow m},k^{W_\psi})\models \phi_x$ 
for some for some element $m$ of $\M$ & 
by semantics\\
 $(\M_\Psi,k^{W_\Psi})\models \theta(\phi_x)$ for some substitution $\theta: \{x\}\to \emptyset$ & 
since $\M_\Psi$ is reachable\\
 $(G,k)\Vdash\theta(\phi_x)$ for some substitution $\theta: \{x\}\to \emptyset$ & 
by induction hypothesis\\
 $(G,k)\Vdash\Exists{x}\phi_x$ & 
by the definition of $\Vdash$
 \end{proofsteps}
\end{description}
\end{proof}
\section{Semantic Forcing}\label{sec:semantic-forcing}
We study a semantic forcing property that will be used to establish the Craig interpolation property for $\FOHLR$. 
It is worth emphasizing that the results of this section apply to any fragment $\frag$ of $\FOHLR$ that is closed under the retrieve operator $@$ and negation $\neg$. 
In particular, $\frag$ may be taken to be $\HPL$ or $\RFOHL$.
\begin{example}[Semantic forcing] \label{ex:semantic-forcing}
Let $\Delta=(\Sigma^\nom,\Sigma^\rigid\subseteq\Sigma)$ be a signature and $\alpha$ a cardinal greater or equal than the power of $\Delta$, that is, $\alpha \geqslant \card(\Sen(\Delta))$.
Let $C=\{C_s\}_{s\in S^\ext}$ be an $S^\ext$-sorted set of new rigid constants for $\Delta$ such that $\card(C_s)=\alpha$ for all sorts $s\in S^\ext$.
The semantic forcing property $\mathbb{P}=(P,\leqslant,\Omega,f)$ is defined as follows:
\begin{enumerate}
\item $P$ is the set of presentations of the form $p=(\Omega_p,\Gamma_p)$,
where 
\begin{enumerate}[label=(\alph*)]
\item $\Omega_p$ is obtained from $\Delta$ by adding a set of constants $C_p\subseteq C$ of cardinality $\card(C_p)<\alpha$,
\item $\Gamma_p$ is a consistent set of $\Omega_p$-sentences.
\end{enumerate}

\item $p\leqslant q$ iff $\Omega_p\subseteq\Omega_q$ and $\Gamma_p\subseteq \Gamma_q$, for all conditions $p,q\in P$.
\item $\Omega$ is the forgetful functor which maps each condition $p\in P$ to $\Omega_p$.
\item  $f(p)\coloneqq \Gamma_p\cap \Sen_b(\Omega_p)$, for all conditions $p \in P$.
\end{enumerate}
\end{example}
Example~\ref{ex:semantic-forcing} sets the fundation for proving CIP in $\FOHLR$.
The conditions are presentations over extensions of $\Delta$ with a number of constants less than the power of $\Delta$.
The functor $f$ maps each condition $p=(\Omega_p,\Gamma_p)$ to the set of locally basic sentences in $\Gamma_p$.
We establish some notational conventions that will be useful in the subsequent developments.
Let $p=(\Omega_p,\Gamma_p)$ and $q=(\Omega_q,\Gamma_q)$ be two conditions as described in Example~\ref{ex:semantic-forcing}.
\begin{itemize}
\item $p\models \phi$ iff $\Gamma_p\models_{\Omega_p}\phi$ for all sentences $\phi\in\Sen(\Omega_p)$.
\item $p\cup q = (\Delta[C_p\cup C_q], \Gamma_p\cup \Gamma_q)$, 
where $C_p$ and $C_q$ are all constants from $C$ that occur in $\Omega_p$ and $\Omega_q$, respectively.
\end{itemize}
\begin{lemma}\label{lemma:semantic-forcing}
Let $\mathbb{P}=\langle P, \leqslant, \Omega, f \rangle $ be a forcing property as described in Example~\ref{ex:semantic-forcing}. 
Let $p\in P$ be a condition.

\begin{enumerate}
\item \label{sf1} 
If  $p\models \lor \Phi$ then $p\cup(\Omega_p,\phi)\in P$ for some $\phi\in\Phi$.
\item \label{sf2} 
If $p \models \at{k}\pos{\lambda}\phi$
then $p\cup(\Omega_p[c],\{\at{k}\pos{\lambda}c,\at{c}\phi\})\in P$ for some $c\in C_{\mathtt{nom}}$.
\item  \label{sf3} 
If $p\models \Exists{x}\phi_x$ 
then $p\cup (\Omega_p[c],\phi_{x\leftarrow c})\in P$ 
for some $c\in C\setminus C_p$,

where $\phi_{x\leftarrow c}$ denotes the sentence obtained from $\phi_x$ by substituting $c$ for $x$.
\end{enumerate}
\end{lemma}
\begin{proof}
Let $p\in P$ be a condition.
\begin{enumerate}
\item Assume that $p\models\lor \Phi$.
We have $(\M,w)\models \Gamma_p$ for some pointed model $(\M,w)$.
Since $\Gamma_p\models\lor\Phi$,
we have $(\M,w)\models \lor\Phi$,
which means $(\M,w) \models \phi$ for some $\phi\in \Phi$.
It follows that $\Gamma_p \cup \{ \phi \} \not\models_{\Omega_p}\bot$.
Hence, $p\cup (\Omega_p,\phi)\in P$.
\item Assume that $p\models \at{k}\pos{\lambda}\phi$.
Let $c\in C_{\mathtt{nom}}\setminus (C_p)_{\mathtt{nom}}$ and show that $\Gamma_p\cup\{\at{k}\pos{\lambda}c,\at{c}\phi\}$ is consistent over $\Omega_p[c]$:
\begin{proofsteps}{20em}
$(\M,w) \models \Gamma_p$ for some pointed model $(\M,w)$ & 
since $\Gamma_p$ is consistent \\
$(\M,k^W)\models \pos{\lambda}\phi$ & 
as $(\M,w)\models\Gamma_p$ and $\Gamma_p\models\at{k}\pos{\lambda}\phi$\\
$(\M,v) \models\phi$ for some $v\in \lambda^W(k^W)$ &
by semantics\\
\rlap{let $\M^{c\leftarrow v}=(W^{c\leftarrow v},M)$ be the expansion of $\M $ to $\Omega_p[c]$ which interprets $c$ as $v$}&\\
$\M^{c\leftarrow v} \models \at{k}\pos{\lambda}c$ and $\M^{c\leftarrow v} \models\at{c}\phi$
& since $v\in \lambda^W(k^W)$ and $\lambda^{W^{c\leftarrow v}}=\lambda^W$\\
$(\M^{c\leftarrow v},w) \models \Gamma_p$
&
since $(\M,w) \models \Gamma_p$ and $ \M^{c\leftarrow v}\red_{\Omega_p} =\M$\\
$\Gamma_p\cup\{\at{k}\pos{\lambda}c, \at{c}\phi\}$ is consistent over $\Omega_p[c]$ 
& since $(\M^{c\leftarrow v},w) \models \Gamma_p\cup\{\at{k}\pos{\lambda}c, \at{c}\phi\}$ 
\end{proofsteps}
Hence, $p\cup (\Omega_p[c],\{\at{k}\pos{\lambda}c,\at{c}\phi\})\in P$.
\item Assume that $p\models\Exists{x}\phi_x$.
Since $\card(C_{p,s})<\alpha$ and $\card(C_s)=\alpha$ for all $s\in S^\rigid$,
there exists $c\in C\setminus C_p$ of the same sort as $x$.
We show that $\Gamma_p\cup\{\phi_{x\leftarrow c}\}$ is consistent over $\Omega_p[c]$:
\begin{proofsteps}{19em}
$(\M,w) \models \Gamma_p$ for some pointed model $(\M,w)$  & 
since $\Gamma_p$ is consistent over $\Omega_p$\\
$(\M,w)\models \Exists{x} \phi_x$ & 
as $(\M,w) \models \Gamma_p$ and $\Gamma_p\models \Exists{x}\phi_x$\\
$(\M^{x\leftarrow m},w)\models_{\Omega_p[x]} \phi_x$ for some element $m$ & by semantics\\
$(\M^{c\leftarrow m},w) \models_{\Omega_p[c]} \phi_{x\leftarrow c}$ &
by the satisfaction condition\\ 
$(\M^{c\leftarrow m},w) \models_{\Omega_p[c]} \Gamma_p$ &
since $\M^{c\leftarrow m} \red_{\Omega_p}=\M $ and $(\M,w)\models\Gamma_p$\\
$\Gamma_p\cup\{\phi_{x\leftarrow c}\}$ is consistent over $\Omega_p[c]$ & 
as $(\M^{c\leftarrow m},w) \models \Gamma_p$ \& $(\M^{c\leftarrow m},w) \models \phi_{x\leftarrow c}$
\end{proofsteps}
It follows that $p\cup (\Omega_p[c],\phi_{x\leftarrow c})\in P$.
\end{enumerate}
\end{proof}
Lemma~\ref{lemma:semantic-forcing} sets the basis for the following important result concerning semantic forcing properties, which says that all sentences of a given condition are forced eventually by some condition greater or equal than the initial one.

\begin{theorem}[Semantic Forcing Theorem] \label{th:semantic-forcing}
Let $\mathbb{P}=\langle P, \leqslant, \Omega,f \rangle $ be the semantic forcing property from Example~\ref{ex:semantic-forcing}.
For all conditions $p\in P$ and all $\Omega_p$-sentences $\at{k}\phi$, we have:
\begin{center}
$ (q,k)\Vdash \phi$ for some $q\geqslant p$ iff $p\cup(\Omega_p,\at{k}\phi)\in P$.
\end{center}
\end{theorem}
\begin{proof}
We proceed by induction on the structure of $\phi$.
\begin{description}
\item[$\phi\in\Sen_0(\Omega_p)$]
Assume that $(q,k)\Vdash\phi$ for some  $q\geqslant p$. 
 We show that $p\cup(\Omega_p,\at{k}\phi)\in P$:
\begin{proofsteps}{21em}
$\at{k}\phi\in\Gamma_q\cap\Sen_b(\Omega_q)$ & 
by the definition of $\Vdash$\\
$\Gamma_q\models \Gamma_p\cup \{\at{k}\phi\}$ & 
since $\Gamma_p\subseteq\Gamma_q$ and $\at{k}\phi\in \Gamma_q$  \\
$(\M,w) \models_{\Omega_q} \Gamma_q$ for some pointed model $(\M,w)$ &
since $q\in P$\\ 
$(\M\red_{\Omega_p},w) \models_{\Omega_p}\Gamma_p\cup\{\at{k}\phi\}$ & since $(\M,w)\models_{\Omega_q} \Gamma_p\cup\{\at{k}\phi\}$\\
$p\cup(\Omega_p,\at{k}\phi)\in P$ & 
since $\Gamma_p\cup\{\at{k}\phi\}$ is consistent over $\Omega_p$
\end{proofsteps} 
Assume that $p\cup (\Omega_p,\at{k}\phi)\in P$. 
Let $q=p\cup(\Omega_p,\at{k}\phi)$. 
By the definition of forcing relation, we have $(q,k)\Vdash \phi$.
\item[$\lor \Phi$] Assume that there exists $q\geqslant p$ such that $(q,k)\Vdash \lor\Phi$. 
We show that $p\cup(\Omega_p,\at{k} \lor\Phi)\in P$:

\begin{proofsteps}{21em}
$(q,k)\Vdash\phi$ for some $\phi\in\Phi$ & 
by the definition of $\Vdash$\\
$p\cup (\Omega_p,\at{k}\phi) \in P$ & 
by induction hypothesis\\
$p\cup (\Omega_p,\at{k} \lor\Phi) \in P$ & since $\at{k}\phi \models \at{k}\lor\Phi$
\end{proofsteps}

Assume that $p\cup (\Omega_p,\at{k}\lor\Phi)\in P$.
We show that $(q,k)\Vdash \lor\Phi$ for some $q\geqslant p$:
\begin{proofsteps}{21em}
$p\cup (\Omega_p,\{ \at{k}\lor\Phi,\at{k}\phi\})\in P$ for some $\phi\in \Phi$ & 
by Lemma~\ref{lemma:semantic-forcing}(\ref{sf1}) \\
$(q,k)\Vdash\phi$ for some $q\geqslant p\cup (\Omega_p, \at{k} \lor\Phi)$ & 
by induction hypothesis\\
$(q,k)\Vdash \lor\Phi$ for some $q\geqslant p$ & by the definition of $\Vdash$
\end{proofsteps}
\item[$\neg\phi$] By the induction hypothesis, for each $q\in P$, the following are equivalent:
 \begin{enumerate}[label=(S\arabic*), ref= S\arabic*]
 
 \item \label{neg-sf1} $(r,k)\Vdash \phi$  for some $r\geqslant q$ iff  $q\cup (\Omega_q, \at{k}\phi)\in P$.
 
 \item \label{neg-sf2} $(r,k)\not\Vdash \phi$  for all  $r\geqslant q$ iff  $q\cup (\Omega_q, \at{k}\phi) \not\in P$.

 \item \label{neg-sf3} $(q,k)\Vdash \neg\phi$ iff  $q\cup(\Omega_q, \at{k}\phi) \not \in P$.
 \end{enumerate}

Assume that $(q,k)\Vdash \neg\phi$  for some $q\geqslant p$. 
We show that $p\cup(\Omega_p,\at{k} \neg\phi)\in P$:
\begin{proofsteps}{21em}
  $q\cup\{\Omega_p, \at{k}\phi\}\not\in P$ & by \ref{neg-sf3}, since $(q,k)\Vdash \neg\phi$\\
  
$(\M,w)\models \Gamma_q$ for some pointed model $(\M,w)$ & 
by the definition of $\Vdash$ \\
$(\M,w) \not\models \at{k}\phi$ & 
since $q\cup(\Omega_q,\at{k}\phi)\not\in P$ \\
$q\cup (\Omega_q,\at{k}\neg\phi)\in P$ & 
since $(\M,w)\models \Gamma_q\cup\{\at{k}\neg\phi\}$\\
$p\cup(\Omega_p,\at{k}\neg\phi)\in P$ & 
since $(\M\red_{\Omega_p},w) \models \Gamma_p\cup\{\at{k}\neg\phi\}$
\end{proofsteps}
Assume that $p\cup(\Omega_p,\at{k}\neg\phi)\in P$.
We show that $(q,k)\Vdash\neg\phi$ for some $q\geqslant p$:
\begin{proofsteps}{21em}
let  $q=p\cup(\Omega_p,\at{k} \neg\phi)$ & \\ 
$q\cup (\Omega_p,\at{k}\phi)\not\in P$ & since $\at{k}\neg\phi \in \Gamma_q$ \\
$(q,k)\Vdash \neg\phi$ & by statement~\ref{neg-sf3} 
\end{proofsteps}
\item[$\at{\ell}\phi$]
This case is straightforward, as $\at{k}\at{\ell}\phi$ is semantically equivalent to $\at{\ell}\phi$.
\item[$\pos{\lambda}\phi$]
Assume that $(q,k)\Vdash\pos{\lambda}\phi$ for some $q\geqslant p$.
We show that $p\cup(\Omega_p,\at{k} \pos{\lambda} \phi) \in P$:
\begin{proofsteps}{20em}
$\at{k}\pos{\lambda}\ell\in f(q)$ and $(q,\ell)\Vdash \phi$ for some $\ell\in F^\nom_q$ & 
from $(q,k)\Vdash \pos{\lambda}\phi$, by Definition~\ref{def:forcing-relation} \\
$r\coloneqq q\cup(\Omega_q,\at{\ell}\phi)\in P$ &
by induction hypothesis, since $(q,\ell)\Vdash \phi$
\\
$ \Gamma_r\models \Gamma_p \cup \{ \at{k}\pos{\lambda}\phi \} $ & 
since $\Gamma_p\subseteq \Gamma_r$ and $\Gamma_r \models \at{k}\pos{\lambda}\phi$\\
$p\cup (\Omega_p,\at{k}\pos{\lambda}\phi)\in P$ & as $\Gamma_p\cup\{\at{k}\pos{\lambda}\phi\}$ is consistent over $\Omega_p$
\end{proofsteps}
Assume that $p\cup (\Omega_p,\at{k}\pos{\lambda}\phi)\in P$.
We show that $(q,k)\Vdash \pos{\lambda}\phi$ for some $q\geqslant p$:

\begin{proofsteps}{19em}
$p\cup(\Omega_p[c],\{ \at{k}\pos{\lambda}\phi, \at{k} \pos{\lambda}c,\at{c} \phi \}) \in P$ for some $c\in C_{\mathtt{nom}} $ & 
by Lemma~\ref{lemma:semantic-forcing}(\ref{sf2})\\
let $r \coloneqq p\cup(\Omega_p[c],\{ \at{k}\pos{\lambda}\phi, \at{k} \pos{\lambda}c\})$ \\
$(r,k)\Vdash\pos{\lambda}c$ & 
since $\at{k}\pos{\lambda}c\in f(r)$\\ 
$(q,c)\Vdash \phi$ for some $q\geqslant r$ & 
by induction hypothesis, as $r\cup\{\at{c}\phi\}\in P$ \\
$(q,k)\Vdash \pos{\lambda} \phi$ & 
as $\at{k}\pos{\lambda}c\in f(q)$ and $(q,c)\Vdash \phi$ 
\end{proofsteps}
\item [$\store{x}\phi_x$] Straightforward as $\at{k}\store{x}\phi_x$ is semantically equivalent to $\at{k}\phi_{x\leftarrow k}$.
\item[$\Exists{x}\phi_x$] 
Assume that $(q,k)\Vdash \Exists{x}\phi_x$ for some $q\geqslant p$.
We show that $p\cup(\Omega_p,\at{k}\Exists{x}\phi_x)\in P$:

\begin{proofsteps}{20em}
$(q,k)\Vdash \phi_{x\leftarrow t} $ for some $t\in T_\Delta$ &
by the definition of $\Vdash$\\
$p\cup(\Omega_p,\at{k}\phi_{x\leftarrow t})\in P$ & 
by induction hypothesis \\ 
$p\cup(\Omega_p,\at{k}\Exists{x}\phi_x)\in P$ & 
since $\at{k}\phi_{x\leftarrow t}\models\at{k}\Exists{x}\phi_x$
\end{proofsteps}

 We assume that $p\cup(\Omega_p,\at{k}\Exists{x}\phi_x)\in P$.
 We show that $(q,k)\Vdash \Exists{x}\phi_x$ for some $q\geqslant p$:

\begin{proofsteps}{27em}
$p\cup(\Omega_p,\at{k}\Exists{x}\phi_x) \cup (\Omega_p[c],\at{k}\phi_{x\leftarrow c})\in P$ 
for some  $c\in  C\setminus C_p$ &  
by Lemma~\ref{lemma:semantic-forcing}(\ref{sf3})\\
    
$(q,k)\Vdash \phi_{x\leftarrow c}$ for some $q\geqslant p\cup(\Omega_p,\at{k}\Exists{x}\phi_x)$ & 
by induction hypothesis \\
    
$(q,k)\Vdash\Exists{x}\phi_x$ & 
by the definition of $\Vdash$
\end{proofsteps}
\end{description}
\end{proof}
The first corollary of Theorem~\ref{th:semantic-forcing} states that the local satisfaction relation coincides with the weak forcing relation.
\begin{corollary}\label{cor:semantic-forcing1}
Let $\mathbb{P}=\langle P, \leqslant, \Omega,f \rangle $ be a semantic forcing property as described in Example~\ref{ex:semantic-forcing}.
For all conditions $p\in P$, all $\Omega_p$-sentences $\at{k}\phi$, we have:
$(p,k)\Vdash^w\phi$ iff $p\models \at{k}\phi$.
\end{corollary}
\begin{proof}
For the forward implication,
assume that $(p,k)\Vdash^w\phi$.
Suppose towards a contradiction that $p\not\models\at{k}\phi$.
\begin{proofsteps}{20em}
$\Gamma_p\cup\{\at{k}\neg\phi\}\not\models_{\Omega_p}\bot$ & 
since $\Gamma_p\not\models_{\Omega_p}\at{k}\phi$\\
let $q\coloneqq p\cup (\Omega_p,\at{k}\neg\phi)$ & \\
$(r,k)\Vdash\phi$ for some $r\geqslant q$ &
since $q\geqslant p$ and $(p,k)\Vdash^w\phi$\\
$q\cup(\Omega_q,\at{k}\phi)\in P$ & 
by Theorem~\ref{th:semantic-forcing}\\
$p\cup(\Omega_p,\{\at{k}\neg\phi,\at{k}\phi\})\in P$ &
since $q= p\cup (\Omega_p,\at{k}\neg\phi)$ \\
contradiction & since $p\cup(\Omega_p,\{\at{k}\neg\phi,\at{k}\phi\})\models \bot$
\end{proofsteps}
For the backward implication,
assume that $p\models\at{k}\phi$.
Let $q\geqslant p$ and show that $(r,k)\Vdash \phi$ for some $r\geqslant q$.
\begin{proofsteps}{20em}
$q\models\at{k}\phi$ & since $q\geqslant p$ and $p\models\at{k}\phi$ \\
$q\cup (\Omega_q,\at{k}\phi)\in P$ & by semantics\\
 $(r,k)\Vdash \phi$ for some $r\geqslant q$ & by Theorem~\ref{th:semantic-forcing}
\end{proofsteps}
\end{proof}
The second corollary of Theorem~\ref{th:semantic-forcing}  shows that each generic set of a given semantic forcing property has a reachable model that satisfies all its conditions.
\begin{corollary} \label{cor:semantic-forcing2}
 Assume a semantic forcing property $\mathbb{P}=\langle P, \leqslant, \Omega,f \rangle $  as described in Example~\ref{ex:semantic-forcing}.
Let $M_G$ be a generic model for a generic set $G\subseteq P$.
Then $\M_G\models\at{k}\phi$,  for all conditions $p\in G$ and all sentences $\at{k}\phi \in \Gamma_p$.
\end{corollary}
\begin{proof}
First, assume 
a condition $p\in G$,
a sentence $\at{k}\phi\in \Gamma_p$.
By Corollary~\ref{cor:semantic-forcing1},
$(p,k)\Vdash^w \phi$.
We show that $(G,k)\Vdash\phi$:
\begin{itemize}
\item[] Suppose towards a contradiction that $(G,k)\not\Vdash\phi$.
Since $G$ is generic, $(p_1,k)\Vdash\neg\phi$ for some $p_1\in G$. 
Let $q\in G$ such that $q\geqslant p$ and $q\geqslant p_1$.
Since $(p,k)\Vdash^w\phi$, there exists $r\geqslant q$ such that $(r,k)\Vdash \phi$.
Since $(p_1,k)\Vdash\neg\phi$ and $r\geqslant p_1$, by Lemma~\ref{lemma:forcing-property}(\ref{fp-2}), $(r,k)\Vdash \neg\phi$, which is a contradiction.
\end{itemize}
Since $M_G$ is generic for $G$, we get $M_G\models\at{k}\phi$.
\end{proof}
\section{Craig Interpolation Property} \label{sec:CIP}
CIP is proved for $\FOHLR$ using the semantic forcing property defined in Example~\ref{ex:semantic-forcing}. 
\subsection{Framework}\label{sec:framework}
We set the framework in which CIP is proved.
\paragraph{Henkin constants}
Let $\Delta_\ac\stackrel{\chi}\leftarrow\Delta\stackrel{\delta}\to\Delta_\bc$ be a span of signature morphisms.
Let $\alpha$ be the smallest regular cardinal that is greater then $\card(\Sen(\Delta))$, $\card(\Sen(\Delta_\ac))$ and $\card(\Sen(\Delta_\bc))$.
\begin{enumerate}
\item Let $\C=\{\C_s\}_{s\in S^\ext}$ be an $S^\ext$-sorted set of new constants for the signature $\Delta$ such that 
\begin{enumerate}
\item $\card(\C_s)=\alpha$ for all sorts $s\in S^\ext$.
\item $\C_{s_1}\cap\C_{s_2}=\emptyset$ for all sorts  $s_1,s_2\in S^\ext$ with $s_1\neq s_2$.
\end{enumerate}
Traditionally, the constants in $\C$ are called Henkin constants.
\item Similarly, we choose sets of fresh Henkin constants $\A$ and $\B$ for $\Delta_\ac$ and $\Delta_\bc$, respectively.
\end{enumerate} 
\paragraph{Forcing properties}
Let $\mathbb{P}_\ac=(P_\ac,\leqslant_\ac,\Omega_\ac,f_\ac)$ denote the semantic forcing property introduced in Example~\ref{ex:semantic-forcing}, constructed over the signature $\Delta_\ac$ using the set of Henkin constants $\A$.
Similarly, we define the semantic forcing property
$\mathbb{P}_\bc=(P_\bc,\leqslant_\bc,\Omega_\bc, f_\bc)$  over $\Delta_\bc$ using the set of Henkin constants $\B$.
When no ambiguity arises, we omit the subscripts $\ac$ and $\bc$ from the notation for the last three components of  $\mathbb{P}_\ac$ and $\mathbb{P}_\bc$.
\paragraph{Enumerations}
For each condition $p\in P_\ac$,
let $\{\at{\kappa(p,n)} \gamma(p,n)\}_{n<\alpha}$ be an enumeration of $\Omega_p$-sentences which have retrieve $@$ as top operator.
Similarly, for each condition $q\in P_\bc$,
let $\{\at{\kappa(q,n)} \gamma(q,n)\}_{n<\alpha}$ be an enumeration of $\Omega_q$-sentences which have retrieve $@$ as top operator.
Let $pair:\alpha\times\alpha\to\alpha$ be any bijective function such that $i,j\leqslant n$ when $pair(i,j)=n$, for all $i,j,n\leqslant\alpha$.
\subsection{Interpolation Theorem}
This section contains the main results of the paper, namely the Interpolation Theorem and a collection of counterexamples that clarify its scope of applicability.
The following definition introduces the key condition underlying our forcing-based method for proving that a pushout of signature morphisms satisfies CIP.
\begin{definition}[Lifting]
A span of signature morphisms
$\Delta_\ac\stackrel{\chi}\leftarrow\Delta\stackrel{\delta}\to\Delta_\bc$ 
lifts quasi-isomorphisms  whenever 
for all pointed $\Delta_\ac$-models $(\M_\ac,w_1)$ and all pointed $\Delta_\bc$-models $(\M_\bc,w_2)$ such that $(\M\red_\chi,w_1)$ and $(\M_\bc\red_\delta,w_2)$ are quasi-isomorphic,
there exist
\begin{enumerate}
\item a $\Delta_\ac$-model $(\N_\ac,v_1)$ quasi-isomorphic with $(\M_\ac,w_1)$, and
\item a $\Delta_\bc$-model $(\N_\bc,v_2)$ quasi-isomorphic with $(\M_\bc,w_2)$
\end{enumerate}
$(\N_\ac\red_\chi,v_1)=(\N_\bc\red_\delta,v_2)$, that is,
$\N_\ac\red_\chi=\N_\bc\red_\delta$ and $v_1=v_2$.
\end{definition}
In Section~\ref{sec:criterion}, we establish conditions under which a span of signature morphisms lifts quasi-isomorphisms. These conditions provide a criterion for interpolation to hold. With all definitions and preliminary results in place, we are now ready to prove the central result of this paper.
\begin{theorem} [Interpolation]\label{th:interpolation}
Consider a pushout of signature morphisms  
$\{\Delta_\ac\stackrel{\chi}\leftarrow\Delta\stackrel{\delta}\to\Delta_\bc,\allowbreak \Delta_\bc\stackrel{\chi_\bc}\to\Delta_\dc\stackrel{\delta_\ac}\leftarrow\Delta_\ac\}$.
\begin{center}
\begin{tikzcd}
\Phi_\bc \ar[r,dotted,no head]& \Delta_\bc \ar[r,"\chi_\bc"] & 
\Delta_\dc & \delta_\ac(\Phi_\ac)\models\chi_\bc(\Phi_\bc)  \ar[l,dotted,no head]\\
& \Delta \ar[r,"\chi",swap] \ar[u,"\delta"] & \Delta_\ac \ar[u,"\delta_\ac",swap] & \Phi_\ac \ar[l,dotted,no head]
\end{tikzcd}
\end{center}
If the span $\Delta_\ac\stackrel{\chi}\leftarrow\Delta\stackrel{\delta}\to\Delta_\bc$ 
lifts quasi-isomorphisms  then the above pushout has CIP, that is,
if $\delta_\ac(\Phi_\ac)\models_{\Delta_\dc}\chi_\bc(\Phi_\bc)$, 
where $\Phi_\ac\subseteq\Sen(\Delta_\ac)$ and $\Phi_\bc\subseteq\Sen(\Delta_\bc)$, 
then $\Phi_\ac\models_{\Delta_\ac}\chi(\Phi)$ and $\delta(\Phi)\models_{\Delta_\bc} \Phi_\bc$ for some $\Phi\subseteq\Sen(\Delta)$.
\end{theorem}
\begin{proof}
For this proof we employ the framework and the notational conventions described in Section~\ref{sec:framework}.
Since $\FOHLR$ is compact and semantically closed under conjunction,
it suffices to consider two sentences
$\phi_\ac\in\Sen(\Delta_\ac)$ and $\phi_\bc\in\Sen(\Delta_\bc)$ such that $\delta_\ac(\phi_\ac)\models_{\Delta_\dc}\chi_\bc(\phi_\bc)$.
Then we show that $\phi_\ac\models_{\Delta_\ac} \chi(\phi)$ and $\delta(\phi)\models_{\Delta_\bc}\phi_\bc$ for some $\Delta$-sentence $\phi$.
Suppose towards a contradiction that 
$\phi_\ac\not\models_{\Delta_\ac}\chi(\phi)$ or
$\delta(\phi)\not\models_{\Delta_\bc}\phi_\bc$
for all  $\phi\in\Sen(\Delta)$.
Since $\FOHLR$ has negations,
\begin{equation} \tag{$\dagger$} \label{eq:assumption}
\phi_\ac\not\models_{\Delta_\ac}\chi(\phi) \text{ or }
\neg\phi_\bc\not\models_{\Delta_\bc}\delta(\neg\phi),
\text{ for all } \phi\in \Sen(\Delta).
\end{equation}
\emph{This means that $\phi_\ac$ and $\neg\phi_\bc$ are consistent on $\Delta$.}
Then we construct two pointed models $(\M,w)$ and $(\N,v)$ over $\Delta_\ac$ and $\Delta_\bc$, respectively, such that:
\begin{center}
 \begin{enumerate*}[label=(\alph*)]
\item $(\M,w)\models\phi_\ac$ and  $(\N,v)\models \neg\phi_\bc$, and
\item $(\M\red_\chi,w)=(\N\red_\delta,v)$.
\end{enumerate*}
\end{center}
By Lemma~\ref{lemma:amalgamation},
there exists a $\Delta_\dc$-model $\M'$ such that $\M'\red_{\delta_\ac}=\M$ and $\M'\red_{\chi_\bc}=\N$.
By the satisfaction condition, 
$(\M',w)\models \{\delta_\ac(\phi_\ac),\chi_\bc(\neg\phi_\bc)\}$,
Therefore, $\delta_\ac(\phi_\ac)\not\models_{\Delta_\dc}\chi_\bc(\phi_\bc)$,
which is a contradiction.
For the remainder of the proof, we focus on constructing $(\M,w)$ and $(\N,v)$ assuming that property \ref{eq:assumption} holds.
\paragraph{Step 1} 
The key idea behind our construction is to define two ascending chains of conditions: \begin{enumerate} \item $p_0\leqslant p_1\leqslant\dots$ in $\mathbb{P}_\ac$, and \item $q_0\leqslant q_1\leqslant\dots$ in $\mathbb{P}_\bc$, 
\end{enumerate} 
such that, for every $n<\alpha$, the following conditions hold: 
\begin{enumerate}[label=(C\arabic*)] 
\item $\Gamma_{p_n}\not\models \chi_n(\phi)$ or $\Gamma_{q_n}\not\models \delta_n(\neg\phi)$ for every $\phi\in \Sen(\Delta[\C_n])$.
\item \label{step1:cond2} 
$ \C_n=\chi_n^{-1}(\A_n)=\delta_n^{-1}(\B_n)$.
\end{enumerate}
\begin{center}
\begin{tikzcd}
\Gamma_{p_n} \ar[r,dotted,no head] & \overbrace{\Delta_\ac[\A_n]}^{\Omega(p_n)}& \ar[l,"\chi_n",swap] \Delta[\C_n] \ar[r,"\delta_n"] & \overbrace{\Delta_\bc[\B_n]}^{\Omega(q_n)} & \Gamma_{q_n} \ar[l,dotted,no head] \\
& \Delta_\ac  \ar[u,hook] & \ar[l,"\chi"] \Delta \ar[r,"\delta",swap]  \ar[u,hook] & \Delta_\bc \ar[u,hook] 
\end{tikzcd}
\end{center}
Here: 
\begin{enumerate}[label=(\alph*)] 
\item $\A_n$ is the set of all Henkin constants occurring in $\Omega(p_n)$; 
\item $\B_n$ is the set of all Henkin constants occurring in $\Omega(q_n)$; 
\item $\C_n=\chi_n^{-1}(\A_n)=\delta_n^{-1}(\B_n)$; and
\item $\chi_n:\Delta[\C_n]\to \Delta_\ac[\A_n]$ and  $\delta_n:\Delta[\C_n]\to \Delta_\bc[\B_n]$ are extensions of $\chi:\Delta\to\Delta_\ac$ and $\delta:\Delta\to \Delta_\bc$, respectively.
\end{enumerate}

We proceed by induction on ordinals.

\paragraph{Induction base ($n=0$)}
Let $a_0\in \A_\mathtt{nom}$, $b_0\in \B_\mathtt{nom}$ and $c_0\in \C_\mathtt{nom}$ be three nominals.
Let $\chi_0:\Delta[c_0]\to\Delta_\ac[a_0]$ be the extension of $\chi$ mapping $c_0$ to $a_0$.
Similarly, let $\delta_0:\Delta[c_0]\to\Delta_\ac[b_0]$ be the extension of $\delta$ mapping $c_0$ to $b_0$.
We define the conditions 
$p_0\coloneqq (\Delta_\ac[a_0],\at{a_0} \phi_\ac)$ and 
$q_0\coloneqq (\Delta_\bc[b_0], \at{b_0}\neg\phi_\bc)$.
Suppose towards a contradiction that 
$\at{a_0}\phi_\ac \models_{\Delta_\ac[a_0]} \chi_0(\phi)$ and 
$\at{b_0}\neg\phi_\bc \models_{\Delta_\bc[b_0]} \delta_0(\neg\phi)$ 
for some $\phi\in\Sen(\Delta[c_0])$.
\begin{center}
\begin{tikzcd}
 & \overbrace{\Delta_\ac[a_0]}^{\Omega(p_0)}& \ar[l,"\chi_0",swap] \Delta[c_0] \ar[r,"\delta_0"] & \overbrace{\Delta_\bc[b_0]}^{\Omega(q_0)} &  \\
\phi_\ac \ar[r,dotted,no head] & \Delta_\ac  \ar[u,hook] & \ar[l,"\chi"] \Delta \ar[r,"\delta",swap]  \ar[u,hook] & \Delta_\bc \ar[u,hook] & \neg\phi_\bc \ar[l,dotted,no head]
\end{tikzcd}
\end{center}
\begin{proofsteps}{20em}
$\store{a_0}\phi_\ac \models_{\Delta_\ac} \store{a_0}\chi_0(\phi)$ & 
since $\at{a_0}\phi_\ac \models_{\Delta_\ac[a_0]} \chi_0(\phi)$ \\
$\phi_\ac\models_{\Delta_\ac} \store{a_0}\chi_0(\phi)$ & 
since $a_0$ does not occur in $\phi_\ac$\\
$\phi_\ac\models_{\Delta_\ac} \chi_0(\store{c_0}\phi)$ & 
since $\chi_0(c_0)=a_0$  \\
$\store{b_0}\neg\phi_\bc \models_{\Delta_\bc} \store{b_0}\delta_0(\neg\phi)$ & 
since $\at{b_0}\neg\phi_\bc \models_{\Delta_\bc[b_0]} \delta_0(\neg\phi)$ \\
$\neg\phi_\bc\models_{\Delta_\bc} \store{b_0}\delta_0(\neg\phi)$ &
 $b_0$ does not occur in $\phi_\bc$\\
$\neg\phi_\bc\models_{\Delta_\bc} \delta_0(\neg\store{c_0}\phi)$ & 
since $\delta_0(c_0)=b_0$ \\
contradiction & by (\ref{eq:assumption}), 
since $\phi_\ac\models_{\Delta_\ac} \chi_0(\store{c_0}\phi)$ and 
$\neg\phi_\bc\models_{\Delta_\bc} \delta_0(\neg\store{c_0}\phi)$ 
\end{proofsteps}
Hence,
$\phi_\ac\not\models_{\Delta_\ac[a_0]} \chi_0(\phi)$ or
$\neg\phi_\bc\not\models_{\Delta_\bc[b_0]} \delta_0(\neg\phi)$ 
for all $\phi\in\Sen(\Delta[c_0])$.

\paragraph{Induction step ($n\implies n+1$)}
By induction hypothesis, we assume a span of presentation morphisms 
\begin{center}
$(\Delta[\A_n],\Gamma_{p_n}) \xleftarrow{\chi_n}\Delta[\C_n]\xrightarrow{\delta_n}(\Delta[\B_n],\Gamma_{q_n})$
such that 
$\Gamma_{p_n}\not\models \chi_n(\phi)$ or
$\Gamma_{q_n}\not\models \delta_n(\neg\phi)$
for all $\phi\in\Sen(\Delta[\C_n])$.
\end{center}
Let $(i,j)\coloneqq pair^{-1}(n)$.
Based on the sentence $\at{\kappa(p_i,j)} \gamma(p_i,j)$,
we define:
\begin{itemize}
\item two signature morphisms 
$\chi_n':\Delta[\C_n']\to \Delta_\ac[\A_n']$ and
$\delta_n':\Delta[\C_n']\to \Delta_\bc[\B_n']$ which extend 
$\chi_n$ and $\delta_n$, respectively;
\item two conditions $p_n'\in P_\ac$ and $q_n'\in P_\bc$ such that 
\begin{enumerate*}[label=(\alph*)]
\item $\Omega_{p_n'}=\Delta_\ac[\A_n']$ and  $\Omega_{q_n'}=\Delta_\bc[\B_n']$; and
\item $\Gamma_{p_n'}$ and $\Gamma_{q_n'}$ are consistent on $\Delta[\C_n']$.
\end{enumerate*}
\end{itemize}
We have two cases:
\begin{enumerate}
\item $(p_n,\kappa(p_i,j))\Vdash\neg\gamma(p_i,j)$: In this case, we define $p_n'$ and $q_n'$ as follows:
\begin{itemize}
\item $p_n'\coloneqq p_n\cup (\Omega(p_n),\at{\kappa(p_i,j)}\neg\gamma(p_i,j))$ and $\chi_n'\coloneqq\chi_n$.
\item $q_n'\coloneqq q_n$ and $\delta_n'\coloneqq \delta_n$. 
\end{itemize}
We show that $\Gamma_{p_n'}$ and $\Gamma_{q_n'}$ are consistent over $\Delta[\C_n']$, 
that is,
$\Gamma_{p_n'}\not \models \chi_n'(\phi)$ or
$\Gamma_{q_n'}\not \models \delta_n'(\neg \phi)$,
for all $\phi\in\Sen(\Delta[\C_n'])$:
\begin{proofsteps}{28em}
$\Gamma_{p_n}\models\at{\kappa(p_i,j)}\neg\gamma(p_i,j)$ &
since $(p_n,\kappa(p_i,j)) \Vdash \neg\gamma(p_i,j)$, 
by Corollary~\ref{cor:semantic-forcing1}\\
$\Gamma_{p_n}$ and $\Gamma_{q_n}$ are consistent over $\Delta[\C_n]$ &
by induction hypothesis\\
$\Gamma_{p_n'}=\Gamma_{p_n}\cup\{\at{\kappa(p_i,j)}\neg\gamma(p_i,j)\}$ and $\Gamma_{q_n'}=\Gamma_{q_n}$ are consistent over $\Delta[\C_n]$ & since $\Gamma_{p_n'}\equiv\Gamma_{p_n}$  
\end{proofsteps}

\item $(p_n,\kappa(p_i,j))\not\Vdash\neg\gamma(p_i,j)$:
There are two sub-cases to consider here.
\begin{enumerate}
\item  $\Gamma_{p_n}\cup\{\at{\kappa(p_i,j)} \gamma(p_i,j)\}\models \chi_n(\phi)$ and $\Gamma_{q_n}\models\delta_n(\neg\phi)$ for some {$\phi\in\Sen(\Delta[\C_n])$}:
\begin{itemize}
\item $p_n'\coloneqq p_n\cup (\Omega(p_n),\at{\kappa(p_i,j)} \neg\gamma(p_i,j))$  and
$\chi_n'\coloneqq\chi_n$.
\item $q_n'\coloneqq q_n$ and $\delta_n'\coloneqq \delta_n$.
\end{itemize}
Suppose a contradiction that 
$\Gamma_{p_n}\cup\{\at{\kappa(p_i,j)} \neg\gamma(p_i,j)\}\models \chi_n(\phi')$ and $\Gamma_{q_n}\models\delta_n(\neg\phi')$ for some {$\phi'\in\Sen(\Delta[\C_n])$}.
\begin{proofsteps}{20em}
$\Gamma_{p_n}\models \chi_n(\phi)\lor \chi_n(\phi')$ and $\Gamma_{q_n}\models \neg\delta_{q_n} (\phi)\land \neg \delta_{q_n}(\phi')$ & by semantics\\
contradiction & by ind. hypothesis, 
since $\Gamma_{p_n}$ and $\Gamma_{q_n}$ are consistent over $\Delta[\C_n]$
\end{proofsteps}
Hence, $\Gamma_{p_n'}$ and $\Gamma_{q_n'}$ are consistent over $\Delta[\C_n]$. 
\item $\Gamma_{p_n}\cup\{\at{\kappa(p_i,j)} \gamma(p_i,j)\}\not\models \chi_n(\phi)$ or $\Gamma_{q_n}\not\models\delta_n(\neg\phi)$ for all {$\phi\in\Sen(\Delta[\C_n])$}:
\begin{enumerate}
\item  $\gamma(p_i,j)$ is of the form $\Exists{x}\psi_x$:

Assume that $s_\ac$ is the sort of the variable $x$.
Let $a \in \A \setminus \A_n$ be a constant of sort $s_\ac$ that does not occur in $p_n$.
For each $s\in S^\ext$ such that $\chi(s)=s_\ac$, we choose a constant $c\in \C\setminus \C_n$ of sort $s$.
Let $\C'$ be the set of all such constants from $\C$ and observe that $\C'$ is infinite if $\chi$ maps infinitely many sorts to $s_\ac$.
Define $\A_n' \coloneqq \A_n \cup \{a\}$ and $\C_n' \coloneqq \C_n \cup \C'$.
Let $\chi_n' : \Delta[\C_n'] \to \Delta[\A_n']$ be the extension of $\chi_n$ that maps each $c \in \C'$ to $a \in \A \setminus \A_n$.
For each constant $c \in \C'$ of sort $s$, we choose a fresh constant $b \in \B \setminus \B_n$ of sort $\delta(s)$.
Let $\B'$ denote the set of all such chosen constants.
We then define
$\B_n'\coloneqq \B_n\cup\B'$ and
$\delta_n':\Delta[\C_n']\to \Delta_\bc[\B_n']$ 
as the extension of $\delta_n$ that maps each constant $c \in \C'$ to the corresponding constant $b \in \B'$ under the correspondence established above.
\begin{itemize}
\item $p_n'\coloneqq p_n\cup(\Delta_\ac[\A_n'],\{ \at{\kappa(p_i,j)}\gamma(p_i,j),\at{\kappa(p_i,j)} \psi_{x\leftarrow a} \})$.
\item  $q_n'\coloneqq (\Delta_\bc[\B_n'],\Gamma_{q_n})$.
\end{itemize}
\begin{center}
\begin{tikzcd}
 \Gamma_{p_n'} \ar[r,dotted,no head] & \overbrace{\Delta_\ac[\A_n']}^{\Omega(p_n')}& \ar[l,"\chi_n'",swap] \Delta[\C_n'] \ar[r,"\delta_n'"] & \overbrace{\Delta_\bc[\B_n']}^{\Omega(q_n')} & \\
\Gamma_{p_n} \ar[r,dotted,no head]& \underbrace{\Delta_\ac[\A_n]}_{\Omega(p_n)}  \ar[u,hook] & \ar[l,"\chi_n"] \Delta[\C_n] \ar[r,"\delta_n",swap]  \ar[u,hook] & \underbrace{\Delta_\bc[\B_n]}_{\Omega(q_n)} \ar[u,hook] & \Gamma_{q_n} \ar[l,dotted,no head]
\end{tikzcd}
\end{center}
Suppose towards a contradiction that 
$\Gamma_{p_n}\cup\{\at{\kappa(p_i,j)} \gamma(p_i,j), \at{\kappa(p_i,j)}\psi_{x\leftarrow a} \}\models_{\Delta_\ac[\A_n']} \chi_n'(\phi)$ and 
$\Gamma_{q_n}\models_{\Delta_\bc[\B_n']}\neg\delta_n'(\phi)$ for some sentence $\phi\in\Sen(\Delta[\C_n'])$.
Let $c_1,\dots,c_m$ be all Henkin constants from $\C_n'\setminus \C_n$ which occur in $\phi$. 

\begin{proofsteps}{20em}
\rlap{$\Gamma_{p_n}\cup\{\at{\kappa(p_i,j)} \gamma(p_i,j), \at{\kappa(p_i,j)}\psi_{x\leftarrow a} \} \models_{\Delta_\ac[\A_n']} \chi_n(\Exists{c_1,\dots,c_m}\phi)$ \hspace{0.7cm} by semantics} & \\
\rlap{$\Gamma_{p_n}\cup\{\at{\kappa(p_i,j)} \gamma(p_i,j), \at{\kappa(p_i,j)} \Exists{a} \psi_{x\leftarrow a} \} \models_{\Delta_\ac[\A_n]} \chi_n(\Exists{c_1,\dots,c_m}\phi)$ \hspace{0.2cm} since $a$ occurs only in $\psi_{x\leftarrow a}$} \\
$\Gamma_{p_n}\cup\{\at{\kappa(p_i,j)} \gamma(p_i,j) \} \models_{\Delta_\ac[\A_n]} \chi_n(\Exists{c_1,\dots,c_m}\phi)$ &
as $\at{\kappa(p_i,j)} \Exists{a} \psi_{x\leftarrow a}\equiv\at{\kappa(p_i,j)} \Exists{x} \psi_x = \at{\kappa(p_i,j)} \gamma(p_i,j)$ \\
let $b_\ell\coloneqq \delta_n'(c_\ell)$ for all $\ell\in\{1,\dots,m\}$ & \\
$\Gamma_{q_n}\models \Forall{b_1,\dots,b_m}\neg\delta_n'(\phi)$ & 
since $\Gamma_{q_n}\models\neg\delta_n'(\phi)$
and $b_1,\dots,b_m$ does not appear in $\Gamma_{q_n}$ \\
$\Gamma_{q_n}\models\delta_n(\Forall{c_1,\dots,c_m}\neg\phi)$ &
since $\delta_{q_n}(\Forall{c_1,\dots,c_m}\neg \phi)\equiv \Forall{b_1,\dots,b_m} \neg\delta_n'(\phi)$\\
contradiction with the assumption for this case & from 3 and 6
\end{proofsteps}
Hence, $\Gamma_{p_n'}$ and $\Gamma_{q_n'}$ are consistent over $\Delta[\C_n']$. 
\item  $\gamma(p_i,j)$ is not of the form $\Exists{x}\psi_x$:
\begin{itemize}
\item $p_n'\coloneqq p_n\cup(\Omega(p_n),\at{\kappa(p_i,j)}\gamma(p_i,j))$ and $\chi_n'\coloneqq\chi_n$.
\item $q_n'\coloneqq q_n$ and $\delta_n'\coloneqq\delta_n$.
\end{itemize}
In this case, it is straightforward to show that 
$\Gamma_{p_n'}$ and $\Gamma_{q_n'}$ are consistent over $\Delta[\C_n']$.
\end{enumerate}
\end{enumerate}
\end{enumerate}
We define $p_{n+1}\in P_\ac$ and $q_{n+1}\in P_\bc$ based on $\at{\kappa(q_i,j)}\gamma(q_i,j)$.
There are two main cases.
\begin{enumerate}
\item $(q_n',\kappa(q_i,j))\Vdash\neg\gamma(q_i,j)$:
\begin{itemize}
\item $q_{n+1}\coloneqq q_n'\cup (\Omega(q_n'),\at{\kappa(q_i,j)}\neg\gamma(q_i,j))$ and $\delta_{n+1}\coloneqq \delta_n'$.
\item $p_{n+1}\coloneqq p_n'$ and $\chi_{n+1}\coloneqq \chi_n'$.
\end{itemize}
\item $(q_n',\kappa(q_i,j))\not\Vdash\neg\gamma(q_i,j)$:
There are two sub-cases to consider here.
\begin{enumerate}
\item  $\Gamma_{q_n'}\cup\{\at{\kappa(q_i,j)} \gamma(q_i,j)\}\models \delta_n'(\phi)$ and 
$\Gamma_{p_n'}\models \chi_n'(\neg\phi)$ for some $\phi\in\Sen(\Delta[\C_n'])$:
\begin{itemize}
\item $q_{n+1}\coloneqq q_n'\cup (\Omega(q_n'),\at{\kappa(q_i,j)} \neg\gamma(q_i,j))$ and $\delta_{n+1}\coloneqq \delta_n'$.
\item $p_{n+1}\coloneqq p_n'$ and $\chi_{n+1}\coloneqq \chi_n'$.
\end{itemize}
\item $\Gamma_{q_n'}\cup\{\at{\kappa(q_i,j)} \gamma(q_i,j)\} \not\models \delta_n'(\phi)$ or 
$\Gamma_{p_n'}\not\models \chi_n'(\neg\phi)$ for all $\phi\in\Sen(\Delta[\C_n'])$:
\begin{enumerate}
\item $\gamma(q_i,j)$ is of the form $\Exists{x}\psi_x$:
Let $s_\bc$ be the sort of the variable $x$ and choose a constant $b \in \B \setminus \B_n'$ of sort $s_\bc$ that does not occur in $q_n'$.
We define $\B_{n+1} \coloneqq \B_n' \cup \{b\}$.
For each sort $s\in S^\ext$ such that $\delta(s)=s_\bc$,
we choose a constant $c\in \C\setminus \C_n'$ of sort $s$.
Let $\C''$ be the set of all such constants chosen from $\C$.
We then define $\C_{n+1} \coloneqq \C_n' \cup \C''$ and 
let $\delta_{n+1} : \Delta[\C_{n+1}] \to \Delta_\bc[\B_{n+1}]$ be the extension of $\delta_n'$ that maps each $c \in \C''$ to $b$.
For each constant $c \in \C''$ of sort $s$, we choose a fresh constant $a \in \A \setminus \A_n'$ of sort $\chi(s)$.
Let $\A''$ denote the set of all such constants, chosen in one-to-one correspondence with $\C''$.
Finally, we  define $\A_{n+1}\coloneqq \A_n'\cup\A''$ and let $\chi_{n+1}:\Delta[\C_{n+1}] \to \Delta_\ac[\A_{n+1}]$ 
be the extension of $\chi_n'$ that maps each constant $c \in \C''$ to the corresponding constant $a \in \A''$ under the established correspondence  above.
\begin{itemize}
\item $q_{n+1}\coloneqq q_n'\cup(\Delta_\bc[\B_{n+1}], \{ \at{\kappa(q_i,j)}\gamma(q_i,j),\at{\kappa(q_i,j)} \psi_{x\leftarrow b} \})$.
\item $p_{n+1} \coloneqq (\Delta_\ac[\A_{n+1}],\Gamma_{p_n'})$.
\end{itemize}
\begin{center}
\begin{tikzcd}
 & \overbrace{\Delta_\ac[\A_{n+1}]}^{\Omega(p_{n+1})}& \ar[l,"\chi_{n+1}",swap] \Delta[\C_{n+1}] \ar[r,"\delta_{n+1}"] & \overbrace{\Delta_\bc[\B_{n+1}]}^{\Omega(q_{n+1})} & \Gamma_{q_{n+1}} \ar[l,dotted,no head]\\
\Gamma_{p_n'} \ar[r,dotted,no head]& \underbrace{\Delta_\ac[\A_n']}_{\Omega(p_n')}  \ar[u,hook] & \ar[l,"\chi_n'"] \Delta[\C_n'] \ar[r,"\delta_n'",swap]  \ar[u,hook] & \underbrace{\Delta_\bc[\B_n']}_{\Omega(q_n')} \ar[u,hook] & \Gamma_{q_n'} \ar[l,dotted,no head]
\end{tikzcd}
\end{center}
\item  $\gamma(q_i,j)$ is not of the form $\Exists{x}\psi_x$:
\begin{itemize}
\item $q_{n+1} \coloneqq q_n'\cup(\Omega(q_n'),\at{\kappa(q_i,j)}\gamma(q_i,j))$ and $\delta_{n+1}\coloneqq\delta_n'$.
\item $p_{n+1}\coloneqq p_n'$ and $\chi_{n+1}\coloneqq \chi_n'$.
\end{itemize}
\end{enumerate}
\end{enumerate}
\end{enumerate}
Similarly, one can show that $\Gamma_{p_{n+1}}$ and $\Gamma_{q_{n+1}}$ are consistent over $\Delta[\C_{n+1}]$.

\paragraph{Limit ordinal $\beta < \alpha$}
Let $\chi_\beta\coloneqq \bigcup_{n<\beta} \chi_n$ and $\delta_\beta\coloneqq \bigcup_{n<\beta} \delta_n$.
By compactness, both $\Gamma_{p_\beta}=\bigcup_{n<\beta} \Gamma_{p_n}$ and $\Gamma_{q_\beta}=\bigcup_{n<\beta} \Gamma_{q_n}$ are consistent.
It follows that $p_\beta\coloneqq \bigcup_{n<\beta} p_n$  and $q_\beta\coloneqq \bigcup_{n<\beta} q_n$ are well-defined.
Since $\alpha$ is regular,
the cardinal $\card(\A_{p_\beta,s})<\alpha$ for all sorts $s\in S^\ext$.

\paragraph{Step 2}
We have constructed a chain $p_0\leqslant p_1 \leqslant p_2\dots $ such that
for all sentences $\at{k}\phi \in \bigcup_{n<\alpha}\Sen(\Omega_{p_n})$ we have:
\begin{enumerate}[label=(P\arabic*)]
\item \label{state:p-1} if $\at{k}\phi\not\in \bigcup_{n<\alpha}\Gamma_{p_n}$ then  $(p_\beta,k)\Vdash\neg\phi$ for some $\beta<\alpha$; and
\item \label{state:p-2} if $\at{k}\phi\in \bigcup_{n<\alpha}\Gamma_{p_n}$ then $(p_\beta,k)\Vdash\phi$ for some $\beta<\alpha$.
\end{enumerate}

For the first statement,
assume that $\at{k}\phi\not\in \bigcup_{n<\alpha}\Gamma_{p_n}$. 
\begin{proofsteps}{25em}
$\at{k}\phi \in \Sen(\Omega_{p_i})$ for some $i<\alpha$ & 
since $\at{k}\phi \in \bigcup_{n<\alpha}\Sen(\Omega_{p_n})$\\
$\at{k}\phi=\at{\kappa(p_i,j)} \gamma(p_i,j)$ for some $j<\alpha$ & 
since $\{\at{\kappa(p_i,j)} \gamma(p_i,j)\}_{j<\alpha}$ is an enumeration of the sentences in $\Sen(\Omega_{p_i})$ with $@$ as top operator\\
$\at{\kappa(p_i,j)} \gamma(p_i,j) \not\in \Gamma_{p_\beta}$, where  $\beta\coloneqq pair(i,j)$ & 
since $\at{\kappa(p_i,j)} \gamma(p_i,j)\not\in \bigcup_{n<\alpha}\Gamma_{p_n}$\\
$\at{\kappa(p_i,j)} \neg \gamma(p_i,j)\in \Gamma_{p_\beta}$ & 
by construction\\
$(p_\beta,\kappa(p_i,j))\Vdash^w\neg\gamma(p_i,j)$ &
by Corollary~\ref{cor:semantic-forcing1}\\
$(p_\beta,k)\Vdash^w\neg\phi$ & since $k=\kappa(p_i,j)$ and $\phi=\gamma(p_i,j)$ \\
$(p_\beta,k)\Vdash\neg\phi$ & by the definition of forcing relation
\end{proofsteps}
We prove the second statement by induction on the structure of $\phi$. 
\begin{description}
\item[$\phi\in \bigcup_{n<\alpha}\Sen_0(\Omega_{p_n})$:] 
In this case, 
$\at{k}\phi\in \Gamma_{p_\beta}\cap \Sen_b(\Omega_{p_\beta})$ for some $\beta<\alpha$.
It follows that $\at{k}\phi\in f(p_\beta)$.
By the definition of forcing relation, $(p_\beta,k)\Vdash\phi$.
\item [$\neg\phi:$]
We have $\at{k}\neg\phi\in \Gamma_{p_\beta}$ for some $\beta<\alpha$.
We have $p_\beta\models \at{k}\neg\phi$.
By Corollary~\ref{cor:semantic-forcing1}, $(p_\beta,k)\Vdash^w\neg\phi$.
By the definition of forcing, $(p_\beta,k)\Vdash\neg\phi$.
\item [$\lor \Phi$:]
We have $\at{k}\lor\Phi\in\Gamma_{p_{\beta_0}}$ for some $\beta_0<\alpha$.
Suppose towards a contradiction that 
$\at{k}\phi\not\in \bigcup_{n<\alpha}\Gamma_{p_n}$  for all $\phi\in \Phi$.
By \ref{state:p-1},
for each $\phi\in \Phi$
there exists $\beta_\phi$ such that 
$(p_{\beta_\phi},k)\Vdash\neg\phi$.
Let $\beta$ be the maximum of $\{\beta_0\} \cup \{\beta_\phi\mid\phi\in \Phi\}$.
By Lemma~\ref{lemma:forcing-property}(\ref{fp-2}), 
$(p_\beta,k)\Vdash\neg\phi$ for all $\phi\in\Phi$.
By Corollary~\ref{cor:semantic-forcing1}, 
$p_\beta\models \at{k}\neg\phi$ for all $\phi\in \Phi$.
By semantics, $p_\beta\models \at{k}\neg\lor\Phi$, 
which is a contradiction with $\at{k}\lor\Phi \in \Gamma_{p_\beta}$.
\item[$\at{\ell}\phi$:]
We have $\at{k}\at{\ell}\phi\in \Gamma_{p_\beta}$ for some $\beta<\alpha$.
Since $\at{k}\at{\ell}\phi$ is semantically equivalent to $\at{\ell}\phi$,
by construction,
$\at{\ell}\phi\in \bigcup_{n<\alpha}\Gamma_{p_n}$.
By induction hypothesis,
$(p_\beta,\ell)\Vdash\phi$ for some $\beta<\alpha$.
Hence, $(p_\beta,k)\Vdash \at{\ell}\phi$.
\item[$\Exists{x}\phi_x$:]
We have $\at{k}\Exists{x}\phi_x\in \Gamma_{p_n}$ for some $n<\alpha$. 
By the construction of $p_0\leqslant p_1\leqslant\dots$, 
we have $\at{k}\phi_{x\leftarrow a}\in\Gamma_{p_n}$ for some constant $a\in \A$,
where $\phi_{x\leftarrow a}$ is the sentence obtained from $\phi_x$ by substituting $a$ for $x$.
By induction hypothesis, $(p_\beta,k)\Vdash \phi_{x\leftarrow a}$ for some $\beta<\alpha$.
By the definition of forcing relation, $(p_\beta,k)\Vdash \Exists{x}\phi_x$.
\item [$\pos{\lambda}\phi$:]
Straightforward, since $\at{k}\pos{\lambda}\phi$ is semantically equivalent to $\at{k}\Exists{x}\pos{\lambda}x \land \at{x}\phi$.
\item [$\store{x}\phi_x$:]
Straightforward, since $\at{k}\store{x}\phi_x$ is semantically equivalent to $\at{k}\phi_{x\leftarrow k}$.
\end{description}
Similarly, one can show that the chain $q_0\leqslant q_1 \leqslant q_2\dots $ satisfies the following properties:
\begin{enumerate}[label=(Q\arabic*)]
\item \label{state:q-1} if $\at{k}\phi\not\in \bigcup_{n<\alpha}\Gamma_{q_n}$ then  $(p_n,k)\Vdash\neg\phi$ for some $n<\alpha$; and
\item \label{state:q-2} if $\at{k}\phi\in \bigcup_{n<\alpha}\Gamma_{q_n}$ then $(q_n,k)\Vdash\phi$ for some $n<\alpha$.
\end{enumerate}

\paragraph{Step 3}
We define the following sets of conditions.
\begin{enumerate} 
\item $G_\ac\coloneqq \{p\in P_\ac\mid p\leqslant p_n \text{ for some }n<\alpha\}$.
By \ref{state:p-1} and \ref{state:p-2}, $G_\ac$ is a generic set.
\item $G_\bc\coloneqq\{q\in P_\bc\mid q\leqslant q_n \text{ for some }n<\alpha\}$.
By \ref{state:q-1} and \ref{state:q-2}, $G_\bc$ is a generic set.
\end{enumerate}
We define two generic models for the generic sets constructed above:
\begin{itemize}
\item Let $\A_\alpha\coloneqq \bigcup_{n<\alpha} \A_n$ and $\B_\alpha\coloneqq \bigcup_{n<\alpha} \B_n$ and $\C_\alpha\coloneqq \bigcup_{n<\alpha} \C_n$ .
\item Define $\chi_\alpha\coloneqq \bigcup_{n<\alpha} \chi_n$.
By \ref{step1:cond2},
for all $x\in \A_\alpha$ there exists $y\in \C_\alpha$ such that $\chi_\alpha(y)=x$.

\item Define $\delta_\alpha\coloneqq \bigcup_{n<\alpha} \delta_n$.
By \ref{step1:cond2},
for all $x\in \B_\alpha$ there exists $y\in \C_\alpha$ such that $\delta_\alpha(y)=x$.
\item Let $\M_\ac=(W_\ac,M_\ac)$ be a reachable generic $\Delta_\ac[\A_\alpha]$-model for $G_\ac$ given by Theorem~\ref{th:gm}.

\item Let $\M_\bc=(W_\bc,M_\bc)$ be a reachable generic $\Delta_\bc[\B_\alpha]$-model for $G_\bc$ given by Theorem~\ref{th:gm}.
\end{itemize}
\begin{center}
\begin{tikzcd}
\M_\ac \ar[r,dotted,no head]&  
\Delta_\ac[\A_\alpha] & \ar[l,"\chi_\alpha",swap] \Delta[\C_\alpha] \ar[r,"\delta_\alpha"]  & \Delta_\bc[\B_\alpha] & 
\M_\bc \ar[l,no head, dotted]\\ 
 & \Delta_\ac \ar[u,hook] & \ar[l, "\chi"] \Delta \ar[r,"\delta",swap] \ar[u,hook] & \Delta_\bc   \ar[u,hook]
\end{tikzcd}
\end{center}
Since $\at{a_0}\phi_\ac\in \Gamma_{p_0}$ and $\at{b_0}\neg\phi_b\in\Gamma_{q_0}$, by Corollary~\ref{cor:semantic-forcing2}, we have:
\begin{equation} \tag{$\ddagger$} \label{eq:ab}
\M_\ac\models \at{a_0}\phi_\ac\text{ and }\M_\bc\models \at{b_0}\neg\phi_\bc.
\end{equation}
\paragraph{Step 4} 
We show that $\M_\ac\red_{\chi_\alpha}\red_\Delta$ and $\M_\bc\red_{\delta_\alpha}\red_\Delta$ are quasi-isomorphic.
\begin{enumerate}[label=(H\arabic*)]
\item \label{H1} 
For all $\at{k}\phi\in \Sen(\Delta[\C_\alpha])$, the following are equivalent: 

\begin{proofsteps}{20em}
$\M_\ac\red_{\chi_\alpha} \models\at{k}\phi$ & \\
$\M_\ac\models \chi_\alpha (\at{k}\phi)$  & by the satisfaction condition\\
$(G_\ac,{\chi_\alpha(k)})\Vdash \chi_\alpha(\phi)$ &
since $\M_\ac$ is a generic model for $G_\ac$\\
$\chi_\alpha(\at{k}\phi)\in \bigcup_{n<\alpha} \Gamma_{p_n}$ & 
by \ref{state:p-1} and \ref{state:p-2}\\
$\delta_\alpha(\at{k}\phi)\in \bigcup_{n<\alpha} \Gamma_{q_n}$ &
since $\bigcup_{n<\alpha} \Gamma_{p_n}$ and $\bigcup_{n<\alpha} \Gamma_{q_n}$ are consistent over $\Delta[\C_\alpha]$\\
$(G_\bc,\delta_\alpha(k)) \Vdash\delta_\alpha(\phi)$ & 
by \ref{state:q-1} and \ref{state:q-2}\\
$\M_\bc\models \delta_\alpha (\at{k}\phi)$ &
since $\M_\bc$ is a generic model for $G_\bc$\\
$\M_\bc\red_{\delta_\alpha} \models\at{k}\phi$ &
by the satisfaction condition
\end{proofsteps}
\item \label{H2}
We show that $\M_\ac$ is reachable by the constants in $\A_\alpha$.
For all nominals $k\in F^\nom$, $\at{k}\Exists{x}x$ is a tautology, 
which implies that $\at{k}\Exists{x} x \in\bigcup_{n<\alpha}\Gamma_{p_n}$. 
By the construction of $p_0\leqslant p_1\leqslant\dots$,
we have $\at{k}a\in \bigcup_{n<\alpha}\Gamma_{p_n}$ for some (witness) nominal $a\in\A_\alpha$.
By \ref{state:p-2}, $(G_\ac,k)\Vdash a$, which means $\M_\ac \models \at{k}a$.
Similarly, for all rigid $\Delta_\ac$-terms $t$, $\at{a_0}\Exists{x}t=x$ is a tautology, 
which implies that $\at{a_0}\Exists{x} t=x \in\bigcup_{n<\alpha}\Gamma_{p_n}$. 
By the construction of $p_0\leqslant p_1\leqslant\dots$,
we have $\at{a_0} (t = a)\in \bigcup_{n<\alpha}\Gamma_{p_n}$ for some (witness) constant $a\in\A_\alpha$.
By \ref{state:p-2}, $(G_\ac,a_0)\Vdash t=a$, which means $\M_\ac\models \at{a_0} (t=a)$.
Since $t$ has the same interpretation across possible worlds, $(\M_\ac,w)\models t=a$ for all $w\in|W_\ac|$.
It follows that $\M_\ac$ is reachable by $\A_\alpha$.
By \ref{step1:cond2}, $\M_\ac\red_{\chi_\alpha}$ is reachable by $\C_\alpha$.
Similarly,
$\M_\bc$ is reachable by the constants in $\B_\alpha$.
By \ref{step1:cond2}, $\M_\bc\red_{\delta_\alpha}$ is reachable by $\C_\alpha$.
\end{enumerate}
We define $h_\alpha:\G_{\M_\ac\red_{\chi_\alpha}} \to \G_{\M_\bc\red_{\delta_\alpha}}$ by:
\begin{itemize}
\item $h_\alpha(c^{\M_\ac\red_{\chi_\alpha}})=c^{\M_\bc\red_{\delta_\alpha}}$ for all constants $c$ in $C_\alpha$, and
\item $h_\alpha(t^{\M_\ac\red_{\chi_\alpha}})=t^{\M_\bc\red_{\delta_\alpha}}$ for all rigid $\Delta$-terms $t$ of flexible sort.
\end{itemize}
By \ref{H1} and \ref{H2}, $h_\alpha:\G_{\M_\ac\red_{\chi_\alpha}} \to \G_{\M_\bc\red_{\delta_\alpha}}$ is an isomorphism.
Let $w_0\coloneqq a_0^{\M_\ac}$ and $v_0\coloneqq b_0^{\M_\bc}$.
It follows that
$h_\alpha(w_0)= h_\alpha(a_0^{\M_\ac})= h_\alpha(\chi_\alpha(c_0)^{\M_\ac})=
h_\alpha(c_0^{\M_\ac\red_{\chi_\alpha}})=c_0^{\M_\bc\red_{\delta_\alpha}}=\delta_\alpha(c_0)^{\M_\bc}=b_0^{\M_\bc}=v_0$.
The reduct  $h\coloneqq h_\alpha\red_\Delta$ is also an isomorphism which maps $w_0$ to $v_0$.
Since $\Delta_\ac\stackrel{\chi}\leftarrow\Delta\stackrel{\delta}\to\Delta_\bc$  lifts quasi-isomorphisms,
there exist 
\begin{enumerate}
\item a pointed $\Delta_\ac$-model $(\M,w)$ quasi-isomorphic to $(\M_\ac\red_{\Delta_\ac},w_0)$, and 
\item a pointed $\Delta_\bc$-model $(\N,v)$ quasi-isomorphic to $(\M_\bc\red_{\Delta_\bc},v_0)$,
\end{enumerate}
such that 
$(\M\red_\chi,w)=(\N\red_\delta,v)$.
Since $\M_\ac\models\at{a_0}\phi_\ac$, we have $(\M_\ac,w_0)\models\phi_\ac$.
By the satisfaction condition,  $(\M_\ac\red_{\Delta_\ac},w_0)\models\phi_\ac$. 
 By Corollary~\ref{cor:gen-Kripke}, $(\M,w)\models\phi_\ac$.
Similarly, since $\M_\bc\models\at{b_0}\neg\phi_\bc$, we have $(\M_\bc,v_0)\models\neg\phi_\bc$.
By the satisfaction condition, $(\M_\bc\red_{\Delta_\bc},v_0)\models\neg\phi_\bc$.
By Corollary~\ref{cor:gen-Kripke}, $(\N,v)\models\neg\phi_\bc$.
\end{proof}
Notice that our framework can handle situations where $\Phi_\ac$ (or $\Phi_\bc$) contains sentences of the form $\Forall{x} \bot$, which essentially state that all models of $\Phi_\ac$ interpret the sort of $x$ as the empty set. This is possible because the constants used to construct models are introduced dynamically only when they are needed.
\subsection{Practical criterion for interpolation} \label{sec:criterion}
To derive a practical criterion for identifying pushout squares of signature morphisms that satisfy CIP, it remains to characterize the conditions under which a span of signature morphisms lifts quasi-isomorphisms. The counterexamples presented in this section are obtained using Corollary~\ref{cor:gen-Kripke}.
\begin{definition} [Protecting flexible symbols] \label{def:flex}
A signature morphism $\chi:\Delta\to \Delta'$ \emph{protects flexible symbols} if the following properties are satisfied:
\begin{description}
\item [(Preservation)] $\chi$  preserves flexible sorts, that is, $\chi(s)\in S'^\flex$ for all $s\in S^\flex$.
\item [(Injectivity)] $\chi$ is injective on both 
\begin{enumerate}[label=(I\arabic*)]
\item \label{inj-a} flexible sorts and 
\item \label{inj-b} flexible function and relation symbols that have at least one flexible sort in the arity.
\end{enumerate}
\item [(Surjectivity)] For all flexible function symbols $\sigma:\ari\to s \in F'^\flex\setminus \chi(F^\flex)$, 
\begin{enumerate}[label=(J\arabic*)]
\item \label{junk-a} if $s$ is flexible then $s$ is outside the image of $\chi$, that is, $s\in S'^\flex\setminus \chi(S^\flex)$, and
\item \label{junk-b} if there exists a flexible sort $st\in S^\flex$ such that $\chi(st)$ occurs in $\ari$ then there exists a rigid term of sort $s$, that is, $T_{\Delta',s}\neq\emptyset$.
\end{enumerate}
\end{description}
\end{definition}
The following result is the key for establishing a criterion for characterizing a Craig interpolation square.
\begin{lemma}[Lifting Lemma] \label{lemma:lifting}
Assume  
\begin{enumerate*}[label=(\alph*)] 
\item a signature morphism $\chi:\Delta\to \Delta'$ which is injective on sorts and protects flexible symbols,
\item a $\Delta$-model $\M=(W,M)$,
\item a $\Delta'$-model $\N'=(V',N')$ and
\item an isomorphism $h:\G_\N \to \G_\M$, where $\N\coloneqq \N'\red_\chi$.
\end{enumerate*}
Then there exist a $\Delta'$-model $\M'$ and an isomorphism $h':\G_{\N'} \to \G_{\M'}$ such that $\M'\red_\chi=\M$ and $h'\red_\chi=h$.
Moreover, if $h(v)=w$ then $h'(v)=w$, for all possible worlds $v\in |V'|$ and $w\in|W|$.
\end{lemma}
\begin{proof}
Without loss of generality we assume that $\G_{\N'}=\N'$. 
By ``surjectivity''\ref{junk-a} condition, 
we have $\G_\N=\N$.
Therefore, $\N$ is isomorphic with $\G_\M$. 
We construct a $\Delta'$-model $\M' = (W', M')$ and a $\Delta'$-homomorphism $h':\N'\to \M'$ through the following steps.
\begin{enumerate}[label=S\arabic*)]
\item Since $h: V\to W$ is a $\Sigma^\nom$-isomorphism and $V'\red_{\chi^\nom}=V$,
by the same arguments used in \cite[Proposition 11]{gai-pop-rob},
there exists a $\Sigma'^\nom$-isomorphism $h': V'\to W'$ such that $h'\red_\chi=h$.
\begin{center}
\begin{tikzcd}
V' \ar[dotted]{d}[left]{\red_{\chi^\nom}} \ar[r,"h'",gray]& \textcolor{gray}{W'} \ar[dotted,no head,gray]{r} & \Sigma'^\nom \\
V  \ar[r,"h"]& W \ar[dotted,no head]{r}& \Sigma^\nom \ar{u}[right]{\chi^\nom}
\end{tikzcd}
\end{center}
\item Since $\N \cong \G_\M= h(\N)$, 
the restriction of $h$ to rigid sorts $\R_h:\R_\N\to \R_\M$ is an isomorphism. 
Since $\N'\red_\chi=\N$, we have $\R_{\N'}\red_{\chi^\rigid}=\R_\N$.
Since $\chi^\rigid:\Sigma^\rigid\to\Sigma'^\rigid$ is injective on sorts,
by the same arguments used in \cite[Proposition 11]{gai-pop-rob},
there exists a $\Sigma'^\rigid$-isomorphism $\R_{h'}:\R_{\N'}\to \R_{\M'}$ such that $\R_{h'}\red_{\chi^\rigid}=\R_h$.
\begin{center}
\begin{tikzcd}
\R_{\N'} \ar[dotted]{d}[left]{\red_{\chi^\rigid}} \ar[r,"\R_{h'}",gray]& \textcolor{gray}{\R_{\M'}} \ar[dotted,no head,gray]{r} & \Sigma'^\rigid \\
\R_\N  \ar[r,"\R_h"]& \R_\M \ar[dotted,no head]{r}& \Sigma^\rigid \ar{u}[right]{\chi^\rigid}
\end{tikzcd}
\end{center}
\item We define $\M'$ on the flexible symbols in the image of $\chi$.

Let $w\in |W'|$ and $v= h^{-1}(w)$.
\begin{itemize}
\item $M'(w)_{\chi(s)} \coloneqq M(w)_s$ for all flexible sorts $s\in S^\flex$.
\item $\chi(\sigma)^{M'(w)}\coloneqq \sigma^{M(w)}$ for all flexible function symbols $\sigma:\ari\to s\in F^\flex$.
By ``injectivity''~\ref{inj-b} condition,
$\chi(\sigma)^{M'(w)}$ is well-defined.
\item $\chi(\pi)^{M'(w)}\coloneqq \pi^{M(w)}$ for all flexible relation symbols $\pi:\ari\in P^\flex$.
By ``injectivity''~\ref{inj-b} condition,
$\chi(\pi)^{M'(w)}$ is well-defined.
\end{itemize}
Let $h'_{v,\chi(s)} \coloneqq h_{v,s}$ for all sorts $s\in S^\flex$, which ensures that $h'\red_\chi=h$.
\item We define $\M'$ on the flexible symbols outside the image of $\chi$. 

Let $w\in|W'|$ and $v= h^{-1}(w)$.
\begin{itemize}
\item $M'(w)_s\coloneqq N'(v)_s$, for all flexible sorts $s\in S'^\flex\setminus\chi(S^\flex)$.
\end{itemize}
Let $h'_{v,s}:M'(w)_s\to N'(v)_s$ be the identity on $N'(v)_s$, that is, $h'_{v,s}\coloneqq 1_{N'(v)_s}$.
\begin{itemize}
\item for all $\sigma:\ari\to s\in F'^\flex \setminus \chi(F^\flex)$, the function $\sigma^{M'(w)} : M'(w)_\ari \to M'(w)_s$ is defined as follows:
\begin{enumerate}
\item $\sigma^{M'(w)}(x)\coloneqq h'_{v,s}(\sigma^{N'(v)}(h'^{-1}_{v,\ari}(x)))$ if $x\in h'_{v,\ari}(N'(v)_\ari)$;
\item $\sigma^{M'(w)}(x)$ is arbitrarily defined if $x\not\in h'_{v,\ari}(N'(v)_\ari)$.
By ``surjectivity''\ref{junk-b} condition, 
$T_{\Delta',s}\neq\emptyset$, which means that $M'(w)_s\neq \emptyset$ and $\sigma^{M'(w)}(x)$ is well-defined.
\end{enumerate}
\item $m\in \pi^{M'(w)}$ iff $h'^{-1}_{v,\ari}(m)\in \pi^{\N'(v)}$ 
for all $(\pi:\ari)\in P'^\flex\setminus\chi(P^\flex)$ and all $m\in M'(w)_\ari$.
\end{itemize}
\end{enumerate}
\end{proof}
The following result, derived as a corollary of Theorem~\ref{th:interpolation} and Lemma~\ref{lemma:lifting}, provides a characterization of pushouts that form Craig Interpolation squares.
\begin{corollary}\label{cor:cip}
Assume a pushout of signature morphisms as depicted in the diagram below. 
\begin{center}
\begin{tikzcd}
\Delta_\bc \ar[r,"\chi_\bc"] & 
\Delta_\dc \\
\Delta \ar[r,"\chi",swap] \ar[u,"\delta"] & \Delta_\ac \ar[u,"\delta_\ac",swap]
\end{tikzcd}
\end{center}
If  $\chi$ or $\delta$ is injective on sorts and protects flexible symbols, then the pushout above has CIP.
\end{corollary}

Now, assume two spans of signature morphisms
\begin{enumerate}
\item $\{\Delta_\ac\xleftarrow{\chi}\Delta\xrightarrow{\delta}\Delta_\bc\}$ such that $\chi$ is injective on sorts and protects flexible symbols, and
\item $\{\Delta_\ac'\xleftarrow{\chi'}\Delta'\xrightarrow{\delta'}\Delta_\bc'\}$ such that $\delta'$ is injective on sorts and protects flexible symbols.
\end{enumerate}
We can form their disjoint union, obtaining the span
\begin{center}
$\{\Delta_\ac\oplus \Delta_\ac'\xleftarrow{\chi\oplus\chi'}\Delta\oplus\Delta'\xrightarrow{\delta\oplus\delta'}\Delta_\bc\oplus\Delta_\bc'\}$
\end{center}
This combined span lifts quasi-isomorphisms, and the corresponding pushout of signature morphisms satisfies the CIP.

The following lemma demonstrates that the ``preservation'' condition from Definition~\ref{def:flex} is a necessary requirement for Corollary~\ref{cor:cip}.
\begin{lemma}[Preservation]\label{lemma:preservation}
The pushout shown in the diagram below fails to satisfy CIP.
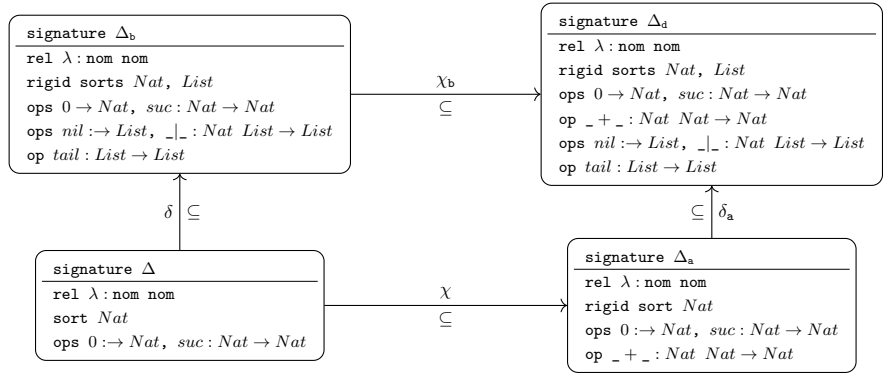
\begin{center}
 \begin{tikzpicture}[scale=.7, transform shape]
  \node(1)[draw,rectangle,rounded corners] {
   \tt\begin{tabular}{l}
    signature $\Delta_\bc$  \\ \hline
    rel $\lambda:\mathtt{nom}~\mathtt{nom}$ \\
    rigid sorts $Nat$, $List$\\
    ops $0\to Nat$, $suc: Nat \to Nat$ \\
    ops $nil:\to List$, $\_|\_:Nat~List\to List$ \\
    op $tail: List\to List$ 
   \end{tabular}}; 
  \node(2)[draw,rectangle,rounded corners, right of = 1, node distance = 10 cm] {
   \tt\begin{tabular}{l}
    signature $\Delta_\dc$ \\ \hline
    rel $\lambda:\mathtt{nom}~\mathtt{nom}$ \\
    rigid sorts $Nat$, $List$\\
    ops $0\to Nat$, $suc: Nat \to Nat$ \\
    op $\_+\_: Nat~Nat\to Nat$\\
    ops $nil:\to List$, $\_|\_:Nat~List\to List$\\
    op $tail: List\to List$
   \end{tabular}};
  \node(3)[draw,rectangle,rounded corners, below of = 1, node distance = 4cm] 
   {\tt 
   \begin{tabular}{l}
   signature $\Delta$ \\ \hline
   rel $\lambda:\mathtt{nom}~\mathtt{nom}$ \\
   sort $Nat$\\
   ops $0:\to Nat$, $suc: Nat\to Nat$
   \end{tabular}};
  \node(4)[draw,rectangle,rounded corners, below of = 2, node distance = 4cm] {
  \tt\begin{tabular}{l}
   signature $\Delta_\ac$ \\ \hline
   rel $\lambda:\mathtt{nom}~\mathtt{nom}$ \\
   rigid sort $Nat$\\
   ops $0:\to Nat$, $suc:Nat\to Nat$\\
   op $\_+\_:Nat~Nat\to Nat$
  \end{tabular}};
   \draw[->] 
    (1) to 
     node[above]{$\chi_\bc$} 
     node[below]{$\subseteq$}
    (2);
   \draw[->] (3) to node[above]{$\chi$} node[below]{$\subseteq$} (4);
   \draw[->] (3) to node[left]{$\delta$} node[right]{$\subseteq$} (1);
   \draw[->] (4) to node[right]{$\delta_\ac$} node[left]{$\subseteq$} (2);
\end{tikzpicture}
\end{center}
\end{lemma}
\begin{proof}
It is important to observe that the sort $Nat$ is flexible in $\Delta$, whereas it is rigid in both $\Delta_\ac$ and $\Delta_\bc$.
Let $\phi_\ac \coloneqq  \Forall{x} (suc(x)= 0)\implies \pos{\lambda} (suc(x)\neq 0)$ and
$\phi_\bc \coloneqq  \Forall{x} (suc(x)= 0)\implies \pos{\lambda} (suc(x)\neq 0)$.
By definition, $\phi_\ac\models_{\Delta_\dc} \phi_\bc$. 
Next, we construct a $\Delta_\ac$-model $\M_\ac$ which satisfies $\phi_\ac$ and a $\Delta_\bc$-model $\M_\bc$ which satisfy $\neg\phi_\bc$ such that their reducts to $\Delta$ are semantically equivalent.

The Kripke structure $\M_\ac=(W_\ac,M_\ac)$ is defined as follows:
\begin{enumerate} 
\item The set of possible worlds is $|W_\ac|\coloneqq \{w_n\mid n \geqslant 2 \}$ and the transition relation is $\lambda^{W_\ac}\coloneqq \{(w_n,w_{n+1}) \mid n\geqslant 2 \}$.
\item For all natural numbers $n\geqslant 2$, the first-order model $M_\ac(w_n)$ assigns 
the sort $Nat$ to the set $\omega$ of natural numbers with 
\begin{enumerate*}[label=(\alph*)]
\item $suc^{M_\ac(w_n)}(x)\coloneqq (x+1)\mod n$, for all natural numbers $x<\omega$, and
\item $x+^{M_\ac(w_n)}y\coloneqq (x+y) \mod n$, for all natural numbers $x,y<\omega$.
\end{enumerate*}
\end{enumerate}

By construction, $\M_\ac\models_{\Delta_\ac}\phi_\ac$.

The Kripke structure $\M_\bc=(W_\bc,M_\bc)$ is defined as follows:
\begin{enumerate}
\item Let $W_\bc\coloneqq W_\ac$. 
Since $\Sigma^\nom_\bc=\Sigma^\nom_\ac$, the frame $W_\bc$ is well-defined.
\item For all $n\geqslant 2$, the first-order model $M_\bc(w_n)$ assigns 
\begin{itemize}
\item the sort $Nat$ to $\{-1\}\cup \omega$ with 
\begin{enumerate*}[label=(\alph*)]
\item  $0^{M_\bc(w_n)}\coloneqq 0$ and 
\item $suc^{M_\bc(w_n)}(x)\coloneqq (x+1) \mod n$ for all $x\in \{-1\}\cup \omega$;
\end{enumerate*}
\item the sort $List$ to the set of lists with elements from $\{-1,0,\dots,n-1\}$ with $nil:\to List$, $\_|\_:Nat~List\to List$ and $tail: List\to List$ interpreted in the usual way.
\end{itemize}
\end{enumerate}

By construction, for all $n\geqslant 2$, we have
$(\M_\bc^{x\leftarrow(-1)},w_n)\models_{\Delta_\bc} suc(x)=0$.
It follows that for all $n\geqslant 2$, we have $(\M_\bc,w_n)\models_{\Delta_\bc} \Exists{x} suc(x)=0 \land \nec{\lambda} (suc(x)=0)$, that is,
$(\M_\bc,w_n)\models_{\Delta_\bc} \neg\phi_\bc$.
Since $\G_{\M_\ac\red_\Delta}=\G_{\M_\bc\red_\Delta}$, for all $n\geqslant 2$ we have $(\M_\ac\red_\Delta,w_n)\equiv(\M_\bc\red_\Delta,w_n)$.

Suppose towards a contradiction that 
$\phi_\ac\models_{\Delta_\ac}\chi(\phi)$ and 
$\delta(\phi)\models_{\Delta_\bc} \phi_\bc$ 
for some $\phi\in\Sen(\Delta)$.
For all natural numbers $n>0$, 
we have:  
\begin{proofsteps}{20em}
$(\M_\ac,w_n)\models_{\Delta_\ac} \phi_\ac$ & 
since $\M_\ac\models_{\Delta_\ac} \phi_\ac$ \\
$(\M_\ac,w_n)\models_{\Delta_\ac} \chi(\phi)$ & 
since $\phi_\ac\models_{\Delta_\ac}\chi(\phi)$ \\
$(\M_\ac\red_\Delta,w_n)\models_\Delta \phi$ & 
by local satisfaction condition \\
$(\M_\bc\red_\Delta,w_n)\models\phi$ & 
since $(\M_\ac\red_\Delta,w_n)\equiv(\M_\bc\red_\delta,w_n)$ \\
$(\M_\bc,w_n)\models_{\Delta_\bc} \delta(\phi)$ & 
by local satisfaction condition \\
$(\M_\bc,w_n)\models_{\Delta_\bc}\phi_\bc$ & 
since $\delta(\phi)\models_{\Delta_\bc} \phi_\bc$\\
 $(\M_\bc, w_n)\models_{\Delta_\bc}\neg \phi_\bc$ &
since $\M_\bc\models_{\Delta_\bc}\neg \phi_\bc$ \\
contradiction & from previous two steps
\end{proofsteps}
\end{proof}
In Lemma~\ref{lemma:preservation}, both morphisms $\chi$ and $\delta$ fail to satisfy the ``Preservation'' condition from Definition~\ref{def:flex}.
\begin{lemma}[Sort injectivity]\label{lemma:sort-inj}
The pushout shown in the diagram below fails to satisfy CIP.
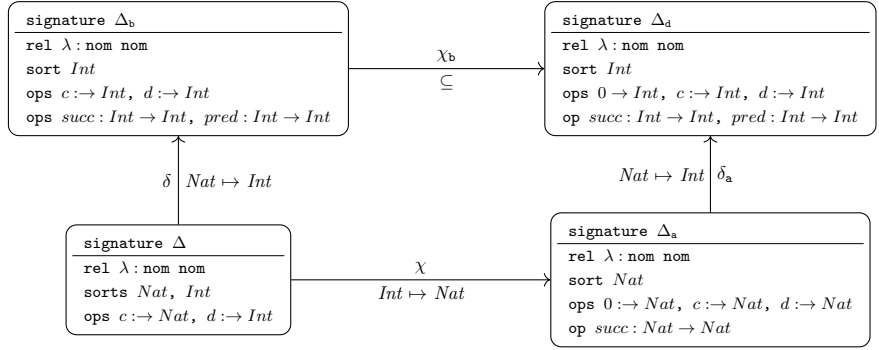
\begin{center}
 \begin{tikzpicture}[scale=.7, transform shape]
  \node(1)[draw,rectangle,rounded corners] {
   \tt\begin{tabular}{l}
    signature $\Delta_\bc$  \\ \hline
    rel $\lambda:\mathtt{nom}~\mathtt{nom}$ \\
    sort $Int$\\
    ops $c:\to Int$, $d:\to Int$
   \end{tabular}}; 
  \node(2)[draw,rectangle,rounded corners, right of = 1, node distance = 10 cm] {
   \tt\begin{tabular}{l}
    signature $\Delta_\dc$ \\ \hline
    rel $\lambda:\mathtt{nom}~\mathtt{nom}$ \\
    sort $Int$\\
    ops $c: \to Int$, $d:\to Int$
   \end{tabular}};
  \node(3)[draw,rectangle,rounded corners, below of = 1, node distance = 4cm] 
   {\tt 
   \begin{tabular}{l}
   signature $\Delta$ \\ \hline
   rel $\lambda:\mathtt{nom}~\mathtt{nom}$ \\
   sorts $Nat$, $Int$\\
   ops $c:\to Nat$, $d:\to Int$ \\ 
   \end{tabular}};
  \node(4)[draw,rectangle,rounded corners, below of = 2, node distance = 4cm] {
  \tt\begin{tabular}{l}
   signature $\Delta_\ac$ \\ \hline
   rel $\lambda:\mathtt{nom}~\mathtt{nom}$ \\
   sort $Nat$\\
   ops $c:\to Nat$, $d:\to Nat$
  \end{tabular}};
   \draw[->] 
    (1) to 
     node[above]{$\chi_\bc$} 
     node[below]{$\subseteq$}
    (2);
   \draw[->] (3) to 
     node[above]{$\chi$} 
     node[below]{ Int $\mapsto$ Nat} (4);
   \draw[->] (3) to node[left]{$\delta$} node[right]{Nat $\mapsto$ Int} (1);
   \draw[->] (4) to node[right]{$\delta_\ac$} node[left]{Nat $\mapsto$ Int} (2);
\end{tikzpicture}
\end{center}
\end{lemma}
\begin{proof}
Let $\phi_\ac\coloneqq (c\neq d)$ and $\phi_\bc\coloneqq (c \neq d)$.
Next, 
we construct a $\Delta_\ac$-model $\M_\ac$ which satisfies $\phi_\ac$ and 
a $\Delta_\bc$-model $\M_\bc$ which satisfies $\neg\phi_\bc$ such that 
$\M_\ac\red_\Delta\equiv \M_\bc\red_\Delta$.

The Kripke structure $\M_\ac=(W_\ac,M_\ac)$ is defined as follows:
\begin{enumerate} 
\item The set of possible worlds is $|W_\ac|\coloneqq \{w_n\mid 0 < n\}$ and the transition relation is $\lambda^{W_\ac}\coloneqq \{(w_n,w_{n+1}) \mid 0 < n\}$.

\item For each natural number $0< n$,
the first-order model $M_\ac(w_n)$ assigns $Nat$ to the set $\mathbb{Z}$ of integers with 
\begin{enumerate*}[label=(\alph*)]
\item $c^{M_\ac(w_n)}\coloneqq n$ and
\item $d^{M_\ac(w_n)}\coloneqq -n$.
\end{enumerate*}
\end{enumerate}

By construction, $\M_\ac\models\phi_\ac$.

The Kripke structure $\M_\bc=(W_\bc,M_\bc)$ is defined as follows:
\begin{enumerate} 
\item Let $W_\bc\coloneqq W_\ac$. 
Since $\Sigma^\nom_\bc=\Sigma^\nom_\ac$, the frame $W_\bc$ is well-defined.
\item For all natural numbers $n>0$, the first-order model $M_\bc(w_n)$ assigns the sort $Nat$ to the set $\mathbb{Z}$ of integers with 
\begin{enumerate*}[label=(\alph*)]
\item $c^{M_\bc(w_n)}\coloneqq n$ and
\item $d^{M_\bc(w_n)}\coloneqq n$.
\end{enumerate*}
\end{enumerate}

By construction, $\M_\bc\models \neg\phi_\bc$.

We define an isomorphism $h:\M_\ac\red_\chi\to \M_\bc\red_\delta$ as follows:
\begin{enumerate}
\item $h:W_\ac\to W_\bc$ is the identity.
\item For all natural numbers $n > 0$, the function $h_{w_n,Nat}:M_\ac(w_n)_{Nat}\to M_\bc(w_n)_{Nat}$ is the identity.
Observe that $h_{w_n,Nat}(c^{M_\ac(w_n)})=h_{w_n,Nat}(n)=n= c^{\M_\bc(w_n)}$, which means that $h_{w_n,Nat}$ is well-defined.
\item For all natural numbers $n>0$, the function $h_{w_n,Int}:M_\ac(w_n)_{Int}\to M_\bc(w_n)_{Int}$ is defined by $h_{w_n,Int}(x)=-x$ for all integers $x\in\mathbb{Z}$.
Since $h_{w_n,Int}(d^{M_\ac(w_n)})=h_{w_n,Int}(-n)=-(-n)= d^{\M_\bc(w_n)}$, the mapping $h_{w_n,Int}$ is well-defined.
\end{enumerate}

Since $\M_\ac\red_\chi$ and $\M_\bc\red_\delta$ are isomorphic, we have $(\M_\ac\red_\chi,w_n)\equiv(\M_\bc\red_\delta,w_n)$ for all $n>0$.

By applying the same reasoning as in the second part of the proof of Lemma~\ref{lemma:preservation}, we conclude that there is no sentence $\phi\in\Sen(\Delta)$ such that
$\phi_\ac\models_{\Delta_\ac}\chi(\phi)$ and 
$\delta(\phi)\models_{\Delta_\bc} \phi_\bc$.
\end{proof}
The counterexample from Lemma~\ref{lemma:sort-inj} is based on the counterexample for $\FOL$ presented in Lemma~\ref{lemma:FOL-CIP}. 
Observe that both $\chi$ and $\delta$ from Lemma~\ref{lemma:sort-inj} are not injective on sorts.
\begin{lemma}[Operation injectivity \& surjectivity]\label{lemma:conf+junk}
The pushout shown in the diagram below fails to satisfy CIP.
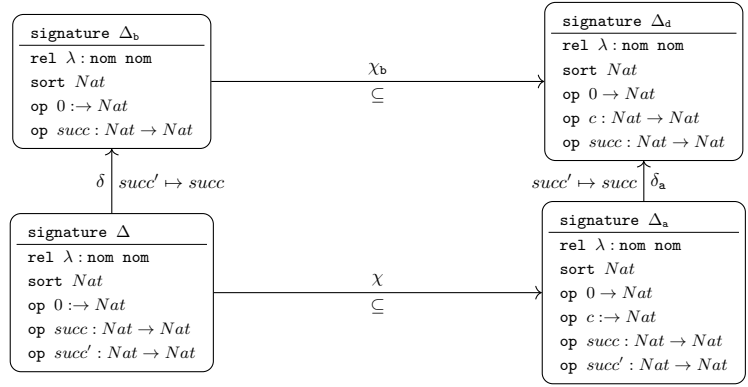
\begin{center}
 \begin{tikzpicture}[scale=.7, transform shape]
  \node(1)[draw,rectangle,rounded corners] {
   \tt\begin{tabular}{l}
   signature $\Delta_\bc$ \\ \hline
   rel $\lambda:\mathtt{nom}~\mathtt{nom}$ \\
   sort $Nat$\\
   op $0:\to Nat$\\ 
   op $succ:Nat\to Nat$\\
  \end{tabular}}; 
  \node(2)[draw,rectangle,rounded corners, right of = 1, node distance = 10 cm] {
   \tt\begin{tabular}{l}
    signature $\Delta_\dc$ \\ \hline
    rel $\lambda:\mathtt{nom}~\mathtt{nom}$ \\
    sort $Nat$\\
    op $0\to Nat$\\ 
    op $c: Nat \to Nat$ \\
    op $succ: Nat\to Nat$
   \end{tabular}};
  \node(3)[draw,rectangle,rounded corners, below of = 1, node distance = 4cm] 
   {\tt 
   \begin{tabular}{l}
   signature $\Delta$ \\ \hline
   rel $\lambda:\mathtt{nom}~\mathtt{nom}$ \\
   sort $Nat$\\
   op $0:\to Nat$\\ 
   op $succ: Nat\to Nat$\\
   op $succ': Nat\to Nat$
   \end{tabular}};
  \node(4)[draw,rectangle,rounded corners, below of = 2, node distance = 4cm] {
  \tt\begin{tabular}{l}
    signature $\Delta_\ac$  \\ \hline
    rel $\lambda:\mathtt{nom}~\mathtt{nom}$ \\
    sort $Nat$\\
    op $0\to Nat$\\ 
    op $c:\to Nat$ \\
    op $succ: Nat \to Nat$\\
    op $succ': Nat \to Nat$
   \end{tabular}};
   \draw[->] 
    (1) to 
     node[above]{$\chi_\bc$} 
     node[below]{$\subseteq$}
    (2);
   \draw[->] (3) to node[above]{$\chi$} node[below]{$\subseteq$} (4);
   \draw[->] (3) to node[left]{$\delta$} node[right]{$succ'\mapsto succ$} (1);
   \draw[->] (4) to node[right]{$\delta_\ac$} node[left]{$succ'\mapsto succ$} (2);
\end{tikzpicture}
\end{center}
\end{lemma}
\begin{proof}
It is important to observe that $Nat$ is a flexible sort in each of the four signatures $\Delta$, $\Delta_\ac$, $\Delta_\bc$, and $\Delta_\dc$. 
Consequently, quantification over variables of sort $Nat$ is not permitted.
Let $\phi_\ac\coloneqq (succ(c)\neq succ'(c))$ and
let $\phi_\bc\coloneqq (succ(0) = 0)$.
Since $\delta_\ac(\phi_\ac)= (succ(c)\neq succ(c))$ is inconsistent over $\Delta_\dc$,
we have $\delta_\ac(\phi_\ac)\models_{\Delta_\dc} \chi_\bc(\phi_\bc)$.
Next, we construct a pointed model $(\M_\ac,w_0)$  over $\Delta_\ac$  which satisfies $\phi_\ac$ and
a pointed model $(\M_\bc,w_0)$ over $\Delta_\bc$ which satisfies $\neg\phi_\bc$ such that 
$(\M_\ac\red_\chi,w_0)\equiv (\M_\bc\red_\delta,w_0)$.

The Kripke structure $\M_\ac=(W_\ac,M_\ac)$ is defined as follows:
\begin{enumerate} 
\item The set of possible worlds is $|W_\ac|\coloneqq \{w_n\mid n < \omega\}$ and the transition relation is $\lambda^{W_\ac}\coloneqq \{(w_n,w_{n+1}) \mid n<\omega\}$.
\item The first-order model $M_\ac(w_0)$ assigns the sort $Nat$ to $\mathbb{Z}$, the set of integers, with 
\begin{enumerate*}[label=(\alph*)]
\item $c^{M_\ac(w_0)}\coloneqq -1$,
\item $succ^{M_\ac(w_0)}(x)$ $\coloneqq x+1$ for all $x\in\mathbb{Z}$, 
\item $succ'^{M_\ac(w_0)}(x)\coloneqq |x| + 1$ for all $x\in\mathbb{Z}$.
\end{enumerate*}
For all $n>0$, the first-order model $M_\ac(w_n)$ assigns the sort $Nat$ to the set $\mathbb{Z}_n$ of integers modulo $n$ with 
\begin{enumerate*}[label=(\alph*)]
\item $0^{M_\ac(w_n)}=[0]$, 
\item $succ^{M_\ac(w_n)}([x])=[x + 1]$ for all $x<\omega$, and
\item $succ'^{M_\ac(w_n)}\coloneqq succ^{M_\ac(w_n)}$.
\end{enumerate*}
\end{enumerate}

By construction, 
$(\M_\ac,w_0)\models_{\Delta_\ac} \phi_\ac$.

The Kripke structure $\M_\bc=(W_\bc,M_\bc)$ is defined as follows:
\begin{enumerate} 
\item Let $W_\bc\coloneqq W_\ac$. 
Since $\Sigma^\nom_\bc=\Sigma^\nom_\ac$, the frame $W_\bc$ is well-defined.
\item The first-order model $M_\bc(w_0)$ assigns the sort $Nat$ to the set $\omega$  of natural numbers interpreting $0:\to Nat$ and $succ: Nat\to Nat$ in the usual way.
For all $n>0$, the first-order model $M_\bc(w_n)$ interprets the sort $Nat$ and the function symbols $0:\to Nat$ and $succ:Nat\to Nat$ exactly as $M_\ac(w_n)$.
\end{enumerate}

By construction, $(\M_b,w_0)\models\neg\phi_\bc$.

Since $\G_{\M_\ac\red_\chi}=\G_{\M_\bc\red_\delta}$, 
by Corollary~\ref{cor:gen-Kripke},
we have $(\M_\ac\red_\chi,w_0)\equiv(\M_\bc\red_\delta,w_0)$.

By applying the same reasoning as in the second part of the proof of Lemma~\ref{lemma:preservation}, we conclude that there is no sentence $\phi\in\Sen(\Delta)$ such that
$\phi_\ac\models_{\Delta_\ac}\chi(\phi)$ and 
$\delta(\phi)\models_{\Delta_\bc} \phi_\bc$.
\end{proof}
In the example presented in Lemma~\ref{lemma:conf+junk}, observe the following:
\begin{enumerate}[label=(\alph*)]
\item The morphism $\chi$ adds new flexible constant $c:\to Nat$  on the flexible sort $Nat$, thereby failing to satisfy  ``surjectivity'' condition~\ref{junk-a} from Definition~\ref{def:flex}.
\item The morphism $\delta$ maps both $succ:Nat\to Nat$ and $succ':Nat\to Nat$ to  $succ:Nat\to Nat$ contravening condition ``injectivity''~\ref{inj-b} from Definition~\ref{def:flex}.
\end{enumerate}
The following lemma highlights a distinction between $\FOHLR$ and its semantic restriction to models with non-empty carrier sets, at least with respect to CIP.
\begin{lemma}[Operation surjectivity] \label{lemma:counter-empty}
The pushout shown in the diagram below fails to satisfy CIP.
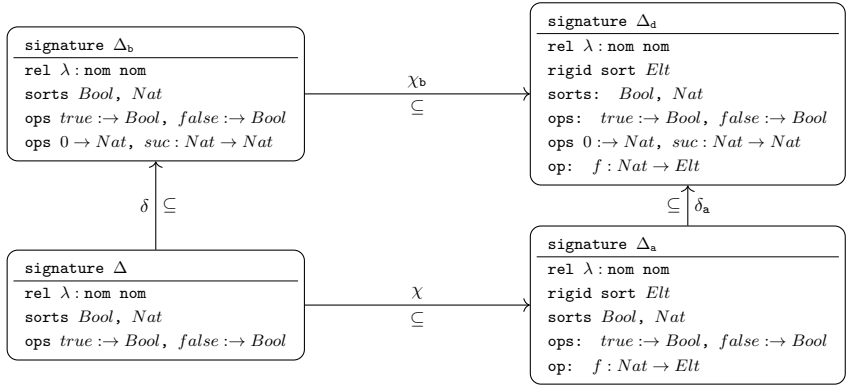
\begin{center}
 \begin{tikzpicture}[scale=.7, transform shape]
  \node(1)[draw,rectangle,rounded corners] {
   \tt\begin{tabular}{l}
    signature $\Delta_\bc$  \\ \hline
    rel $\lambda:\mathtt{nom}~\mathtt{nom}$ \\
    sorts $Bool$, $Nat$\\
    ops $true:\to Bool$, $false:\to Bool$\\
    ops $0\to Nat$, $suc: Nat \to Nat$ 
   \end{tabular}}; 
  \node(2)[draw,rectangle,rounded corners, right of = 1, node distance = 10 cm] {
   \tt\begin{tabular}{l}
    signature $\Delta_\dc$ \\ \hline
    rel $\lambda:\mathtt{nom}~\mathtt{nom}$ \\
    rigid sort $Elt$\\
    sorts: $Bool$, $Nat$\\
    ops: $true:\to Bool$, $false:\to Bool$\\
    ops $0:\to Nat$, $suc: Nat\to Nat$\\
    op: $f: Nat\to Elt$
   \end{tabular}};
  \node(3)[draw,rectangle,rounded corners, below of = 1, node distance = 4cm] 
   {\tt 
   \begin{tabular}{l}
   signature $\Delta$ \\ \hline
   rel $\lambda:\mathtt{nom}~\mathtt{nom}$ \\
   sorts $Bool$, $Nat$\\
   ops $true:\to Bool$, $false:\to Bool$
   \end{tabular}};
  \node(4)[draw,rectangle,rounded corners, below of = 2, node distance = 4cm] {
  \tt\begin{tabular}{l}
   signature $\Delta_\ac$ \\ \hline
   rel $\lambda:\mathtt{nom}~\mathtt{nom}$ \\
   rigid sort $Elt$\\
   sorts $Bool$, $Nat$\\
   ops: $true:\to Bool$, $false:\to Bool$\\
   op: $f:Nat\to Elt$
  \end{tabular}};
   \draw[->] 
    (1) to 
     node[above]{$\chi_\bc$} 
     node[below]{$\subseteq$}
    (2);
   \draw[->] (3) to node[above]{$\chi$} node[below]{$\subseteq$} (4);
   \draw[->] (3) to node[left]{$\delta$} node[right]{$\subseteq$} (1);
   \draw[->] (4) to node[right]{$\delta_\ac$} node[left]{$\subseteq$} (2);
\end{tikzpicture}
\end{center}
\end{lemma}
\begin{proof}
It is important to observe that the sorts $Nat$ and $Bool$ are flexible in each of the four signatures $\Delta$, $\Delta_\ac$, $\Delta_\bc$, and $\Delta_\dc$. 
We define the following sentences: 
\begin{itemize} 
\item $\phi_\ac\coloneqq \Forall{x}\bot$, defined over $\Delta_\ac$, which asserts that the sort $Elt$ is empty.
\item  $\phi_\bc\coloneqq  (true \neq false) \land \pos{\lambda} (true = false)$, defined over $\Delta_\bc$, which asserts that the inequality $true\neq false$ is not preserved along transitions.
\end{itemize}
Since $\phi_\ac$  is inconsistent over $\Delta_\dc$, 
we have $\phi_\ac \models_{\Delta_\dc} \phi_\bc$.
Next, we construct a $\Delta_\ac$-model $\M_\ac$ which satisfies $\phi_\ac$ and 
a $\Delta_\bc$-model $\M_\bc$ which satisfies $\neg\phi_\bc$ such that $
\M_\ac\red_\Delta\equiv \M_\bc\red_\Delta$.

The Kripke structure $\M_\ac=(W_\ac,M_\ac)$ is defined as follows:
\begin{enumerate} 
\item The set of possible worlds is $|W_\ac|\coloneqq \{w_n\mid n<\omega\}$ and the transition relation is $\lambda^{W_\ac}\coloneqq \{(w_n,w_{n+1}) \mid n<\omega\}$.
\item For all $n<\omega$, the first-order model $M_\ac(w_n)$ assigns 
\begin{enumerate}
\item the sorts $Nat$ and $Elt$ to the the empty set, and
\item the sort $Bool$ to $\{true,false\}$ with $true^{M_\ac(w_n)}=true$ and $false^{M_\ac(w_n)}=false$.
\end{enumerate}
\end{enumerate}

By construction, $\M_\ac\models\phi_\ac$.

The Kripke structure $\M_\bc=(W_\bc,M_\bc)$ is defined as follows:
\begin{itemize}
\item It interprets all symbols in the signature $\Delta$, except for $Nat$, in the same way as $\M_\ac$.
\item For all $n<\omega$, the restriction of $M_\bc(w_n)$ to the signature of natural numbers is $\mathbb{Z}_n$, the model of integers modulo~$n$.
\end{itemize}

By construction,  $\M_\bc\models_{\Delta_\bc}\neg\phi_\bc$.

Since $Nat$ is flexible in the signature $\Delta$ and has no terms, 
we have $\G_{\M_\ac\red_\Delta}=\G_{\M_\bc\red_\Delta}$.
By Corollary~\ref{cor:gen-Kripke}, for all $n<\omega$, the pointed models $(\M_\ac\red_\Delta,w_n)$ and $(\M_\bc\red_\Delta,w_n)$ are indistinguishable by $\Delta$-sentences,
that is, $(\M_\ac\red_\Delta,w_n)\equiv (\M_\bc\red_\Delta,w_n)$.

By applying the same reasoning as in the second part of the proof of Lemma~\ref{lemma:preservation}, we conclude that there is no sentence $\phi\in\Sen(\Delta)$ such that
$\phi_\ac\models_{\Delta_\ac}\chi(\phi)$ and 
$\delta(\phi)\models_{\Delta_\bc} \phi_\bc$.
\end{proof}
In the example presented in Lemma~\ref{lemma:counter-empty}, observe the following:
\begin{enumerate}[label=(\alph*)]
\item The morphism $\chi$ adds a new flexible function symbol $f:Nat\to Elt$ into the source signature $\Delta$. Since there are no terms of sort $Elt$, the ``surjectivity'' condition~\ref{junk-b} is violated.
\item The morphism $\delta$ adds new flexible function symbols $0:\to Nat$ and $suc:Nat\to Nat$ on the flexible sort $Nat$, thereby contravening  ``surjectivity'' condition~\ref{junk-a} from Definition~\ref{def:flex}.
\end{enumerate}
\section{Conclusions}\label{sec:conclusion}
Recall that the signatures of $\RFOHL$, defined in Example~\ref{ex:RFOHL}, form a subcategory of $\FOHLR$ signatures. 
It is straightforward to verify that this subcategory is closed under pushouts. Consequently, Theorem~\ref{th:interpolation} applies to $\RFOHL$. 

$\FOHLR$ and $\FOHLS$, defined in Example~\ref{ex:FOHLS}, share the same signatures and Kripke structures. According to \cite[Lemma 2.20]{gai-acm}, $\FOHLS$ and $\FOHL$ have the same expressive power. The relationship between $\FOHLR$ and $\FOHLS$ is analogous to that between first-order logic and unnested first-order logic, which allows only terms of depth at most one~\cite{DBLP:books/daglib/0080659}. Therefore, a square of $\FOHLR$ signature morphisms satisfies  CIP in $\FOHLR$ if and only if it satisfies CIP in $\FOHLS$.

Similarly, the signatures of $\HPL$, defined in Example~\ref{ex:HPL}, form a subcategory of $\FOHLR$ signatures, which is closed under pushouts. Consequently, Theorem~\ref{th:interpolation} applies to $\HPL$ as well. However, since our results rely on quantification over possible worlds, this work does not encompass the interpolation result from \cite{ArecesBM01}, which applies to hybrid propositional logic without quantification.
However, the results presented in this paper can be extended to hybrid logics without quantification over possible worlds by refining the notion of generated substructures, in line with \cite[Definition 3.6]{ArecesBM01}.
\bibliographystyle{plainurl}
\bibliography{hyb-int}

@article{ArecesB01,
  author    = {Carlos Areces and
               Patrick Blackburn},
  title     = {{Bringing them all Together}},
  journal   = {Journal of Logic and Computation},
  volume    = {11},
  number    = {5},
  pages     = {657--669},
  year      = {2001}
}

@article{ArecesBM03,
  author    = {Carlos Areces and
               Patrick Blackburn and
               Maarten Marx},
  title     = {Repairing the interpolation theorem in quantified modal logic},
  journal   = {Ann. Pure Appl. Log.},
  volume    = {124},
  number    = {1-3},
  pages     = {287--299},
  year      = {2003}
}

@article{ArecesBM01, 
title={Hybrid logics: characterization, interpolation and complexity}, 
volume={66}, 
number={3}, 
journal={Journal of Symbolic Logic}, 
publisher={Cambridge University Press}, 
author={Areces, Carlos and Blackburn, Patrick and Marx, Maarten}, 
year={2001}, 
pages={977--1010}
}

@inproceedings{DBLP:conf/wollic/BlackburnMMH19,
  author    = {Patrick Blackburn and
               Manuel A. Martins and
               Mar{\'{\i}}a Manzano and
               Antonia Huertas},
  title     = {Rigid First-Order Hybrid Logic},
  booktitle = {Logic, Language, Information, and Computation - 26th International Workshop, WoLLIC 2019, Utrecht, The Netherlands, July 2-5, 2019, Proceedings},
  address   = {Utrecht},
  pages     = {53--69},
  year      = {2019},
  crossref  = {DBLP:conf/wollic/2019}
}

@proceedings{DBLP:conf/wollic/2019,
  editor    = {Rosalie Iemhoff and
               Michael Moortgat and
               Ruy J. G. B. de Queiroz},
  title     = {Logic, Language, Information, and Computation - 26th International Workshop, WoLLIC 2019, Utrecht, The Netherlands, July 2-5, 2019, Proceedings},
  series    = {Lecture Notes in Computer Science},
  volume    = {11541},
  publisher = {Springer},
  year      = {2019}
}

@article{DBLP:journals/jacm/BergstraHK90,
  author       = {Jan A. Bergstra and
                  Jan Heering and
                  Paul Klint},
  title        = {Module Algebra},
  journal      = {J. {ACM}},
  volume       = {37},
  number       = {2},
  pages        = {335--372},
  year         = {1990},
  url          = {https://doi.org/10.1145/77600.77621},
  doi          = {10.1145/77600.77621}
}

@article{journals/ipl/Borzyszkowski00,
  author = {Tomasz Borzyszkowski},
  title = {Generalized interpolation in {CASL}},
  journal = {Information Processing Letters},
  volume = {76},
  number = {1-2},
  pages = {19--24},
  year = {2000},
  doi = {10.1016/S0020-0190(00)00120-4}
}

@article{Borzyszkowski2002,
  author    = {Borzyszkowski, Tomasz},
  title     = {Logical Systems for Structured Specifications},
  journal   = {Theoretical Computer Science},
  volume    = {286},
  number    = {2},
  pages     = {197--245},
  year      = {2002}
}

@article{cohen63,
  author = {Cohen, Paul J.},
  title = {The Independence of the Continuum Hypothesis},
  journal = {Proceedings of the National Academy of Sciences of the United States of America},
  volume = {50},
  number = {6},
  year = {1963},
  month = {December},
  pages = {1143-1148},
  doi = {10.1073/pnas.50.6.1143},
  _bib2doi_finished = {true},
}

@article{cohen64,
  author = {Cohen, Paul J.},
  title = {The Independence of the Continuum Hypothesis, {II}},
  journal = {Proceedings of the National Academy of Sciences of the United States of America},
  volume = {51},
  number = {1},
  year = {1964},
  month = {January},
  pages = {105-110},
  doi = {10.1073/pnas.51.1.105},
  _bib2doi_finished = {true},
}

@article{Craig1957-CRALRA,
  author = {William Craig},
  doi = {10.2307/2963593},
  journal = {Journal of Symbolic Logic},
  number = {3},
  pages = {250--268},
  publisher = {Association for Symbolic Logic},
  title = {Linear Reasoning. {A} New Form of the {Herbrand-Gentzen} Theorem},
  volume = {22},
  year = {1957},
  timestamp = {Sun, 28 May 2017 13:21:51 +0200},
  biburl = {https://dblp.org/rec/journals/jsyml/Craig57.bib},
  bibsource = {dblp computer science bibliography, https://dblp.org},
  _bib2doi_selected = {dblp:/rec/journals/jsyml/Craig57.bib},
  _bib2doi_confirmed = {true},
}

@inproceedings{cod-H,
  author    = {Mihai Codescu},
  title     = {Hybridisation of Institutions in {HETS} (Tool Paper)},
  booktitle = {8th Conference on Algebra and Coalgebra in Computer Science, {CALCO} 2019, June 3-6, 2019, London, United Kingdom},
  pages     = {17:1--17:10},
  year      = {2019},
  crossref  = {DBLP:conf/calco/2019}
}

@proceedings{DBLP:conf/calco/2019,
  editor    = {Markus Roggenbach and
               Ana Sokolova},
  title     = {8th Conference on Algebra and Coalgebra in Computer Science, {CALCO}
               2019, June 3-6, 2019, London, United Kingdom},
  series    = {LIPIcs},
  volume    = {139},
  publisher = {Schloss Dagstuhl - Leibniz-Zentrum f{\"{u}}r Informatik},
  year      = {2019}
}

@book{Diaconescu2008,
  author = {R{\u{a}}zvan Diaconescu},
  title = {Institution-Independent Model Theory},
  series = {Studies in Universal Logic},
  publisher = {Birkh{\"{a}}user Cham},
  edition={Second},
  year={2025}
}

@article{dia-msc,
author  = {Diaconescu,R\u{a}zvan and Madeira,Alexandre},
title   = {{Encoding Hybridised Institutions into First-Order Logic}},
journal = {Mathematical Structures in Computer Science},
volume  = {26},
number  = {5},
year    = {2016},
pages   = {745--788}
}

@article{dia-ult,
author = {R\u{a}zvan Diaconescu},
title = {{Institution-independent Ultraproducts}},
journal = {Fundamenta Informatic{\ae}},
volume = 55,
number = {3-4},
pages = {321--348},
year = 2003
}

@article{DBLP:journals/logcom/Diaconescu17,
  author       = {Razvan Diaconescu},
  title        = {{Implicit Kripke semantics and ultraproducts in stratified institutions}},
  journal      = {J. Log. Comput.},
  volume       = {27},
  number       = {5},
  pages        = {1577--1606},
  year         = {2017},
  doi          = {10.1093/LOGCOM/EXW018}
}

@article{gog-ins,
  title = {Institutions: Abstract Model Theory for Specification and Programming},
  author = {Joseph A. Goguen and Rod M. Burstall},
  journal = {J. {ACM}},
  doi = {10.1145/147508.147524},
  number = {1},
  volume = {39},
  pages = {95--146},
  year = {1992}
}

@article{journals/sigplan/GoguenM82,
  author = {Joseph Goguen and Jos{\'{e}} Meseguer},
  title = {Completeness of many-sorted equational logic},
  journal = {{ACM} {SIGPLAN} Notices},
  volume = {17},
  number = {1},
  pages = {9--17},
  year = {1982},
  doi = {10.1145/947886.947887},
  timestamp = {Wed, 20 Sep 2023 08:58:12 +0200},
  biburl = {https://dblp.org/rec/journals/sigplan/GoguenM82.bib},
  bibsource = {dblp computer science bibliography, https://dblp.org},
  _bib2doi_selected = {dblp:/rec/journals/sigplan/GoguenM82.bib},
  _bib2doi_confirmed = {true},
}

@article{rob71,
  author = {Abraham Robinson},
  title = {Forcing in model theory},
  journal = {Symposia Mathematica},
  volume = {5},
  pages = {69--82},
  year = {1971},
  _bib2doi_finished = {true},
}

@article{rob,
title = {A Result on Consistency and its Application to the Theory of Definition},
journal = {Indagationes Mathematicae (Proceedings)},
volume = {59},
pages = {47-58},
year = {1956},
issn = {1385-7258},
doi = {https://doi.org/10.1016/S1385-7258(56)50008-X},
url = {https://www.sciencedirect.com/science/article/pii/S138572585650008X},
author = {Abraham Robinson}
}

@article{gai-com,
  author = {Daniel G\u{a}in\u{a} and Marius Petria},
  title = {Completeness by Forcing},
  journal = {Journal of Logic and Computation},
  volume = {20},
  number = {6},
  pages = {1165--1186},
  year = {2010},
  doi = {10.1093/LOGCOM/EXQ012}
}

@inproceedings{go-icalp24,
  author       = {Go Hashimoto and
                  Daniel G\u{a}in\u{a} and
                  Ionut \c{T}u\c{t}u},
  editor       = {Karl Bringmann and
                  Martin Grohe and
                  Gabriele Puppis and
                  Ola Svensson},
  title        = {{Forcing, Transition Algebras, and Calculi}},
  booktitle    = {51st International Colloquium on Automata, Languages, and Programming,
                  {ICALP} 2024, July 8-12, 2024, Tallinn, Estonia},
  series       = {LIPIcs},
  volume       = {297},
  pages        = {143:1--143:17},
  publisher    = {Schloss Dagstuhl - Leibniz-Zentrum f{\"{u}}r Informatik},
  year         = {2024},
  doi          = {10.4230/LIPICS.ICALP.2024.143}
}

@InProceedings{hashimoto2025,
  author =	  {Hashimoto, Go and G\u{a}in\u{a}, Daniel},
  title =	  {{Model-Theoretic Forcing in Transition Algebra}},
  booktitle = {50th International Symposium on Mathematical Foundations of Computer Science (MFCS 2025)},
  pages =	  {55:1--55:18},
  series =	  {Leibniz International Proceedings in Informatics (LIPIcs)},
  ISBN =	  {978-3-95977-388-1},
  ISSN =	  {1868-8969},
  year =	  {2025},
  volume =	  {345},
  editor =	  {Gawrychowski, Pawe{\l} and Mazowiecki, Filip and Skrzypczak, Micha{\l}},
  publisher = {Schloss Dagstuhl -- Leibniz-Zentrum f{\"u}r Informatik},
  address =	  {Dagstuhl, Germany},
  doi =		  {10.4230/LIPIcs.MFCS.2025.55}
}

@book{DBLP:books/daglib/0030198,
  author    = {Wilfrid Hodges},
  title     = {A Shorter Model Theory},
  publisher = {Cambridge University Press},
  year      = {1997},
  series    = {Cambridge Tracts in Mathematics},
  volume    = {87}
}

@article{TOCL2025,
author = {Badia, Guillermo and G\u{a}in\u{a}, Daniel and Knapp, Alexander and Kowalski, Tomasz and Wirsing, Martin},
title = {{Hybrid-Dynamic Ehrenfeucht-Fra\"{\i}ss\'{e} Games}},
publisher = {Association for Computing Machinery},
address = {New York, NY, USA},
volume = {26},
number = {4},
doi = {10.1145/3750046},
journal = {ACM Trans. Comput. Logic},
month = sep,
year = {2025},
numpages = {25}
}

@article{gai-her,
  author    = {Daniel G\u{a}in\u{a}},
  title     = {Foundations of logic programming in hybrid logics with user-defined
               sharing},
  journal   = {Theor. Comput. Sci.},
  volume    = {686},
  pages     = {1--24},
  year      = {2017}
}

@article{gai-pop-rob,
  author = {Daniel G\u{a}in\u{a} and Andrei Popescu},
  title = {An Institution-Independent Proof of the {R}obinson Consistency Theorem},
  journal = {Studia Logica},
  volume = {85},
  number = {1},
  pages = {41--73},
  year = {2007},
  doi = {10.1007/S11225-007-9022-4}
}

@inproceedings{GBK22,
  author = {Daniel G\u{a}in\u{a} and Guillermo Badia and Tomasz Kowalski},
  editor = {David Fern{\'{a}}ndez{-}Duque and Alessandra Palmigiano and Sophie Pinchinat},
  title = {Robinson consistency in many-sorted hybrid first-order logics},
  booktitle = {Advances in Modal Logic, AiML 2022, Rennes, France, August 22-25, 2022},
  pages = {407--428},
  publisher = {College Publications},
  year = {2022}
}

@article{gai-acm,
  author = {Daniel G\u{a}in\u{a}},
  title = {Forcing and Calculi for Hybrid Logics},
  journal = {Journal of the Association for Computing Machinery},
  volume = {67},
  number = {4},
  pages = {1--55},
  year = {2020},
  doi = {10.1145/3400294}
}

@article{DBLP:journals/apal/GainaBK23,
  author       = {Daniel G\u{a}in\u{a} and
                  Guillermo Badia and
                  Tomasz Kowalski},
  title        = {Omitting types theorem in hybrid dynamic first-order logic with rigid symbols},
  journal      = {Ann. Pure Appl. Log.},
  volume       = {174},
  number       = {3},
  pages        = {103212},
  year         = {2023},
  doi          = {10.1016/J.APAL.2022.103212}
}

@article{GainaBK23,
  author = {Daniel G\u{a}in\u{a} and Guillermo Badia and Tomasz Kowalski},
  title = {Omitting types theorem in hybrid dynamic first-order logic with rigid symbols},
  journal = {Annals of Pure and Applied Logic},
  volume = {174},
  number = {3},
  pages = {103212},
  year = {2023},
  doi = {10.1016/J.APAL.2022.103212}
}

@incollection{tar-bit,
author    =  {Andrzej Tarlecki},
title     =  {{Bits and Pieces of the Theory of Institutions}},
year      =  1986,
booktitle =  {Proceedings, Summer Workshop on Category Theory and Computer Programming, Lecture Notes in Computer Science},
location  =  {University of Surrey, Guildford, U.K.},
publisher =  {Springer},
volume    =  {240},
pages     =  {334--360},
editor    = {David Pitt and Samson Abramsky and Axel Poign\'e and David Rydeheard}
}

@article{DBLP:journals/tcs/Tarlecki85,
  author       = {Andrzej Tarlecki},
  title        = {On the Existence of Free Models in Abstract Algebraic Institutions},
  journal      = {Theor. Comput. Sci.},
  volume       = {37},
  pages        = {269--304},
  year         = {1985},
  doi          = {10.1016/0304-3975(85)90094-5}
}

@inproceedings{Petria07,
  author = {Marius Petria},
  editor = {Till Mossakowski and Ugo Montanari and Magne Haveraaen},
  title = {An Institutional Version of {G}{\"{o}}del's Completeness Theorem},
  booktitle = {Algebra and Coalgebra in Computer Science, Second International Conference, {CALCO} 2007, Bergen, Norway, August 20-24, 2007, Proceedings},
  series = {LNCS},
  volume = {4624},
  pages = {409--424},
  publisher = {Springer},
  year = {2007},
  doi = {10.1007/978-3-540-73859-6\_28}
}

@article{DBLP:journals/mscs/SannellaT14,
  author       = {Donald Sannella and
                  Andrzej Tarlecki},
  title        = {Property-oriented semantics of structured specifications},
  journal      = {Math. Struct. Comput. Sci.},
  volume       = {24},
  number       = {2},
  year         = {2014},
  url          = {https://doi.org/10.1017/S0960129513000212},
  doi          = {10.1017/S0960129513000212}
}

@article{Tarlecki2024Fragility,  
author    = {Andrzej Tarlecki},  
title     = {On the Fragility of Interpolation},  
journal   = {The Journal of Symbolic Logic},  
volume    = {FirstView},  
pages     = {1--38},  
year      = {2024},  
doi       = {10.1017/jsl.2024.19},  
publisher = {Cambridge University Press}
}

@article{dia-qvh,
author  = {R\u{a}zvan Diaconescu},
title   = {{Quasi-varieties and initial semantics for hybridized institutions}},
journal = {Journal of Logic and Computation},
volume  = {26},
number  = {3},
pages   = {855--891},
year    = {2016}
}

@inproceedings{martins2011,
  author    = {Manuel A. Martins and
               Alexandre Madeira and
               Razvan Diaconescu and
               Lu{\'{\i}}s Soares Barbosa},
  title     = {Hybridization of Institutions},
  booktitle = {Algebra and Coalgebra in Computer Science - 4th International Conference, {CALCO} 2011, Proceedings},
  address   = {Winchester},
  pages     = {283--297},
  year      = {2011},
  crossref  = {DBLP:conf/calco/2011}
}

@proceedings{DBLP:conf/calco/2011,
  editor    = {Andrea Corradini and
               Bartek Klin and
               Corina C{\^{\i}}rstea},
  title     = {Algebra and Coalgebra in Computer Science - 4th International Conference, {CALCO} 2011, Proceedings},
  series    = {Lecture Notes in Computer Science},
  volume    = {6859},
  publisher = {Springer},
  year      = {2011}
}

@book{DBLP:books/daglib/0080659,
  author    = {Heinz{-}Dieter Ebbinghaus and
               J{\"{o}}rg Flum and
               Wolfgang Thomas},
  title     = {Mathematical logic {(2.} ed.)},
  series    = {Undergraduate Texts in Mathematics},
  publisher = {Springer},
  year      = {1994}
  }

@inproceedings{DBLP:conf/fsttcs/MaibaumSV84,
  author       = {T. S. E. Maibaum and
                  M. R. Sadler and
                  Paulo A. S. Veloso},
  editor       = {Mathai Joseph and
                  R. K. Shyamasundar},
  title        = {Logical Specification and Implementation},
  booktitle    = {Foundations of Software Technology and Theoretical Computer Science,
                  Fourth Conference, Bangalore, India, December 13-15, 1984, Proceedings},
  series       = {Lecture Notes in Computer Science},
  volume       = {181},
  pages        = {13--30},
  publisher    = {Springer},
  year         = {1984},
  url          = {https://doi.org/10.1007/3-540-13883-8\_62},
  doi          = {10.1007/3-540-13883-8\_62}
}

@article{DBLP:journals/ipl/Veloso96,
  author       = {Paulo A. S. Veloso},
  title        = {On Pushout Consistency, Modularity and Interpolation for Logical Specifications},
  journal      = {Inf. Process. Lett.},
  volume       = {60},
  number       = {2},
  pages        = {59--66},
  year         = {1996},
  url          = {https://doi.org/10.1016/S0020-0190(96)00146-9},
  doi          = {10.1016/S0020-0190(96)00146-9}
}

@article{DBLP:journals/ipl/VelosoM95,
  author       = {Paulo A. S. Veloso and
                  T. S. E. Maibaum},
  title        = {On the Modularization Theorem for Logical Specifications},
  journal      = {Inf. Process. Lett.},
  volume       = {53},
  number       = {5},
  pages        = {287--293},
  year         = {1995},
  url          = {https://doi.org/10.1016/0020-0190(94)00203-B},
  doi          = {10.1016/0020-0190(94)00203-B}
}
\end{document}